\documentclass{aa}  

\usepackage{natbib}
\usepackage{graphicx}
\usepackage{txfonts}
\usepackage{amsmath}
\usepackage{amssymb}
\usepackage{siunitx}
\usepackage{breqn}
\usepackage{geometry}
\usepackage[colorlinks=true,linkcolor=blue,citecolor=blue,urlcolor=blue,bookmarks=false,hypertexnames=True]{hyperref} 
\usepackage{hyperref}
\usepackage{cleveref}
\usepackage{array,multirow}
\def\h0{$H_0$} 
\def\hunit{\si{km\  s^{-1}\ Mpc^{-1}}}
\def\sne{SNe Ia }
\def\msun{$M_{\odot}$ }
\def\shoes{SH0ES }
\def\st{$s_{BV}$ }

\defcitealias{riess2016}{R16} 
\defcitealias{freedman2001}{F01}
\defcitealias{Udalski1999}{U99}
\begin{document}

    \title{A new measurement of the Hubble constant using Type Ia supernovae calibrated with surface brightness fluctuations}

%   \subtitle{I. Overviewing the $\kappa$-mechanism}

     \author{Nandita Khetan \inst{1,2},
           Luca Izzo \inst{3},
           Marica Branchesi \inst{1,2,4},
           Rados{\l}aw Wojtak\inst{3},
           Michele Cantiello \inst{4},\\
           Chandrashekar Murugeshan \inst{5},
           Adriano Agnello \inst{3},
           Enrico Cappellaro \inst{6},
           Massimo Della Valle \inst{7},
           Christa Gall \inst{3},\\
           Jens Hjorth \inst{3},
           Stefano Benetti \inst{6},
           Enzo Brocato \inst{4,8},
           Jamison Burke \inst{9,10},
           Daichi Hiramatsu \inst{9,10},
           D. Andrew Howell \inst{9,10},\\
           Lina Tomasella \inst{6},
           Stefano Valenti \inst{11}
           }

  \institute{Gran Sasso Science Institute, 
              Viale F. Crispi 7, I-67100 LAquila (AQ), Italy
              \and
              INFN, Laboratori Nazionali del Gran Sasso, I-67100 Assergi, Italy
              \and
              DARK, Niels Bohr Institute, University of Copenhagen, Jagtvej 128, 2200 Copenhagen, Denmark
              \and
              INAF, Osservatorio Astronomico d’Abruzzo, Via Mentore Maggini, Teramo, I-64100, Italy
              \and 
              Centre for Astrophysics and Supercomputing, Swinburne University of Technology, Hawthorn, Victoria 3122, Australia
              \and
              INAF, Osservatorio Astronomico di Padova, Vicolo dell'Osservatorio 5, 35122- Padova, Italy
              \and
              INAF, Osservatorio Astronomico di Capodimonte, Salita Moiariello 16, 80131- Naples, Italy
              \and
              INAF, Osservatorio Astronomico di Roma, Via Frascati 33, 00040, Monte Porzio Catone (RM), Italy
              \and
              Las Cumbres Observatory, Goleta, California 93117, USA
              \and
              Department of Physics, University of California, Santa Barbara, California 93106, USA
              \and
              Department of Physics, University of California, 1 Shields Avenue, Davis, CA 95616-5270, USA\\ 
              \email{nandita.khetan@gssi.it}
             }
\authorrunning{N. Khetan et al.}
\titlerunning{SNe Ia calibration with SBF}
%\titlerunning{$H_0$ measurement using SNe Ia calibrated with SBF}
%   \date{Received September 15, 1996; accepted March 16, 1997}

% \abstract{}{}{}{}{} 
% 5 {} token are mandatory
 
   \abstract
  % {To calibrate SNe Ia with SBF}
  % {Calibrate SNe Ia with SBF and measure H0}
 %  {Tripp calibration. We present a new and independent method of determining the local value of the Hubble constant based on the use of Surface Brightness Fluctuations (SBF) to calibrate the Supernovae Type Ia (SNe Ia). We find a value of H0 XXX $\pm$ XXX. This value agrees at XX sigma with CMB and XX sigma with Riess. SBF is blah blah -- it comlements the cepheids in terms of host type...host mass matching the avg host mass of HUbble flow sample}
  % {results here}
  % {conclusion here}
{We present a new calibration of the peak absolute magnitude of Type Ia supernovae (SNe Ia) based on the surface brightness fluctuations (SBF) method, aimed at measuring the value of the Hubble constant. We build a sample of calibrating anchors consisting of 24 SNe hosted in galaxies that have SBF distance measurements. Applying a hierarchical Bayesian approach, we calibrate the SN Ia peak luminosity and extend the Hubble diagram into the Hubble flow by using a sample of 96 \sne in the redshift range $0.02 < z < 0.075$, which was extracted from the Combined Pantheon Sample. We estimate a value of $H_0 = 70.50 \pm 2.37\textrm{(stat)} \pm 3.38\textrm{(sys)}$ \hunit\ (i.e., 3.4\% stat, 4.8\% sys), 
which is in agreement with the value obtained using the tip of the red giant branch calibration. It is also consistent, within errors, with the value obtained from SNe Ia calibrated with Cepheids or the value inferred from the analysis of the cosmic microwave background. We find that the \sne distance moduli calibrated with SBF are on average larger by  0.07 mag than those calibrated with Cepheids. Our results point to possible differences among SNe in different types of galaxies, which could originate from different local environments and/or progenitor properties of SNe Ia. Sampling different host galaxy types, SBF offers a complementary approach to using Cepheids, which is important in addressing possible systematics.
As the SBF method has the ability to reach larger distances than Cepheids, the impending entry of the Vera C. Rubin Observatory and JWST into operation will increase the number of \sne hosted in galaxies where SBF distances can be measured, making SBF measurements attractive for improving the calibration of SNe Ia, as well as in the estimation of \h0.}


   \maketitle
%
%-------------------------------------------------------------------

%%%%%%%%%%%%%%%%% Add relevant sections here %%%%%%%%%%%%%%%%%%
\section{Introduction}
\label{sec:intro1.1}

The standard cosmological model, also known as the $\Lambda$ cold dark matter ($\Lambda$CDM) model, represents the only model that is consistent with a wide set of observations from different epochs of the Universe. This concordance model describes our Universe as flat, accelerating, and primarily composed of radiation, baryons, dark matter, and dark energy, with the latter two components being the most dominant, albeit elusive, constituents at the present time. One of the fundamental parameters governing the $\Lambda$CDM model is the Hubble constant (\h0), which sets the absolute scale of the Universe and can be measured both at early epochs, by the size of the sound horizon from the cosmic microwave background (CMB, \citealt{planck2015, bennett2013}), and in the local present-time Universe using luminosity distance indicators \citep{freedman2001,Sandage2006, riess2016}. Comparing the absolute scale at the two opposite ends of the expanding Universe provides a stringent test of the standard cosmological paradigm. \par

With extensive ongoing efforts, \h0 measurements are achieving remarkable accuracy and precision at both of these extremes. In the local Universe, one of the most reliable measurements of \h0 comes from supernovae type Ia (SNe Ia), which however rely on primary distance indicators for their zero-point calibration (e.g., Cepheids and geometrical distances). The last couple of decades have witnessed gradual improvements in \h0 measurements using \sne \citep[e.g.,][]{freedman2009, freedman2012, riess2011, riess2016, dhawan2018, phillips2019}, with the most recent estimate by \cite{riess2019}, who obtained $ H_0 = 74.03 \pm 1.42$ \hunit\ using the Cepheid calibration of \sne (\shoes program). Other powerful astrophysical probes measuring \h0 include time-delay gravitational lensing \citep[e.g.,][]{suyu2017, birrer2019}, 
the Tully-Fisher relation \citep[e.g.,][]{sorce2013}, surface brightness fluctuations \citep[SBF; e.g.,][]{Jensen2001,Cantiello2018}, and the distance measurement using gravitational wave signals from binary compact objects \citep{Abbott2017}, to name a few.

The latest estimate of the Hubble constant based on CMB observations by the Planck satellite is $H_0 = 67.4 \pm 0.5$ \hunit\ \citep{planck2018}. Another way of estimating \h0 is through measurements of fluctuations in the matter density called baryon acoustic oscillations \citep[BAOs,][]{cole2005, eisenstein2005, aubourg2015, alam2017}. The absolute calibration of BAOs is based on prior knowledge of the sound horizon size, which depends on the early-time physics, in turn making it dependent on the CMB. \citet{macaulay2018} measured a value of $ H_0 = 67.77 \pm 1.30$ \hunit\ using BAOs and \sne from the Dark Energy Survey (DES), where the absolute distance measurements from the BAOs were used to calibrate the intrinsic magnitude of the SNe Ia. This "inverse" distance ladder approach, where the distance calibration is done through CMB or other high-redshift observations, is not a direct method; it requires a cosmological model to build on.

The majority of the local independent methods, and combinations thereof, used to estimate \h0 stand in tension with the \h0 values inferred from the CMB analysis, with discrepancies ranging between $4\sigma$ and $6\sigma$ \citep{verde2019}. However, a recent calibration of \sne using the tip of the red giant branch (TRGB) method in the color-magnitude diagram of host galaxies of \sne provided $H_0 = 69.8 \pm 0.8 (\pm1.1\%\ \rm stat) \pm 1.7 (\pm2.4\%\ sys)$ \hunit\ \citep{trgb}. Using a different calibration of the TRGB in the Large Magellanic Cloud (LMC), \cite{Yuan2019} estimated $H_0 = 72.4 \pm 2.0$ \hunit. After a reanalysis of the LMC TRGB extinction, \cite{Freedman2020} confirmed their earlier estimate of $H_0 = 69.6 \pm 0.8 (\pm1.1\%\ \rm stat) \pm 1.7 (\pm2.4\%\ sys)$ \hunit, which sits between the Planck-CMB value and the one resulting from \sne calibrated using Cepheids.

If the difference between local \h0 measurements and the Planck-CMB measurement is statistically confirmed by future independent observations and analyses, it would hint at a possible inadequacy of the standard  $\Lambda$CDM model and in turn imply the existence of some "new physics" beyond it, which would include new  species of relativistic particles, nonzero curvature, dark radiation, or even a modification of the equations of general relativity \citep[e.g.,][]{bernal2016,morstell2018,verde2019,Knox2020}. However, many modifications of the $\Lambda$CDM model appear in conflict with other existing cosmological tests and worsen the model fit to the observed CMB power spectrum \citep{Arendse2020,Hill2020}. Neither new physics nor identifiable systematics are currently able to resolve the tension. In this scenario, new and independent \h0 estimates and a credible quantification of the systematic uncertainties (instrumental and astrophysical) are necessary. 

Among Hubble flow distance indicators, \sne are the most reliable probes for \h0 measurement. In the cosmic distance ladder approach, SN Ia distances generally rely on some primary distance measurements in the nearby Universe, such as Cepheids \citep{riess2019} or the aforementioned TRGB. The ladder approach to estimating \h0 consists of three main steps: (1) absolute calibration of the primary distance indicator with geometric anchors, for example using parallaxes for Milky Way Cepheids \citep{GAIA2018} and/or for LMC Cepheids, detached eclipsing binaries \citep[DEBs;][]{Pietrzyski2019}, and masers \citep{riess2016}; (2) calibrating the luminosity of nearby \sne using the distance from the primary indicator to host galaxies of SNe Ia, and (3) using the calibrated relation between SN Ia light curve properties and luminosity to measure distances to \sne in the Hubble flow. Therefore, to obtain accurate distances and \h0 estimates, it is necessary to control the various systematic errors arising in each of the above steps in order to gather a statistically significant sample of galaxies that host \sne in the local Universe to be used as calibrators, as well as to have accurate distance estimates to the galaxies of this calibrating sample via primary distance methods. \par
Presently, the ``yardstick" measurement, which highlights the Hubble tension in finding a higher \h0 with respect to early-Universe estimates, is based on the calibration that uses the Cepheid distance scale, in particular with Weisenheit magnitudes \citep{Madore1991}. This calibration currently relies on a sample of 19 nearby galaxies that host \sne and whose distances are measured with Cepheids (\shoes sample, Supernovae \h0 for the Equation of State of Dark energy, \citealp{riess2005}, \citealp{riess2019}). On the other hand, the \h0 estimate based on a sample of 18 SNe Ia whose distances to their host galaxies have been measured with the  TRGB method shows a lower value of the Hubble constant, which is in agreement at the $1.2 \sigma$ level with the Planck estimate \citep{Freedman2020}. In the wake of these results, it is imperative to exploit different methods to estimate precise distances  in the local Universe in order to confirm or resolve the Hubble tension. 

In this context, this work aims at exploring the use of the SBF distance method as an anchor for measuring distances to host galaxies of \sne in the Hubble flow. \cite{ajhar} compared SBF distances of galaxies hosting a SN Ia with their SN distance calibrated using Cepheids and homogenized the SBF and SN distance scales. Based on this result, we propose, for the first time, to calibrate SN Ia luminosity using SBF distances to their host galaxies, with the main goal of estimating the Hubble constant.

The possibility to measure accurate distances to early-type galaxies (and sometimes bulges of spiral galaxies) in the nearby Universe with SBF was first introduced by \citet{tonry1998}. Detailed descriptions about the SBF methods are given in \citet{blakeslee1999, biscardi2008, blakeslee2009}. The SBF method determines the intrinsic variance in a galaxy image resulting from stochastic variations in the numbers and luminosities of the stars falling within the individual pixels of the image. The measured variance is normalized by the local galaxy surface brightness and then converted to the apparent SBF magnitude $\overline{m}$. The distance modulus, $\mu = \overline{m}-\overline{M}$, is then obtained knowing the absolute SBF magnitude ${\overline{M}}$, which depends on the stellar population properties \citep{blakeslee2001,mei2005,MC18}. The $\overline{M}$ zero-point is tied directly to the Cepheid distance scale \citep[empirical calibration, e.g.,][]{tonry} or derived from stellar population models \citep[theoretical calibration, e.g.,][]{Brocato1998,cantiello2003,raimondo2005,martin2006,biscardi2008}. The stellar populations dominated by evolved stars (with the red giant branch mostly contributing to the flux variance) make early-type galaxies ideal for estimating distances through SBF measurements \citep{blakeslee2012}. The SBF technique enables us to measure distances with a precision of $5-10\%$ up to $\sim$ 100 Mpc with the current observatories, such as the Hubble Space Telescope \citep{jensen2015}.
\par

Although the SBF distances are calibrated themselves using Cepheids, and represent a secondary calibrator method for SNe Ia, they offer potential advantages and useful insights in terms of identifying possible systematic effects associated with the luminosity calibration. While SNe calibrated with Cepheids, such as the ones in the \shoes sample, are all hosted in late-type galaxies, the SBF distance measurements are available mainly for early-type host galaxies, making SBF calibrator sample complementary, in terms of SN hosts, to the Cepheid sample. The comparison among the SBF and Cepheid calibrations enables us to identify possible systematics for luminosities of SNe Ia in different host galaxy types \citep{Kang2019}. Furthermore, early-type galaxies have generally less dust when compared with late-type galaxies, and hence the host extinction, which remains a challenge for SN light curve analysis \citep{tripp,burns2014,Brout2020}, is reduced. 
The distance range covered by SBF measurements significantly exceeds the one covered by the TRGB and Cepheid measurements, which helps to augment the number of calibrators. In terms of observational advantages, SBF can be measured on images that do not require the high resolution and depth necessary to resolve stellar photometry as for TRGB and Cepheids. Furthermore, this method does not require periodic observations of the galaxies. \par

The paper is organized as follows. In Section \ref{sec2-data}, we describe the nearby SN samples used as calibrators and the distant cosmological sample used for \h0 measurement. In Section~\ref{sec3-method}, we explain the details of our analysis, and in Section \ref{sec4-results}, we present the Hubble diagram and our estimates of the Hubble constant. In Section~\ref{sec5-masscorrection}, we  evaluate the influence of the host galaxy type on the SN Ia standardization by applying a galaxy stellar-mass correction. Section~\ref{sec6-comparison} compares the distance measurements obtained by applying the SBF and Cepheid calibrations. We discuss our results in Section \ref{sec7-discussion}, and we draw our conclusions and discuss future prospects in Section~\ref{sec8-conclusion}.

%%%%%%%%%%%%%%%%%%%%%%%%%%%%%%%%%%%%%%%%%%%%%%%%%%%%%%%%%%%%%%%%%%%%%%%%%%%%%%%%%%%%%%%%%%%%%%%%%

\section{Data}
\label{sec2-data}

The choice and number of calibrators are key factors in defining the zero-point of the calibrating relation and eventually the accuracy of Hubble flow distance estimates, given that they determine the uncertainty on the \h0 value. In order to appropriately calibrate the peak luminosity of \sne with SBF distance indicators, we first identified all the galaxies that host a SN Ia and have a distance measurement evaluated through the SBF technique. Then, we filtered this sample according to specific SN Ia data quality criteria, which are reported below.

\subsection{SBF calibration sample}
\label{sec:SBF sample}

In order to build the SBF calibration sample, we cross-matched the major published SBF distance catalogs
\citep{tonry,ajhar,jensen,mei2003,Cantiello07,blakeslee09,cantiello2013,MC18} with the SN catalog from \cite{snespace} available on the Open Supernova Catalog webpage\footnote{\url{https://sne.space/}}. 
The preliminary cross-matched sample consists of 45 galaxies. For the galaxies in this preliminary sample that have multiple SBF distance estimates, we selected the most recent SBF distance estimate, favoring the use of the Hubble Space Telescope (HST) data when available. 

Taking into account the importance of the quality of the data of SN light curves (LC) for a good calibration, we applied LC quality cuts in order to avoid any systematics caused by their observed properties. Our fiducial calibration sample consists of \sne with: (1) data in B and V bands with high cadence observations especially around the maximum and within the first 40 days after the peak brightness, in order to accurately sample the LC evolution and constrain the peak magnitude; (2) regular light curve shape (removing fast decliners with $s_{BV} < 0.5$, see Section~\ref{sec3-method} for the definition of the color-stretch parameter $s_{BV}$); and (3) low reddening (color $m_B - m_V < 0.3$ mag). Throughout this paper (except where differently indicated), $m_B - m_V$  refers to the pseudocolor derived from the maximum flux in the $B$ and the $V$ bands. Among the 45 \sne in the preliminary sample, 24 of them pass the above selection criteria and form our final SBF calibration sample.

For the photometry of the SNe in our calibration sample, the optical ($B$ and $V$ band) light curves are taken from the published data assembled in \cite{snespace}. The individual references for the photometric data of each object are given in Table \ref{tab:LCref} of Appendix \ref{app:LCfitting}. Table \ref{tab:tabel1} shows the SBF calibration sample listing the 24 \sne selected to have high quality photometric data and standard light curve evolution. It lists the SN name ({\it column 1}), the host galaxy name ({\it column 2}),  the SBF distance modulus along with the associated error ({\it columns 3 and 4}), the galaxy morphological type ({\it column 5}), the reference for the galaxy SBF distance ({\it column 6}) and the host galaxy stellar mass along with the error ({\it column 7 and 8}). The stellar mass for each SN host galaxy used in this work is computed using the Ks-band magnitude from the 2MASS survey \citep{2mass} as described in Appendix \ref{app:hostmass}.

\begin{table*}
\centering
\caption{Calibration sample of \sne hosted in galaxies that have SBF distance modulus measurements. The 24 \sne listed here form the SBF calibration sample used in this work. 
}
\label{tab:tabel1}
% table1.tex created at UT 2019-11-06 17:35:31
% A table with adjusted row and column spacings
% \setlength sets the horizontal (column) spacing
% \arraystretch sets the vertical (row) spacing
\begingroup
\setlength{\tabcolsep}{9pt} % Default value: 6pt
\renewcommand{\arraystretch}{1.2} % Default value: 1
\begin{tabular}{c c c c c c c c}
\hline\hline
Supernova & Host Galaxy & $\mu_{SBF}$ & $\sigma_{SBF}$ & Morpholgy & Distance Reference & $\log M_{\ast}$ & $\sigma_{\log M_{\ast}}$ \\
  & & (mag) & (mag) & & & (\msun) & (\msun) \\
\hline
%SN2012fr   & NGC1365  &  31.510 &   0.030 & SB(s)b       & \cite{blakeslee2009} &       11.022 &      0.048 \\
SN2000cx   & NGC524   &  31.921 &   0.212 & SA0(rs)      & \cite{tonry} &       10.929 &      0.090 \\
%SN1991T    & NGC4527  &  31.09 &   0.05 & SAB(s)bc     & \cite{blakeslee2009} &        10.488 &      0.042 \\
SN1994D    & NGC4526  &  31.320 &   0.120 & SAB0(s)      & \cite{MC18} &       10.996 &      0.055 \\
SN2007on   & NGC1404  &  31.526 &   0.072 & E1           & \cite{blakeslee2009} & 10.932 &      0.035 \\
SN2012cg   & NGC4424  &  31.020 &   0.180 & SB(s)a       & \cite{MC18} &        9.706 &      0.083 \\
SN1980N    & NGC1316  &  31.590 &   0.050 & SAB0(s)pec    & \cite{cantiello2013} & 11.514 &      0.032 \\
%SN1981B    & NGC4536  &  31.090 &   0.050 & SAB(rs)bc    & \cite{blakeslee2009} &       10.349 &      0.025 \\
SN2003hv   & NGC1201  &  31.566 &   0.304 & SA0(r)       & \cite{tonry} &       10.565 &      0.064 \\
SN2008Q    & NGC524   &  31.921 &   0.212 & SA0(rs)      & \cite{tonry} &       10.929 &      0.090 \\
SN1970J    & NGC7619  &  33.582 &   0.151 & E            & \cite{mei2003} &       11.340 &      0.073 \\
SN1983G    & NGC4753  &  31.919 &   0.197 & I0           & \cite{tonry} &       11.148 &      0.064 \\
SN2014bv   & NGC4386  &  32.190 &   0.494 & SAB0         & \cite{tonry} &       10.480 &      0.064 \\
SN2015bp   & NGC5839  &  31.737 &   0.314 & SAB0(rs)     & \cite{tonry} &       9.979 &      0.137 \\
SN2016coj  & NGC4125  &  31.922 &   0.258 & E6 pec       & \cite{tonry} &       11.083 &      0.064 \\
SN1981D    & NGC1316  &  31.590 &   0.050 & SAB0(s)pec    & \cite{cantiello2013} & 11.514 &      0.032 \\
SN1992A    & NGC1380  &  31.632 &   0.075 & SA0          & \cite{blakeslee2009} & 10.931 &      0.032 \\
SN2018aoz  & NGC3923  &  31.795 &   0.101 & E4-5         & \cite{Cantiello07} &       11.204 &      0.065 \\
SN2011iv   & NGC1404  &  31.526 &   0.072 & E1           & \cite{blakeslee2009} & 10.932 &      0.035 \\
SN2006dd   & NGC1316  &  31.590 &   0.050 & SAB(s)pec    & \cite{cantiello2013} & 11.514 &      0.032 \\
SN1992bo   & E352-057 &  34.270 &   0.150 & SB0(s)pec    & \cite{ajhar} &       10.395 &      0.071 \\
SN1997E    & NGC2258  &  33.500 &   0.150 & SA0(r)       & \cite{ajhar} &       11.199 &      0.069 \\
SN1995D    & NGC2962  &  32.600 &   0.150 & SAB0(rs)     & \cite{ajhar} &       10.597 &      0.069 \\
SN1996X    & NGC5061  &  32.260 &   0.190 & E0           & \cite{ajhar} &       11.057 &      0.086 \\
SN1998bp   & NGC6495  &  33.100 &   0.150 & E            & \cite{ajhar} &       10.462 &      0.069 \\
SN2017fgc  & NGC474  &  32.536 &   0.133 & SA0(s)       & \cite{Cantiello07} &       10.568 &      0.061 \\
SN2020ue  & NGC4636  &  30.830  &   0.130   & E0           & \cite{tonry} &      10.803    &   0.061 \\
\hline
\end{tabular}
\endgroup
\end{table*}

All the SBF distances used in the present paper are tied to a common empirical zero-point based on the results of the \textit{HST} Key Project (KP) Cepheid distances by \cite{freedman2001} [hereafter, \citetalias{freedman2001}] as described in \cite{Blakeslee2002}. \citetalias{freedman2001} adopt the Cepheid Period--Luminosity ($P$--$L$) relations by \cite{Udalski1999}, and the metallicity corrections to the Cepheid distances by \cite{Kennicutt1998}. The SBF zero-point calibration adopted by \cite{tonry} was based on a previous estimate evaluated using six galaxies with KP Cepheid distances~\citep{ferrarese2000} and needed a revision. A general correction formula for the published distances in \cite{tonry} is provided by \cite{blakeslee2010}. This formula includes both the zero-point and second-order bias correction, which takes into account the variation of the data quality. All the distances from \cite{tonry} in this work are corrected according to this formula bringing them to the same zero-point of the other SBF measurements in the present sample.

It is worth noting that our SBF distances are calibrated to the Cepheid zero-point based on Cepheid distances by \citetalias{freedman2001}, who adopted an LMC distance modulus of $ 18.50 \pm 0.10$ mag. \cite{riess2016} for the Cepheid distances of the SNe Ia in the \shoes sample (described in the next subsection) used a LMC distance of $18.493 \pm 0.008 (stat) \pm 0.047 (sys)$ mag based on 8 DEBs \citep{Pietrzyski2013}. The most recent value by \cite{Pietrzyski2019}, which is anchored on 20 DEB stars in the LMC, gives $\mu_{LMC} = 18.477 \pm 0.004 (stat) \pm 0.026 (sys)$ mag and has been adopted by \cite{riess2019} for the Cepheid distances and by \cite{trgb} for the TRGB estimates.

The statistical errors on SBF distances for all the objects listed in \ref{tab:tabel1} are taken as reported in the corresponding papers. The uncertainties include the propagation of the errors on both the intercept and the slope of the SBF calibration reported in Eq.\ 1 of \cite{tonry}. The systematic error (not included in the SBF distance error) has been estimated including the uncertainty in the tie of the SBF distances to the Cepheid distance scale, which is conservatively evaluated to be 0.1 mag \citep[see e.g.,][]{freedman2010,Cantiello2018}. This error is dominated by the uncertainty on the LMC distance \citep{tonry}.

\subsection{\textit{SH0ES} calibration sample}

In order to compare the SN distances and the \h0 estimated using the SBF calibration with those estimated from Cepheid calibration, we also take the \shoes sample of 19 galaxies from \citet{riess2016,riess2019} as a second calibrator set. These galaxies host a SN Ia and have their distances estimated using Cepheid variable stars. Their distance moduli (and associated uncertainties) are taken from Table~5 of \citet{riess2016} [hereafter, \citetalias{riess2016}], and the photometric data of the \sne  are retrieved from the Open Supernova Catalog.

The Cepheid distances taken from \citetalias{riess2016} are calibrated using the near-infrared (NIR) and optical Cepheid $P$--$L$ relations. Using only the optical relation they find $H_0 = 71.56 \pm 2.49$ \hunit, which is $\sim$2\% smaller than the NIR-based estimate of $H_0 = 73.24 \pm 1.74$ \hunit. The optical Cepheid $P$--$L$ relations used by \citetalias{riess2016} are described in \citet{Hoffmann2016}, where it is also shown that their optical $P$--$L$ relations are in very good agreement with the $P$--$L$ relations of \citet{Udalski1999}, adopted in \citetalias{freedman2001}. Moreover, three of the 19 \shoes calibrators are also present in \citetalias{freedman2001}, namely NGC 1365, NGC 4536 and NGC 4639. For these three galaxies, the differences between the distance modulus from \citetalias{riess2016} and the metallicity corrected one from \citetalias{freedman2001} are 0.04, 0.04, and $-0.18$ mag, respectively. Although it is not a statistically significant comparison, they show no systematic offset.

Multiple SN analyses show correlations between luminosity of the SN Ia and the host galaxy mass \citep[e.g.,][]{kelly2010,lampeitl2010,sullivan2010,gupta2011}, age, metallicity and star formation rate \citep[SFR; e.g.,][]{hayden2013,rigault2013,rigault2015,roman2018}.
Thus, the SN luminosity dependence on host galaxy properties is another important ingredient  that should be taken into account for any SN studies. Differences between the properties of the host galaxies of the calibrating sample and that of the Hubble-flow sample may introduce systematic errors in the value of \h0 \citep[see e.g.,][]{trgb}. Different host environments can also affect the extinction suffered by the SN luminosity, which is another issue to be addressed in SN Ia cosmology \citep{Brout2020}. \par

\begin{figure}
\centering
\hspace*{-0.5cm}
\includegraphics[width=8.0cm,height=13.0cm]{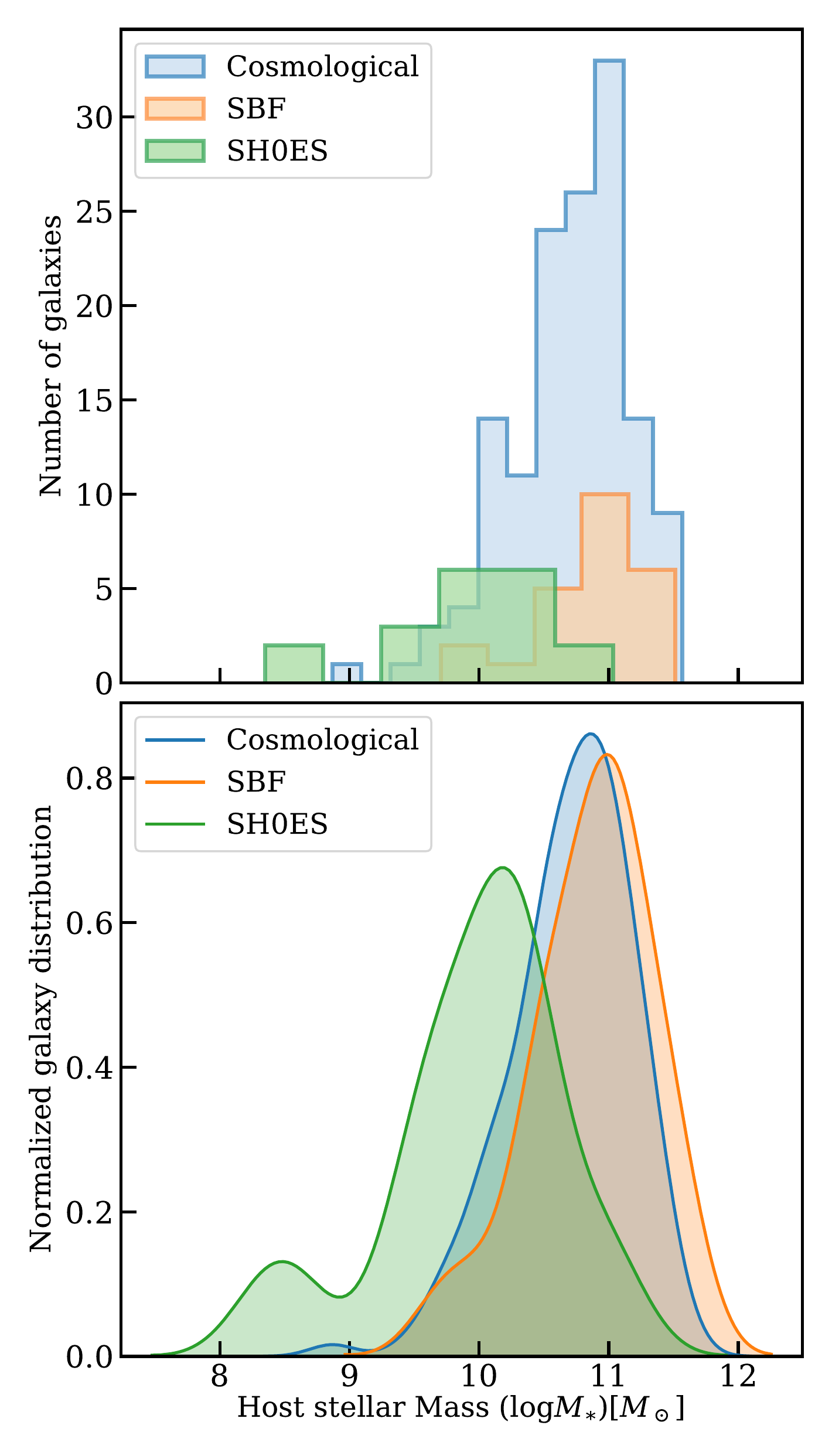}
\caption{Number of galaxies (top panel) and the normalized density distribution (bottom panel) as a function of the host galaxy stellar mass for the SBF and the \shoes calibration samples. We also plot the host stellar mass distribution for the cosmological sample. The KS-test P value for the two calibration samples gives $6.03\times10^{-6}$, indicating that the stellar mass distributions of the host galaxies of \sne are different for the two samples.}
\label{fig:mass_dist}
\end{figure}

The \shoes sample is mainly composed of late-type spiral galaxies as Cepheids are relatively young stars. On the other hand, the SBF sample is biased toward early-type galaxies (92$\%$ of the SBF sample) dominated by old stellar populations. Figure~\ref{fig:mass_dist} shows the histograms (top panel) and the density distributions (bottom panel) of the stellar mass of the host galaxies belonging to the SBF and \shoes calibration samples. The \shoes sample has a substantially lower mean galaxy stellar-mass (log stellar masses in units of $M_{\odot}$) of 9.97 than the SBF sample (10.87). We note that only one SN, namely SN2012cg hosted in  NGC 4424, is common among the two calibration samples. For comparison, Figure~\ref{fig:mass_dist} also shows the distribution of the host-galaxy stellar masses for the cosmological sample described in the next section.

\subsection{Cosmological sample}
\label{sec2.3-cosmosample}
For measuring the Hubble constant, we build a statistically significant sample of \sne extending into the Hubble flow. We extract our cosmological sample from the Combined Pantheon Sample \citep{pantheon}, which consists of 1048 spectroscopically confirmed \sne coming from various local and high-redshift supernova surveys. All the SNe are cross-calibrated with the Pan-STARRS (PS1) survey in order to have a common photometric calibration \citep{scolnic2015}. We select all the \sne spanning a redshift range of  $0.009 < z < 0.075$, with good quality photometric data to appropriately sample the LC, and with 2MASS Ks-band magnitude to evaluate the galaxy stellar-mass. We also exclude fast decliners and very red \sne from the sample (as described in Section~\ref{sec3-method}). This gives us a sample of 140 \sne, referred to as the full cosmological sample hereafter. \par

The main contributions to the data sample come from the Harvard Smithsonian Center for Astrophysics CfA1-CfA4 \citep{riess1999,jha2006,hicken2009a, hicken2009b,hicken2012} and the Carnegie Supernova Project \citep[CSP,][]{contreras2010,folatelli2010,stritzinger2011} survey. The optical photometric data of the SNe of the cosmological sample are assembled using data stored in a dedicated repository\footnote{\url{http://snana.uchicago.edu/downloads.php}\\}, and are analyzed in the same way as the data of the \sne belonging to the calibration samples.

Since the lower redshift range of this cosmological sample starts with $z = 0.009$, where peculiar velocities can have a significant impact on the recessional velocities of the galaxies, we apply a more stringent redshift cut removing all galaxies below $z = 0.02$ (similar to \citetalias{riess2016}) in order to mitigate the contamination from peculiar velocities. This sample cut leaves 96 SNe in the cosmological set and is referred to as the redshift-cut cosmological sample throughout this work. Our main results will be based on the use of the redshift-cut cosmological sample, although we will also estimate \h0 using the full cosmological sample for comparison with studies such as \cite{trgb}.

To summarize, this work has two calibration sets: the SBF sample, which is our main sample consisting of 24 SNe Ia, and the \shoes sample from \citetalias{riess2016} that has 19 SNe Ia. The calibrations derived from these two are applied to the full cosmological sample consisting of 140 \sne and its subsample of 96 \sne with the redshift cut.

%%%%%%%%%%%%%%%%%%%%%%%%%%%%%%%%%%%%%%%%%%%%%%%%%%%%%%%%%%%%%%%%%%%%%%%%%%%%%%%%%%%%%%%%%%%%%%%%%

\section{Analysis}
\label{sec3-method}
While the exact nature of SN Ia progenitors remains uncertain, the regularity of their observed properties enables us to use them for measuring precise distances. This relies on the empirical evidence that their intrinsic luminosity is correlated with the rate of decline of their light curves and hence they can be standardized \citep{phillipsM,Riess1996,perlmutter1997}. Considering this and further implementing correction terms that take into account the absolute luminosity dependence on the color and the host galaxy, \sne can be used as standard candles for cosmological studies. 

\subsection{Lightcurve fitting}
\label{LCfit}
In order to evaluate the SN luminosity from the observable LC properties, we fit the LCs of all the \sne in this study belonging to the two calibration samples and the cosmological sample. We estimate their apparent magnitudes at maximum in the B and V bands, along with the LC shape. For this, we use the \texttt{SNooPy} (SuperNovae in Object Oriented Python) LC fitter \citep{snoopy}. \texttt{SNooPy} corrects the photometry data for Milky Way extinction using the dust maps of \citet{mwextinction}, and applies the K-corrections that are computed using the SED template sequence developed by \cite{hsiao}. We fit the LC using the ``max-model" method\footnote{For details see the online documentation of SNooPy:\href{https://csp.obs.carnegiescience.edu/data/snpy/documentation}{https://csp.obs.carnegiescience.edu/data/snpy/documentation}}, which gives us the epoch and the magnitude (for each filter) of the LC maximum, and the LC shape parameter $s_{BV}$, called the ``color-stretch" parameter \citep{burns2014}. For five SNe Ia in our calibration sample, pre-maximum LC observations are missing. However while fitting the LCs using SN templates, \texttt{SNooPy} is able to reproduce the LC shape at epochs where there are no data as described in \cite{snoopy}. Provided that the peak is well sampled, it ensures reliable modeling of the SNe without pre-max data. \par
\begin{figure}
\centering
\hspace*{-0.5cm}
\includegraphics[width=8.0cm,height=13.5cm]{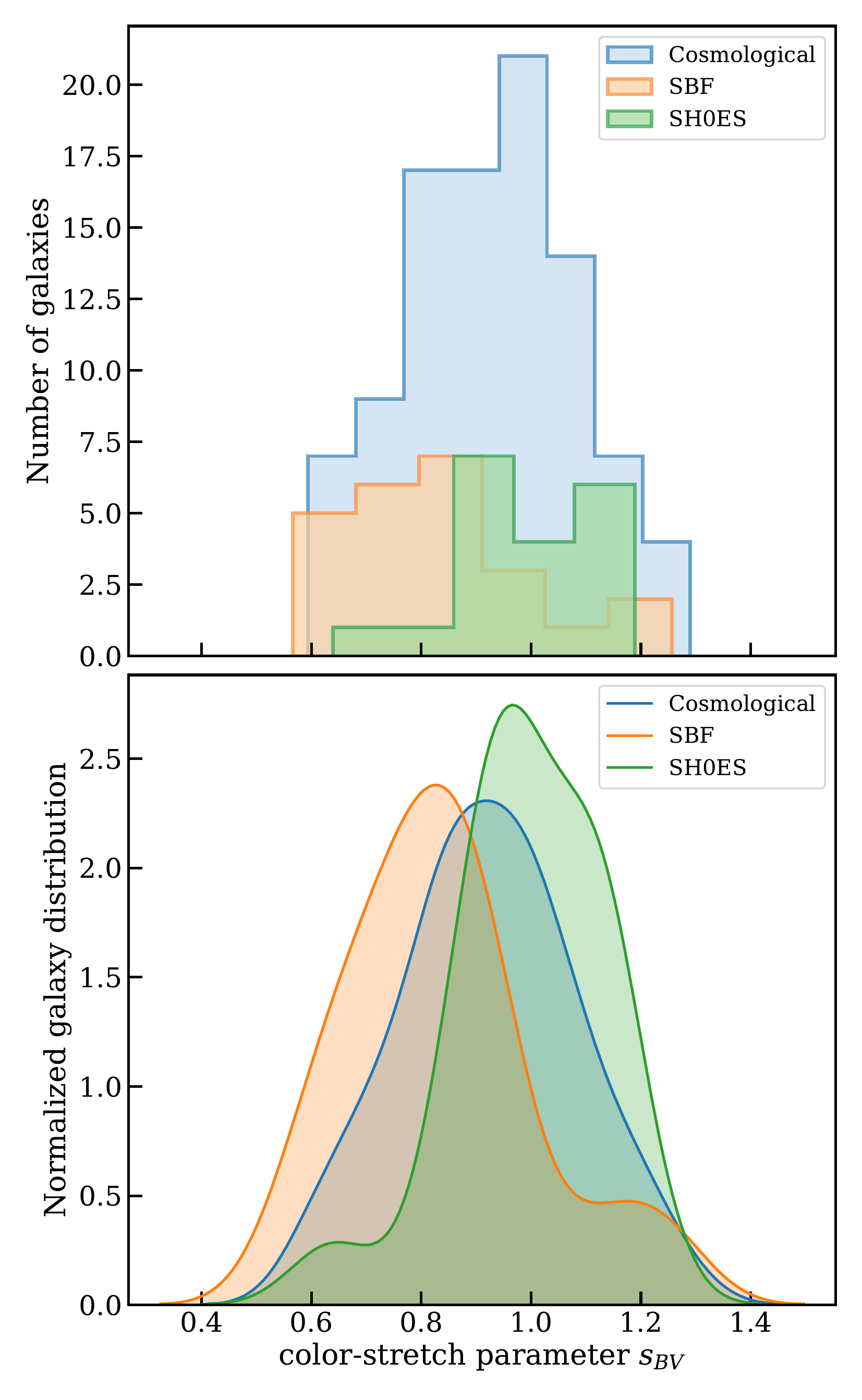}
\caption{Number of galaxies (top panel) and the normalized density distribution (bottom panel) of the \st parameter values of SNe Ia of the three data samples used in this work: the SBF calibration sample, the \shoes calibration samples, and the cosmological sample.}
\label{fig:sbv_dist}
\end{figure}

The color-stretch parameter takes into account the color ($m_B-m_V$) evolution of the \sne and is calculated by getting the time between B maximum and ($m_B-m_V$) maximum (the typical value for which is 30 days) and dividing this time by 30. Here, $m_B-m_V$ refers to real color evolution of the SNe Ia, and it is different from the pseudo-color mentioned before. We use $s_{BV}$ as the LC shape parameter, instead of the more commonly used $\Delta m_{15}$ (magnitude decline between the LC maximum and 15 days later) because $s_{BV}$ properly captures the behavior of fast decliners and it is appropriately sensitive to the extinction \citep{burns2014,burns2018}. For ``normal" \sne \st is about 1, while for fast decliners \st is typically smaller than 0.5. We exclude from our fiducial calibration sample fast decliner \sne (\st $<$ 0.5) and highly reddened \sne (pseudo-color $m_B - m_V > 0.3$ mag). Figure \ref{fig:sbv_dist} shows the histograms (top panel) and the density distributions (bottom panel) of the \st values of the \sne belonging to the two calibration samples. The two calibration samples show a different distribution of \st values while the cosmological sample \st distribution lies between the two. The calibration sample contains two "transitional" objects, namely SN 2007on and SN 2011iv both located in NGC 1404, characterised by peak magnitudes similar to normal SNe Ia but with a relatively faster rate of decline \citep{gall2018}. Since the use of the color-stretch parameterization in their LC fitting should ensure reliable modeling of their evolution, we include them in our fiducial calibration sample maintaining the threshold \st $>$ 0.5 as done in previous works, such as \citet{trgb} and \citet{burns2018}. However, we also perform the analysis excluding them from the sample to investigate their influence on the final results (see Section~\ref{sec4-results}). \par

Table \ref{tab:tabel2} lists the LC parameters for the SBF calibration sample: the maximum brightness in B and V bands along with the errors ({\it Columns 2--5}), the \st parameter and relative error ({\it Columns 6 and 7}), the color ($m_B-m_V$) at maximum ({\it Column 8}) and the absolute magnitude in B band ($M_B$) calculated as $m_B - \mu_{SBF}$ ({\it Column 9}), where $\mu_{SBF}$ is taken from Table~\ref{tab:tabel1}. The B-band light curve fits of the SNe in the SBF sample are shown in Appendix \ref{app:LCfitting} in Table \ref{tab:lcfits}. We analyze the LCs of the SNe of the \shoes calibration sample in the same way. The estimated LC parameters of the \shoes SNe are given in Table \ref{tab:tabel3}. The LC parameters of SNe in the cosmological sample are also obtained following the same fitting analysis.

\begin{table*}[h!]
\centering
\caption{Best-fit lightcurve parameters of the \sne of the SBF sample estimated using \texttt{SNooPy}: $m_B$ and $m_V$ ({\it columns 2 and 4}) are the apparent magnitudes at maximum in the B and V bands, and $s_{BV}$ ({\it column 6}) is the stretch color parameter ({\it column 7}). The color  $m_B-m_V$ ({\it column 8}) is computed as the difference between the maximum brightness in B and V bands, and $M_B$ ({\it column 9}) is the absolute magnitude in B band, computed as $m_B - \mu_{SBF}$, where $\mu_{SBF}$ is taken from Table~\ref{tab:tabel1}.}
\label{tab:tabel2}
% table2.tex created at UT 2019-11-07 17:52:32
\begingroup
\centering
\setlength{\tabcolsep}{14pt} % Default value: 6pt
\renewcommand{\arraystretch}{1.2} % Default value: 1
\begin{tabular}{c c c c c c c c c}
\hline\hline
Supernova & $m_B$ & $\sigma_{m_B}$ & $m_V$ & $\sigma_{m_V}$ & $s_{BV}$ & $\sigma_{s_{BV}}$ & $m_B - m_V$ &   $M_B$ \\
 & (mag) & (mag) & (mag) & (mag) & & & (mag) & (mag) \\
\hline
%SN2012fr   &  12.000 &   0.006 &  11.968 &   0.004 &   1.231 &   0.006 &   0.033 &  -19.510$\\
SN2000cx   &  13.134 &   0.007 &  13.069 &   0.006 &   0.907 &   0.006 &   0.065 &  $-18.788$\\
%SN1991T    &  11.587 &   0.007 &  11.402 &   0.005 &   1.189 &   0.010 &   0.185 &  -19.503\\
SN1994D    &  11.769 &   0.007 &  11.827 &   0.005 &   0.784 &   0.006 &  -0.058 &  $-19.551$\\
SN2007on   &  13.046 &   0.004 &  12.931 &   0.004 &   0.566 &   0.005 &   0.114 &  $-18.480$\\
SN2012cg   &  12.116 &   0.008 &  11.930 &   0.008 &   1.101 &   0.019 &   0.186 &  $-18.904$\\
SN1980N    &  12.459 &   0.011 &  12.334 &   0.012 &   0.848 &   0.011 &   0.125 &  $-19.131$\\
%SN1981B    &  11.863 &   0.016 &  11.788 &   0.021 &   0.639 &   0.038 &   0.075 &  -19.227\\
SN2003hv   &  12.455 &   0.049 &  12.544 &   0.036 &   0.764 &   0.020 &  -0.089 &  $-19.111$\\
SN2008Q    &  13.459 &   0.014 &  13.512 &   0.010 &   0.804 &   0.022 &  -0.053 &  $-18.463$\\
SN1970J    &  14.865 &   0.037 &  14.619 &   0.037 &   0.916 &   0.029 &   0.246 &  $-18.717$\\
SN1983G    &  12.789 &   0.102 &  12.614 &   0.071 &   1.189 &   0.059 &   0.175 &  $-19.131$\\
SN2014bv   &  13.999 &   0.018 &  13.809 &   0.013 &   0.640 &   0.021 &   0.190 &  $-18.191$\\
SN2015bp   &  13.697 &   0.014 &  13.664 &   0.016 &   0.681 &   0.018 &   0.033 &  $-18.040$\\
SN2016coj  &  13.205 &   0.021 &  12.978 &   0.013 &   0.891 &   0.011 &   0.227 &  $-18.717$\\
SN1981D    &  12.486 &   0.048 &  12.327 &   0.046 &   0.852 &   0.041 &   0.159 &  $-19.104$\\
SN1992A    &  12.530 &   0.004 &  12.500 &   0.004 &   0.777 &   0.006 &   0.030 &  $-19.102$\\
SN2018aoz  &  12.515 &   0.009 &  12.590 &   0.008 &   0.841 &   0.007 &  -0.075 &  $-19.280$\\
SN2011iv   &  12.446 &   0.008 &  12.389 &   0.009 &   0.652 &   0.014 &   0.057 &  $-19.080$\\
SN2006dd   &  12.270 &   0.003 &  12.287 &   0.003 &   0.950 &   0.003 &  -0.017 &  $-19.320$\\
SN1992bo   &  15.758 &   0.013 &  15.746 &   0.011 &   0.712 &   0.013 &   0.011 &  $-18.512$\\
SN1997E    &  15.171 &   0.009 &  15.082 &   0.007 &   0.795 &   0.012 &   0.090 &  $-18.329$\\
SN1995D    &  13.379 &   0.033 &  13.253 &   0.015 &   1.256 &   0.025 &   0.126 &  $-19.221$\\
SN1996X    &  13.075 &   0.024 &  13.081 &   0.017 &   0.893 &   0.022 &  -0.006 &  $-19.185$\\
SN1998bp   &  15.368 &   0.013 &  15.071 &   0.014 &   0.597 &   0.025 &   0.297 &  $-17.732$\\
SN2017fgc  &  13.619 &   0.019 &  13.345 &   0.014 &   0.957 &   0.018 &   0.273 &  $-18.917$\\
SN2020ue   &  11.970 &   0.011 &  12.071 &   0.008 &  0.718  &   0.012 &  0.101 & 
$-18.860$\\
\hline

%\multicolumn{8}{c}{Fast decliners and highly reddened SNe }\\
%\hline
%SN1998bu   &  12.141 &   0.008 &  11.797 &   0.006 &   0.969 &   0.007 &   0.344 \\
%SN2017ejb  &  15.411 &   0.014 &  15.091 &   0.018 &   0.447 &   0.017 &   0.320 \\
%SN2006X    &  15.271 &   0.010 &  13.979 &   0.009 &   0.945 &   0.008 &   1.293 \\
%SN1991bg   &  14.638 &   0.009 &  13.859 &   0.008 &   0.269 &   0.003 &   0.779 \\
%SN1986G    &  12.087 &   0.011 &  11.153 &   0.012 &   0.600 &   0.018 &   0.934 \\
%SN2011eh   &  15.454 &   0.066 &  14.781 &   0.076 &   0.481 &   0.070 &   0.673 \\
%SN2005ke   &  14.588 &   0.008 &  13.932 &   0.009 &   0.414 &   0.010 &   0.656 \\
%SN2003gs   &  14.210 &   0.057 &  13.488 &   0.045 &   0.270 &   0.054 &   0.722 \\
%\hline
\end{tabular}
\endgroup
\end{table*}

\begin{table*}[h]
\centering
\caption{Best-fit parameters of the \shoes sample estimated using \texttt{SNooPy}. $m_B$ and $m_V$ ({\it columns 2 and 4}) are the apparent magnitudes at the light curve maximum in the B and V bands, respectively. $s_{BV}$ ({\it column 6}) is the stretch color parameter ({\it column 7}). The color is given as $m_B-m_V$ is computed as difference between the maximum brightness in B and V bands and $M_B$ ({\it column 9}) is the absolute magnitude in B band, computed as $m_B - \mu_{ceph}$, where $\mu_{ceph}$ is taken from Table~5 of \citetalias{riess2016}.}
\label{tab:tabel3}
% table3.tex created at UT 2019-11-10 18:22:13
\begingroup
\setlength{\tabcolsep}{14pt} % Default value: 6pt
\renewcommand{\arraystretch}{1.2} % Default value: 1
\begin{tabular}{c c c c c c c c c}
\hline\hline
Supernova & $m_B$ & $\sigma_{m_B}$ & $m_V$ & $\sigma_{m_V}$ & $s_{BV}$ & $\sigma_{s_{BV}}$ & $m_B - m_V$ & $M_B$ \\
 & (mag) & (mag) & (mag) & (mag) & & & (mag) & (mag) \\
\hline
SN1995al   &  13.339 &   0.010 &  13.172 &   0.010 &   1.075 &   0.018 &   0.167 &  $-19.159$\\
SN2011by   &  12.889 &   0.009 &  12.821 &   0.009 &   0.947 &   0.007 &   0.068 &  $-18.698$\\
SN2012fr   &  11.976 &   0.006 &  11.943 &   0.004 &   1.122 &   0.009 &   0.033 &  $-19.331$\\
SN1981B    &  11.863 &   0.016 &  11.788 &   0.021 &   0.639 &   0.038 &   0.075 &  $-19.043$\\
SN2003du   &  13.492 &   0.004 &  13.548 &   0.004 &   1.011 &   0.004 &  -0.056 &  $-19.427$\\
SN2005cf   &  13.250 &   0.004 &  13.246 &   0.004 &   0.947 &   0.004 &   0.004 &  $-19.013$\\
SN2011fe   &   9.930 &   0.004 &   9.947 &   0.003 &   0.937 &   0.003 &  -0.017 &  $-19.205$\\
SN2013dy   &  12.757 &   0.004 &  12.554 &   0.003 &   1.136 &   0.010 &   0.203 &  $-18.742$\\
SN2002fk   &  13.205 &   0.023 &  13.209 &   0.017 &   1.189 &   0.034 &  -0.004 &  $-19.318$\\
SN1998aq   &  12.322 &   0.006 &  12.414 &   0.004 &   0.940 &   0.004 &  -0.092 &  $-19.415$\\
SN2007af   &  13.164 &   0.003 &  13.058 &   0.003 &   0.919 &   0.003 &   0.106 &  $-18.622$\\
SN1994ae   &  13.064 &   0.051 &  12.933 &   0.041 &   1.125 &   0.157 &   0.131 &  $-19.008$\\
SN2012cg   &  12.116 &   0.008 &  11.930 &   0.008 &   1.101 &   0.019 &   0.186 &  $-18.964$\\
SN2015F    &  12.823 &   0.009 &  12.695 &   0.010 &   0.865 &   0.007 &   0.128 &  $-18.688$\\
SN1990N    &  12.650 &   0.008 &  12.574 &   0.006 &   0.976 &   0.006 &   0.076 &  $-18.882$\\
SN2007sr   &  12.741 &   0.058 &  12.568 &   0.042 &   1.022 &   0.023 &   0.173 &  $-18.549$\\
SN2012ht   &  12.393 &   0.004 &  12.576 &   0.005 &   0.854 &   0.004 &  -0.183 &  $-19.515$\\
SN2009ig   &  13.478 &   0.008 &  13.372 &   0.007 &   1.134 &   0.023 &   0.106 &  $-19.019$\\
SN2001el   &  12.831 &   0.007 &  12.601 &   0.005 &   0.949 &   0.006 &   0.230 &  $-18.480$\\
\hline
\end{tabular}
\endgroup
\end{table*}

\subsection{Luminosity calibration}
\label{sebsec:tripp}
Having obtained the light curve parameters, we proceed to derive the SN Ia calibration relation separately for the SBF and the \shoes samples. \cite{phillipsM} gave the relation between SNe Ia luminosity and their LC shape, and later \cite{tripp} added the color correction. 
We use this two-parameter luminosity relation, including a term relating the peak luminosity of the SN Ia to the LC shape (represented by the color-stretch parameter), and a second term for the color correction accounting for the dust reddening in the host galaxy. The apparent B band peak magnitude ($m_B$) of a \sne is thus modeled as:
\begin{dmath}
\label{eq:Tripp}
    \ \   m_B \ = \ P^N(s_{BV} - 1)\  +\  R (m_B - m_V)  + \mu_{calib}\ ,
\end{dmath}
where $P^N$ is a polynomial of order $N$ as a function of $(s_{BV} - 1)$, which gives the luminosity-decline rate relation, $R$ is the extinction correction coefficient that correlates the peak magnitude with the color $(m_B - m_V)$ at maximum, and $\mu_{calib}$ is the distance modulus for the host galaxy ($\mu_{SBF}$ taken from Table~\ref{tab:tabel1} for the SBF sample, and $\mu_{ceph}$ taken from Table~5 of \citetalias{riess2016} for the \shoes sample).

Besides the Milky Way extinction correction (already included in the LC fitting procedure as described in Section~\ref{LCfit}), there are three other potential sources of reddening that need to be corrected for: (1) intergalactic dust, (2) interstellar dust of the host galaxy, and (3) the intrinsic color of \sne \citep{burns2014,foley1,viewingangle}. To know the properties of these different sources, which may vary from SN to SN, and then disentangle their different effects requires sophisticated color modeling \citep[see, e.g.,][]{burns2014}. Since cosmological analyses do not aim at studying the details of dust properties, we make no distinction between the intrinsic and the extrinsic sources of color variation, and combine the extinction from these different effects into the one correlation term $R$, as done by previous works such as \cite{jla}, \cite{freedman2009}, and \cite{conley1}. \par

In order to identify the optimum model for our SBF calibrators data, we first perform the analysis using the quadratic polynomial ($N = 2$) and then with only the linear polynomial ($N = 1$) in $P$. The comparison of the luminosity-stretch ($s_{BV}$) relation obtained by the two model forms shows that the second-order term does not improve the fits. The $R^2$ score (coefficient of determination\footnote{The $R^2$ score is defined as ($1 - u/v$), where $u$ is the residual sum of squares $\sum{(y_{true} - y_{pred})}^2$ and $v$ is the total sum of squares $\sum{(y_{true} - \overline{y_{true}})}^2$}) of both the fits was calculated to be 0.96 and mean squared error (MSE) as 0.06. The value of $P^2$ parameter was estimated to be $0.35 \pm 1.33$, which  makes it consistent with zero and hence its weight in the model is null. Therefore, we use the calibration relation with the linear term in $P^N$:
\begin{equation}
\label{eq:lintripp}
     m_B = P^0 + P^1(s_{BV} - 1) + R(m_B - m_V) + \mu_{calib} \ ,
\end{equation}

\subsection{The Hubble constant}
In order to estimate \h0, we use a purely kinematic cosmological model that gives  the luminosity distance as a function of redshift \citep{1972Weinberg, visser2004}. The parametrization assumes a Robertson--Walker metric in a flat space for the geometry of the Universe and it is based on the Taylor series expansion of the Hubble--Lemaitre law, with the presence of two additional parameters, $q_0$ and $j_0$, where $q_0 = -\ddot{a}\dot{a}^{-2} a$ is the cosmic deceleration and $ j_0 = -\dddot{a} \dot{a}^{-3} a^2$ is the third derivative of the scale factor, called cosmic jerk. For a flat Universe, the expansion of the luminosity distance to the third order in $z$ is given as:
\begin{equation}
    d_L(z) = \frac{cz}{H_0} \Bigg\{ 1 + \frac{1}{2}(1-q_0)z - \frac{1}{6} (1 - q_0 - 3q_0^2 + j_0)z^2 + O(z^3) \Bigg\}.
\end{equation}
 
Neglecting $O(z^3)$ and higher order term, the corresponding distance modulus is given as

 \begin{equation}
 \label{eq:mu}
     \mu(z, H_0) = 5 \log_{10} \frac{cz}{H_0}\Bigg\{ 1 + \frac{1}{2}(1-q_0)z - \frac{1}{6} (1 - q_0 - 3q_0^2 + j_0)z^2 \Bigg\} + 25.
 \end{equation}

For the cosmological sample, we use equation \ref{eq:lintripp} except that the independent distance moduli $\mu_{calib}$ are replaced with distance moduli as a function of \h0 and the redshift as given by equation \ref{eq:mu}. \h0 is left as a free parameter in the analysis. Hence the intercept term $P^0$ is anchored only to the independent distances of the calibration sample, and it dictates the uncertainty in the estimated Hubble constant value. We fix the value of the deceleration parameter to $q_0 = -0.55$ and the jerk $j_0 = 1$ \citep{planck2018, jla}. In the redshift range of our cosmological sample ($0.009 < z < 0.075$), any  assumption about the expansion history of the Universe does not significantly affect the final estimate of \h0 \citep{dhawan2020}. Hence, fixing the values of $q_0$ and $j_0$ does not bias our \h0 estimates.

\subsection{Bayesian inference}
\label{sec:bayesian}
We perform a hierarchical Bayesian regression using the data of both the calibration sample and the cosmological sample to estimate the free model parameters using Markov Chain Monte Carlo (MCMC) sampling. The modeling here combines two sub-models, one for the calibration sample and one for the cosmological sample, and estimates the posterior distributions for the parameters of interest from both the SN Ia populations simultaneously. Bayes' theorem gives the posterior of the model parameters as
\begin{equation}
\label{eq:bayes}
 P(\Theta | D) \propto P(D | \Theta) P(\Theta),
\end{equation}
where $D$ denotes the vector for the observed SN data (the LC fit parameters: $m_B, m_V, s_{BV}$), and $\Theta$ denotes the vector for the model parameters, namely  $P^0 , P^1$, $R$ and \h0. Each individual SN can be assumed to have its own model parameters $\Theta_i$ (i.e., $P^0_i, P^1_i, R_i$) but they cannot be sufficiently constrained on a SN-by-SN basis and their uncertainties propagate to the \h0 inference on the population level. This makes it necessary that the model assumptions and priors for each SN are propagated hierarchically when inferring the parameters from a population of SNe. Hence, the distribution of each $\Theta_i$ is assumed to be Gaussian with a mean of $\Theta$ making it the hyper-parameter vector. Then the hyper-parameters  $P^0 , P^1$, and $R$ describe the distribution (Gaussian width) of the model parameters of each individual supernova and their priors are the hyper-priors of the model.

Then, marginalising over all $\Theta_i$, the likelihood probability distribution $P(D | \Theta)$ is given as the combined distribution for the calibrator and cosmological SNe,
\begin{equation}
    P(D | \Theta) = \prod_{i=1}^{N_{calib}} P(D_i , \mu_{calib,i} | \Theta)  \prod_{j=1}^{N_{cosmo}} P(D_j, z_j | \Theta, H_0),
\end{equation}
where $i$ is the index for the $N_{calib}$ SNe of the calibration sample (24 for SBF and 19 for SH0ES) and $j$ for the $N_{cosmo}$ SNe of the cosmological sample. $\mu_{calib, i}$ are the independent distance estimates of the calibrating set and $D_{i/j}$ is the data set on an individual SN of the calibrator ($i$) or the cosmological sample ($j$).
 $P^0 , P^1$, $R$ and \h0 are kept as free parameters and are determined simultaneously. The redshift $z_j$ of each SN belonging to the cosmological sample is in the CMB rest frame, and is taken from the online repository as described in Section~\ref{sec2.3-cosmosample}.
Assuming normally distributed errors and treating the B band peak magnitude as the target variable, the log likelihoods for the two samples can be written as
\begin{equation}
    \ln \mathcal{L}_{calib} = -\frac{1}{2}\sum_{i=1}^{N_{calib}} \frac{(m_B^i - m_B^T)^2}{\sigma_{calib,i}^2} - \frac{1}{2} \sum_{i=1}^{N_{calib}} \ln 2 \pi \sigma_{calib,i}^2 
\end{equation}
and
\begin{equation}
\ln \mathcal{L}_{cosmo} = -\frac{1}{2}\sum_{j=1}^{N_{cosmo}} \frac{(m_B^j - m_B^T)^2}{\sigma_{cosmo,j}^2} - \frac{1}{2}\sum_{i=j}^{N_{cosmo}} \ln 2 \pi \sigma_{cosmo,j}^2 \ , 
\end{equation}
while the combined log likelihood is
\begin{equation}
    \ln \mathcal{L} =  \ln \mathcal{L}_{calib} + \ln \mathcal{L}_{cosmo}.
\end{equation}
Here, $m_B^{i/j}$ is the observed B band magnitude for each supernova, and $m_B^T$ is the true magnitude  given by equation \ref{eq:lintripp} for the calibration sample (and replacing $\mu_{calib}$ with $\mu(z, H_0)$ in that equation for the cosmological sample). The variances $\sigma_{calib/cosmo}$ are computed as quadrature sum of the photometric errors and the SBF distance uncertainties. Additionally, in order to account for hidden systematic uncertainties, we include two separate intrinsic scatter parameters, one each for the two samples. These two terms, namely $\sigma_{int,calib}$ and $\sigma_{int,cosmo}$, are added to the variance of their respective sample and are left as free parameters in the analysis accounting for any extra dispersion observed in the measured distance moduli. It should be noted here that we do not include the correlations between light curve fit parameters since they are negligible, while correlations between model parameters are taken into account in the analysis. Hence, the full variance for a given calibrator object $i$ is
\begin{multline}
    \sigma_{calib,i}^2 = \sigma_{m_B,i}^2 + \sigma_{\mu_{SBF},i}^2 + (P^1 \sigma_{s_{BV},i})^2 + (R \sigma_{m_B-m_V,i})^2 \\
    - 2R\sigma_{m_B,i}^2 + \sigma_{int,calib}^2 \ ,
\end{multline}
and the total variance for an object $j$ in cosmological sample is
\begin{multline}
    \sigma_{cosmo,j}^2 = \sigma_{m_B,j}^2 + (P^1 \sigma_{s_{BV},j})^2 + (R \sigma_{m_B-m_V,j})^2 \\
    - 2R\sigma_{m_B,j}^2 + \sigma_{int,cosmo}^2.
\end{multline}

Lastly, the term $P(\Theta)$ in equation \ref{eq:bayes} are the priors on our model parameters. We adopt uniform priors for all the parameters except the intrinsic scatter terms for which we assume a Half Cauchy distribution.
The MCMC sampling is implemented using the "No U-Turn Sampler" (NUTS) provided in the \texttt{PyMC3}\footnote{See:\href{https://docs.pymc.io/}{https://docs.pymc.io/}} \citep{pymc3}, which is a \texttt{python} probabilistic programing package. Using the observed data as input, we estimate simultaneously the posteriors for the correlation parameters $P^0 , P^1, R$ and the Hubble constant $H_0$ along with the two intrinsic scatters. All the best-fit values provided in this work are the posterior means and the errors in the parameters are the standard deviation of their posterior.
The entire data sets for the three samples and the full analysis codes used in this paper are available in a GitHub repository\footnote{\url{https://github.com/nanditakhetan/SBF_SNeIa_H0}}.

%%%%%%%%%%%%%%%%%%%%%%%%%%%%%%%%%%%%%%%%%%%%%%%%%%%%%%%%%%%%%%%%%%%%%%%%%%%%%%%%%%%%%%%%%%%%%%%%%

\section{Results}
\label{sec4-results}
\subsection{SBF calibration}
\begin{figure}[h]
%\centering
\includegraphics[width=9.0cm,height=4.7cm]{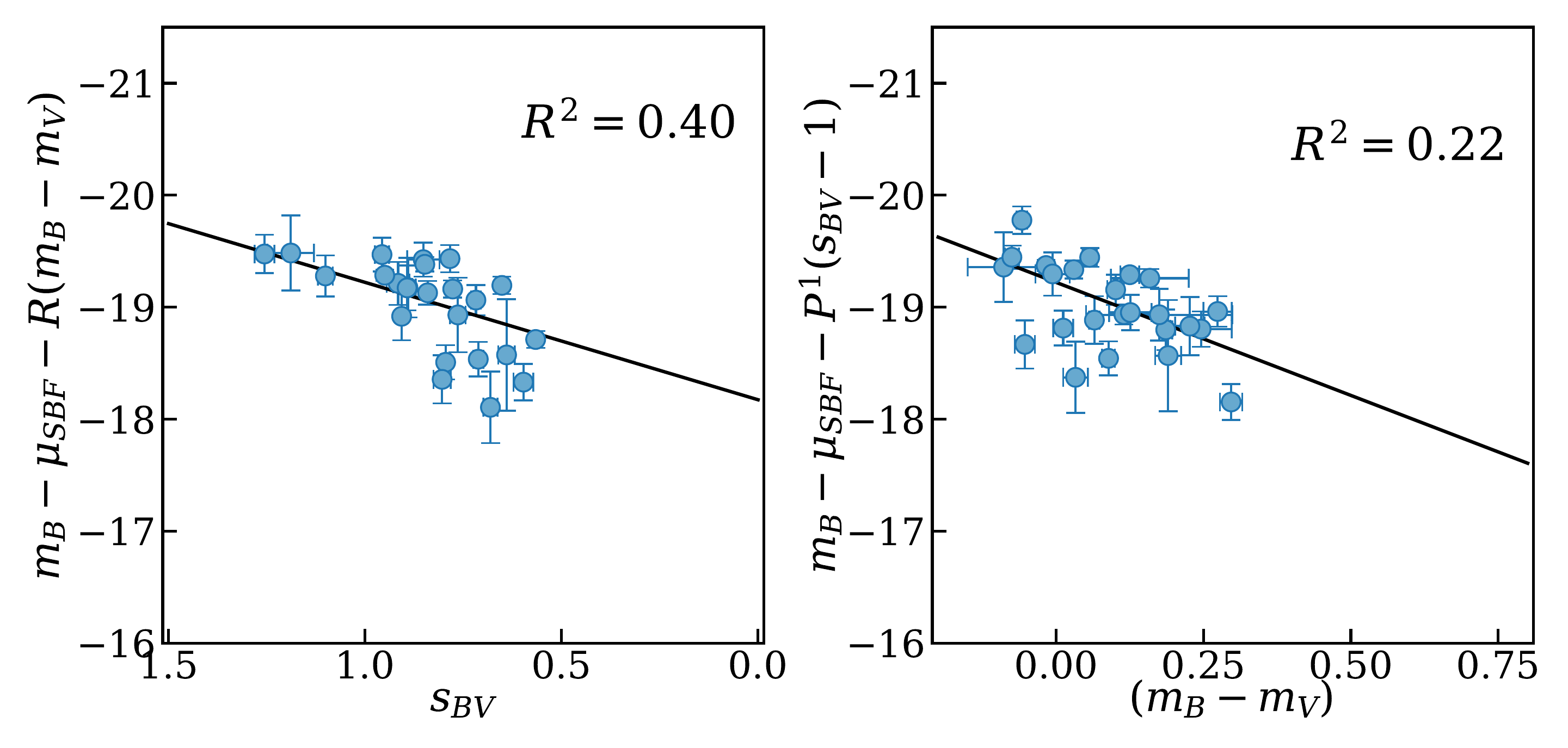}
\caption{Luminosity correlation plots for the SBF sample: the absolute magnitude ($m_B-\mu_{SBF}$) corrected for the color vs.\ the LC stretch parameter (left panel), and the absolute magnitude corrected for the LC stretch vs.\ color (right panel). The correlation parameters are evaluated using the Bayesian analysis described in Section \ref{sec:bayesian}. The solid black lines show the best-fit model obtained with the MCMC sampling. The $R^2$ score (coefficient of determination) of the fit is shown in the top right.}
\label{fig:tripp_sbf}
\end{figure}

Using the SBF calibration sample of 24 \sne and the redshift-cut cosmological sample of 96 \sne ($z>0.02$), we evaluate the posterior distributions of the luminosity correlation parameters and the Hubble constant. The analysis is also performed on the full cosmological sample of 140 SNe Ia.
Table \ref{tab:tripp} gives the mean posterior values for the correlation parameters and the individual intrinsic scatter for both the SBF calibration sample and the cosmological samples. Figure \ref{fig:tripp_sbf} shows the luminosity relations for the SBF calibration sample with respect to the stretch parameter (left panel) and the color (right panel).

\begin{table*}[h]
\centering
\caption{Mean posterior values for the luminosity correlation parameters and associated errors for the full cosmological sample and the redshift-cut sample. The last two columns on the right give the intrinsic scatter of the calibration sample and the cosmological samples. The top part of the table shows the results obtained using the SBF calibration sample, and the bottom part the results obtained using the \shoes calibration sample.}
\setlength{\tabcolsep}{14pt} % Default value: 6pt
\renewcommand{\arraystretch}{1.5} % Default value: 1
\begin{tabular}{cccccccc}
\hline \hline
Redshift range& $N_{sample}$  &  $P_0$  &  $P_1$  &  $R$  & $\sigma_{int,calib}$ & $\sigma_{int,cosmo}$ \\ 
              &  (mag) & (mag)  &   (mag)    &                &      (mag) & (mag)   \\
\hline
\multicolumn{7}{c}{\textbf{SBF Calibration}} \\
\hline
$0.009 < z < 0.075$ &140   & $-19.23 \pm 0.07$  & $-1.07 \pm$ 0.11 & 2.03 $\pm$ 0.16 & 0.29 & 0.18 \\
$0.02 < z < 0.075$ &96    & $-19.22 \pm 0.07$  & $-1.05 \pm$ 0.12 & 2.01 $\pm$ 0.17 & 0.29 & 0.15  \\  
\hline 
\multicolumn{7}{c}{\textbf{Cepheid Calibration}} \\
\hline
$0.009 < z < 0.075$ &140   & $-19.16 \pm 0.05$  & $-1.00 \pm$ 0.11 & 2.15 $\pm$ 0.16 & 0.17 & 0.18 \\
$0.02 < z < 0.075$ &96    & $-19.16 \pm 0.05$  & $-0.99 \pm$ 0.12 & 2.16 $\pm$ 0.17 & 0.17 & 0.15  \\
\hline
\label{tab:tripp}
\end{tabular}
\end{table*}

\begin{figure*}[h!]
\centering
\hspace*{-0.5cm}
\includegraphics[width=14.0cm,height=14.0cm]{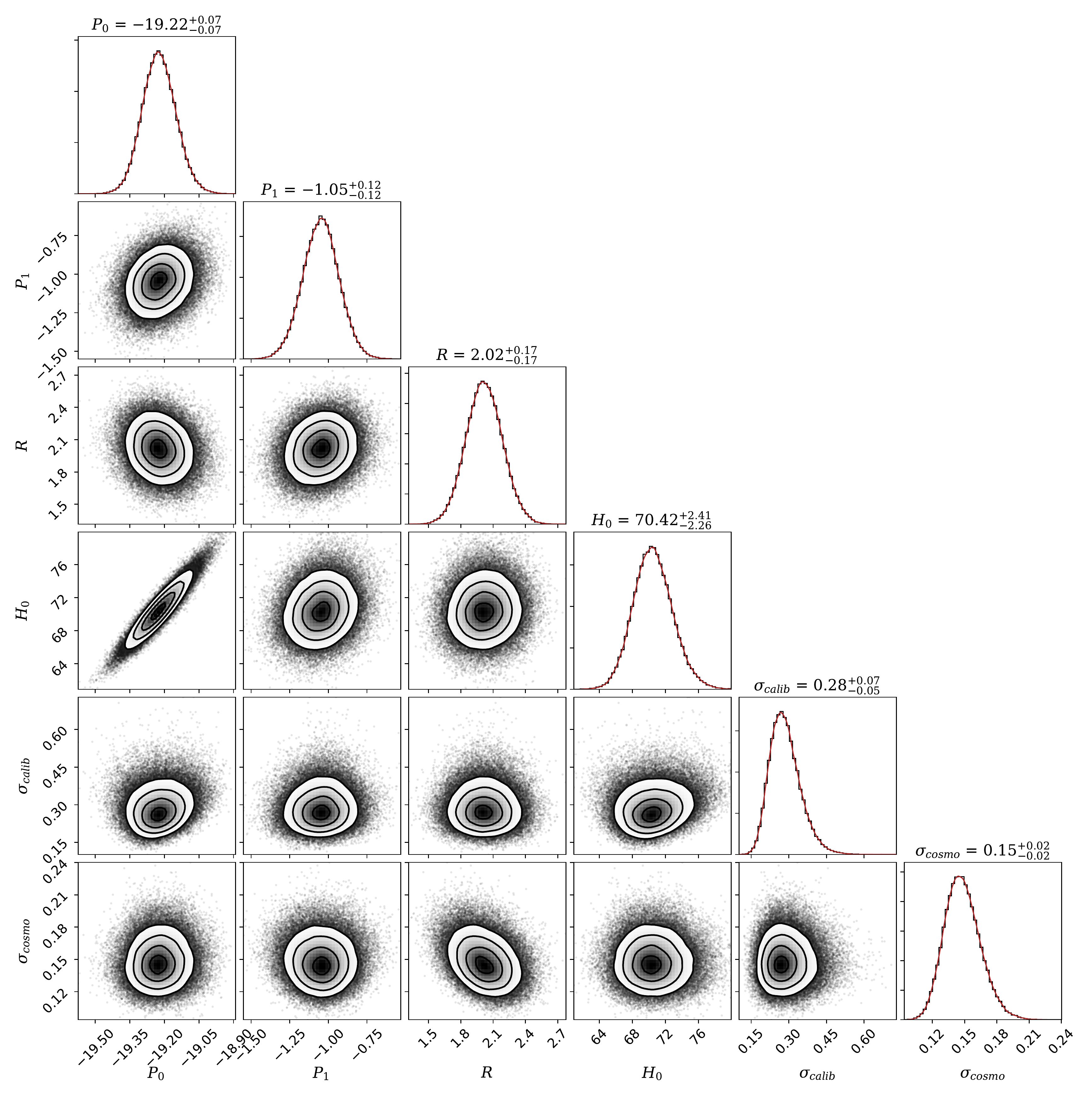}
\caption{Corner plot showing posterior distributions for the parameters $P_0, P_1, R$ and \h0 along with the intrinsic scatter obtained using the SBF sample (24 \sne) with the redshift-cut cosmological sample (96 \sne). The title on each histogram shows the median value of the respective posterior distribution. The luminosity correction does not include any dependence on host galaxy stellar mass.}
\label{fig:corner_noHM_sbf}
\end{figure*}

The best-fit value of the Hubble constant obtained using the SBF calibration on the redshift-cut cosmological sample is $H_0 = 70.50 \pm 2.37$\ \hunit. It is slightly lower, $H_0 = 69.18 \pm 2.33$\ \hunit\ when the full cosmological sample is used. The computed \h0 values are listed in Table \ref{tab:H0}. A corner plot showing posterior samples for SBF calibration is given in Figure \ref{fig:corner_noHM_sbf}.

In order to investigate the influence of the adopted threshold of the $s_{BV}$ parameter ($s_{BV} > 0.5$) on our results, we evaluate the \h0 removing the two transitional objects SN2007on and SN2011iv (see Section~\ref{sec3-method}) from our SBF calibration sample. The net effect is a small increase of 0.7\% in \h0. Furthermore, assuming a more conservative definition of fast decliners in the SBF calibration sample by removing all SNe with $s_{BV} < 0.7$ (5 \sne including the two above transitional \sne), the resultant \h0 value is lower by 1.8\%.

\subsection{Cepheid Calibration}
\begin{figure}[h]
%\centering
\includegraphics[width=9.0cm,height=4.5cm]{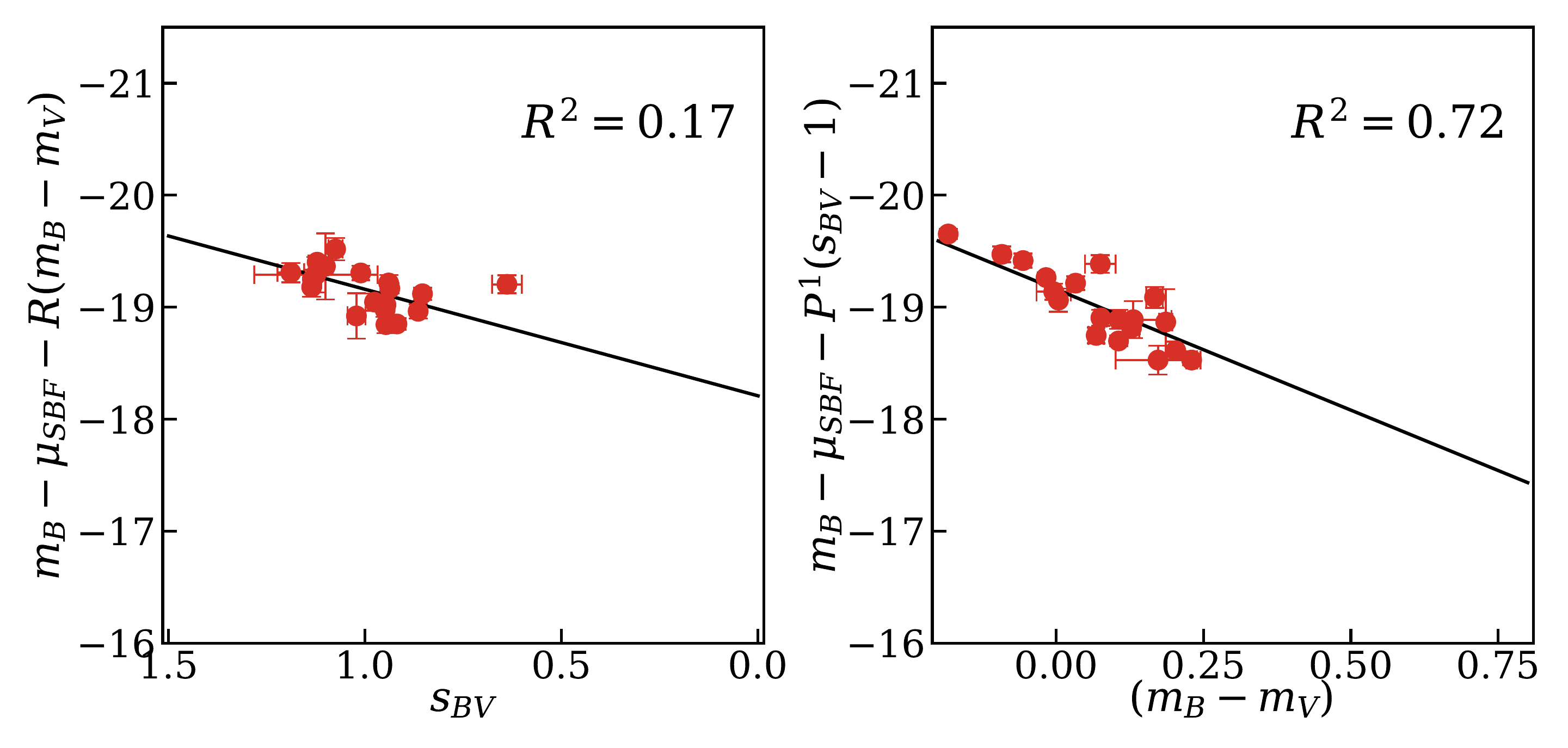}
\caption{Luminosity correlation parameters for the \shoes sample: the absolute magnitude ($m_B-\mu_{SBF}$) corrected for the color vs.\ the LC stretch parameter (left panel), the absolute magnitude corrected for the LC stretch vs.\ color (right panel). The correlation parameters are evaluated using the Bayesian analysis. The solid black lines show the best-fit model obtained with the MCMC sampling. The $R^2$ score (coefficient of determination) of the fit is shown in the top right.}
\label{fig:tripp_shoes}
\end{figure}

For the Cepheid calibration, we used the \shoes sample as calibration set and estimated the correlation parameters and the Hubble constant value following the same analysis as used for the SBF calibration. The estimated parameter values are listed in the lower part of the Table \ref{tab:tripp}. It is worth noting the difference between the $P_0$ values for the SBF and Cepheid calibration and the slightly higher R value for the \shoes calibration. The luminosity relations for the \shoes calibration sample are shown in Figure \ref{fig:tripp_shoes}.

\begin{figure*}[h!]
\centering
\hspace*{-0.5cm}
\includegraphics[width=14.0cm,height=14.0cm]{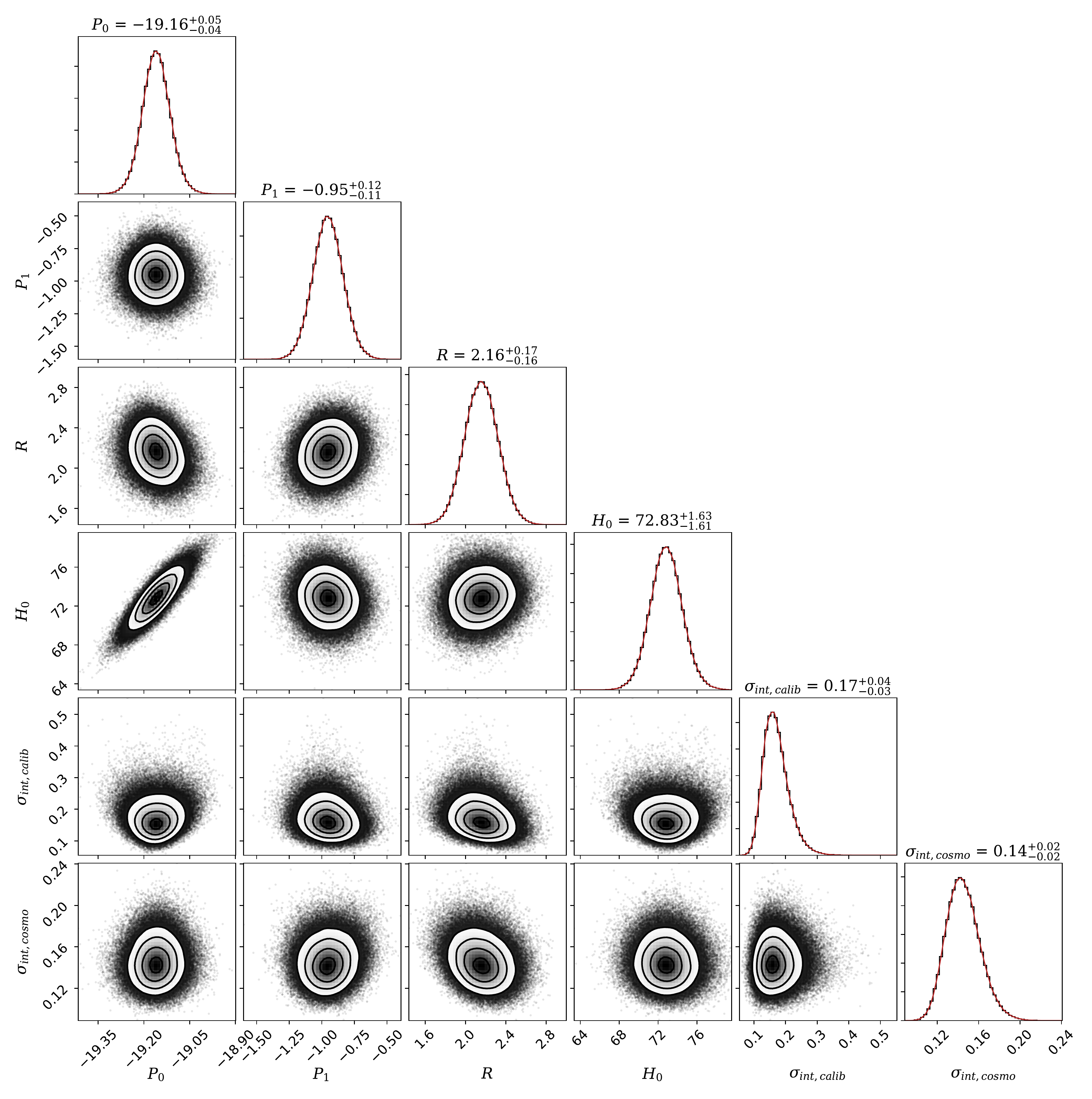}
\caption{Corner plot showing posterior distributions for the parameters $P_0, P_1, R$ and \h0 along with the intrinsic scatter obtained using the \shoes sample with the redshift-cut cosmological sample. The title on each histogram shows the median value of the respective posterior distribution. The luminosity correction does not include any dependence on host galaxy stellar mass.}
\label{fig:corner_noHM_shoes}
\end{figure*}

Applying the \shoes Cepheid calibration to the redshift-cut cosmological sample, the mean value for \h0 is found to be $H_0 = 72.84 \pm 1.66 $ \hunit, and it decreases to $H_0 = 71.51 \pm 1.66 $ \hunit\ when using the full cosmological sample. These values are listed in Table \ref{tab:H0}. The value of the Hubble constant evaluated for the redshift-cut cosmological sample is fully consistent with the measurement of \citetalias{riess2016} ($H_0 = 73.24 \pm 1.74 $). The posterior samples for the \shoes calibration analysis are given in Figure \ref{fig:corner_noHM_shoes}. 

For fitting the SN Ia LCs, we prefer to use \texttt{SNooPy} instead of SALT2 \citep{Mosher2014} since the latter has not been trained on fast-declining SNe, this results in poorly constrained LC shape parameter $(x_1)$ for faster \sne, as shown in Figure 1 of \cite{burns2018}. Our SBF calibration sample contains 11 of 26 SNe with $ 0.5 < s_{BV} < 0.8$, motivating our choice for using SNooPy. The fact that the $H_0$ value obtained with the \shoes calibration is in perfect agreement with \citetalias{riess2016} confirms that using \texttt{SNooPy} instead of SALT2 (as used by \citetalias{riess2016}) provides consistent results.

In order to make the SBF and \shoes calibration samples completely independent, we performed the analysis removing the one object in common between the SBF and \shoes samples, SN2012cg. The \h0 results from both the SBF and \shoes calibrations remain the same.

As noted in Section~\ref{sec:SBF sample}, the SBF distances of our calibration sample are based on the LMC distance modulus of $ 18.50 \pm 0.10$ mag, as in \citetalias{freedman2001}, and the \shoes sample distances are based on the LMC distance modulus of $18.493 \pm 0.008$ mag \citetalias{riess2016}. Referring to the most recent estimate of LMC distance, which is $18.477 \pm 0.004$ by \cite{Pietrzyski2019}, we evaluated the \h0 by scaling down the SBF distances and the \shoes distances by $0.023$ mag and $0.016$ mag, respectively. Using this recent value of LMC distance, the \h0 value calibrated with SBF increases by 1\%, becoming $71.16 \pm 2.38$\ \hunit. The \h0 value calibrated with the \shoes calibration increases by 0.8\%, becoming $73.38 \pm 1.66$\ \hunit. However, a truly updated revision of the SBF calibration, errors, and, hence, distances, would require a comprehensive update of Cepheid distances (P--L relations and zero-point) and the SBF measurement for the six Cepheid-host calibrating sample, which goes beyond the scope of this paper.

\begin{figure*}[h!]
\centering
\hspace*{-0.5cm}
\includegraphics[width=12.0cm,height=9.0cm]{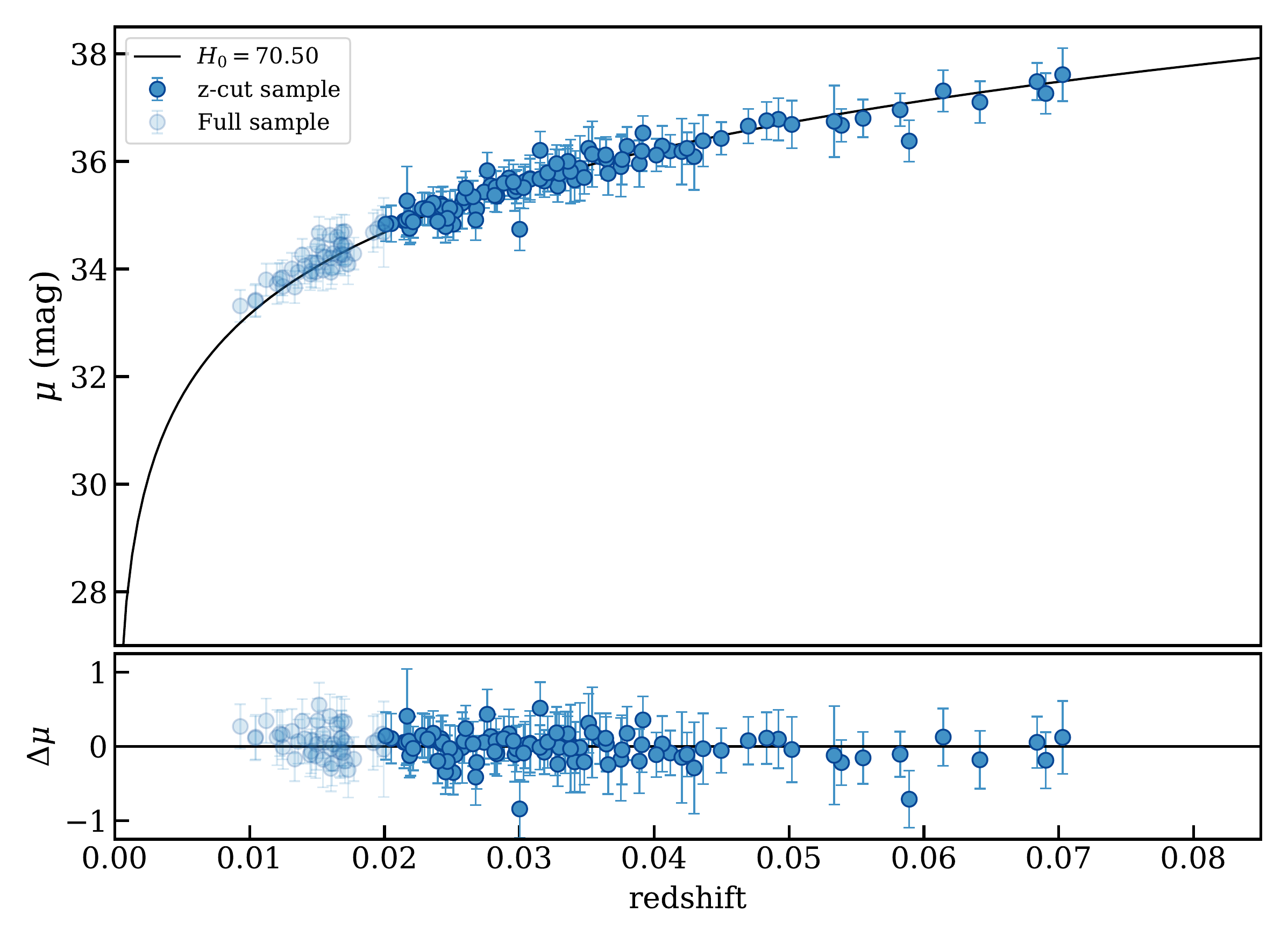}
\caption{Hubble diagram for the cosmological samples. The distance moduli of the \sne are computed using the SBF calibration. The solid black line in the top panel corresponds to the \h0 value estimated for the redshift-cut cosmological sample of the 96 \sne with $z > 0.02$. The lower panel shows the residual plot.}
\label{fig:H01}
\end{figure*}

Figure \ref{fig:H01} shows the Hubble diagram for the  cosmological samples plotting the distance moduli versus the redshift. The plotted distance moduli are computed using the luminosity calibration relation obtained with the SBF sample. The solid line shows the best-fit model derived from the Bayesian regression. Residuals from the best-fit are shown in the bottom panel.

\begin{table*}[h!]
\centering
\caption{\h0 values for the full cosmological sample and the redshift-cut cosmological sample that is obtained excluding the \sne with $z < 0.02$.  The \h0 values are given for both the SBF (central column) and the Cepheid (right column) calibrations.}
\label{tab:H0}
\setlength{\tabcolsep}{20pt} % Default value: 6pt
\renewcommand{\arraystretch}{1.3} % Default value: 1
\begin{tabular}{cccccccccc}
\hline\hline
 \multirow{2}{5em}{Sample cut } & \multicolumn{2}{c}{SBF Calibration} & \multicolumn{2}{c}{Cepheid Calibration} \\
  \cline{2-5}
 &  $H_0$ &  $\sigma_{H_0}$ & $H_0$ &  $\sigma_{H_0}$ \\
 \hline
 $0.009 < z < 0.075$  &  69.18 & 2.33 & 71.51 & 1.66 \\
  $0.02 < z < 0.075$  & 70.50 & 2.37 & 72.84 & 1.66  \\
  \hline

\end{tabular}
\end{table*}

\subsection{$H_0$ systematic uncertainties}
\label{subsect:systematic}

The systematic uncertainty on \h0 is calculated by combining in quadrature the systematic error on SBF measurements and those from the SN LC fitting estimated by \texttt{SNooPy}. The adopted systematic errors are shown in Table \ref{tab:H0error}. Our final \h0 value and its uncertainties, obtained using the SBF calibration (24 SNe Ia, without including any host galaxy dependence), is $ H_0 = 70.50 \pm 2.37 \ (\pm 3.4\%$ stat)\ $\pm 3.38\ (\pm4.8 \%$ sys) \hunit. 

\begin{table}
\centering
\caption{Adopted systematic uncertainties on \h0}
\label{tab:H0error}
\setlength{\tabcolsep}{12pt} % Default value: 6pt
\renewcommand{\arraystretch}{1.2} % Default value: 1
\begin{tabular}{cccccccccc}
\hline\hline
Uncertainty & magnitude & \% error \\
\hline
SBF tie to Cepheid ZP & 0.1 mag & 4.6\% \\
B-band fit & 0.012 mag & 0.55 \% \\
V-band fit & 0.019 mag & 0.87 \% \\
\st estimate & 0.03 mag & 1.4 \% \\
\hline
Total & 0.106 mag & 4.8 \% \\
\hline
 
\end{tabular}

\end{table}

Deriving the Hubble Constant using SBF is a five step approach, which starts with a geometric distance (e.g., LMC), followed by calibration of the Cepheid period--luminosity relation, calibration of the absolute SBF magnitude by tying it to distances based on Cepheids (from the HST Key project), calibration of the SN Ia absolute magnitude (using the \sne listed  in Table~\ref{tab:tabel1}), and finally ending with the determination of the \h0. In this five step approach, the largest source of systematic uncertainty comes from tying the SBF distance scale to the Cepheid zero point (4.6\%), estimated to be 0.1 mag. This uncertainty can be reduced with a recalibration of the Cepheid period-luminosity--metallicity (or color) relationships and the LMC zeropoint using \textit{Gaia} parallaxes \citep{MC18}. This could halve the systematic error. Another possibility is to use the theoretical calibration of ${\overline{M}}$, which makes SBF an independent primary calibrator for the distance ladder approach. The systematic error in \h0 estimated using SBF \sne as calibrators has room for improvement.\par

The last three terms reported in Table \ref{tab:H0error} are the systematic errors in the three LC fit parameters evaluated by \texttt{SNooPy}. These errors are insensitive to the quantity and quality of the LC data. They arise from the use of templates for LC fitting, which do not perfectly represent the observed data. %as opposed to non-parametric functions like splines. 
They are evaluated as rms in the difference between the true and  template-fit values averaged over the training set.  

%%%%%%%%%%%%%%%%%%%%%%%%%%%%%%%%%%%%%%%%%%%%%%%%%%%%%%%%%%%%%%%%%%%%%%%%%%%%%%%%%%%%%%%%%%%%

\section{Host type dependence}
\label{sec5-masscorrection}
Considering the observational evidence that SN Ia luminosity correlates with the host galaxy type and its properties \citep{ Hamuy1996,howell2001,Neill2009,Pruzhinskaya2020,Ponder2020}, an additional term is typically added to the luminosity correction formula \citep[see e.g.,][]{betoule2014,rigault2015,riess2016,trgb}, which takes into account the host galaxy stellar mass $M_*$. This stellar-mass term is considered a proxy of other galaxy properties such as the SFR, metallicity, and/or stellar population \citep{sullivan2010,kelly2010}, possibly associated with different local environment and/or progenitors of the \sne.

We explore here the effect of adding the mass-based correction term (hereafter HM) to the luminosity relation of equation \ref{eq:lintripp} and evaluate its influence on the \h0 estimate. We adopt two recipes for the mass correction: (1) a linear correction and (2) a step correction. The luminosity relation including the HM term is
\vspace{0.1cm}
\begin{equation}
    \label{eq:mass}
   \ \ \ \   m_B \ = \ P^0 + P^1(s_{BV} - 1) + R(m_B - m_V) + \text{HM} + \mu_{calib},
\end{equation}
\noindent where the two recipes of HM can be written as,
%\vspace{0.1cm}
\begin{equation}
\label{eq:step}
\text{Step correction}: \\
    \text{HM}  = 
    \begin{cases}
    \alpha_{step},\ \ \ \text{for } \log_{10}M_{\ast}/M_{\odot} \ < \ M_{step} \\
    0, \ \ \ \ \ \ \ \  \text{otherwise}
    \end{cases}
\end{equation}
\vspace{-0.6cm}
\begin{equation}
\label{eq:linear}
\text{Linear correction}: \\
    \text{HM} = \alpha_{linear}(\log_{10}M_{\ast}/M_{\odot} - M_0).
\end{equation}
The step correction adds a value $\alpha_{step}$ to the SN absolute magnitude for all host galaxies with stellar masses below an arbitrary value $M_{step}$ and a zero correction above it. The linear correction assumes a linear correlation of the luminosity with the host galaxy stellar mass, $\log_{10} M_{\ast}$, given in units of solar mass, $M_{\odot}$, and $M_0$ is again an arbitrary mass zero-point. Details regarding the estimate of the host galaxy stellar mass are given in Appendix \ref{app:hostmass}.

\begin{figure*}[h!]
\centering
\includegraphics[width=12.0cm, height=7cm]{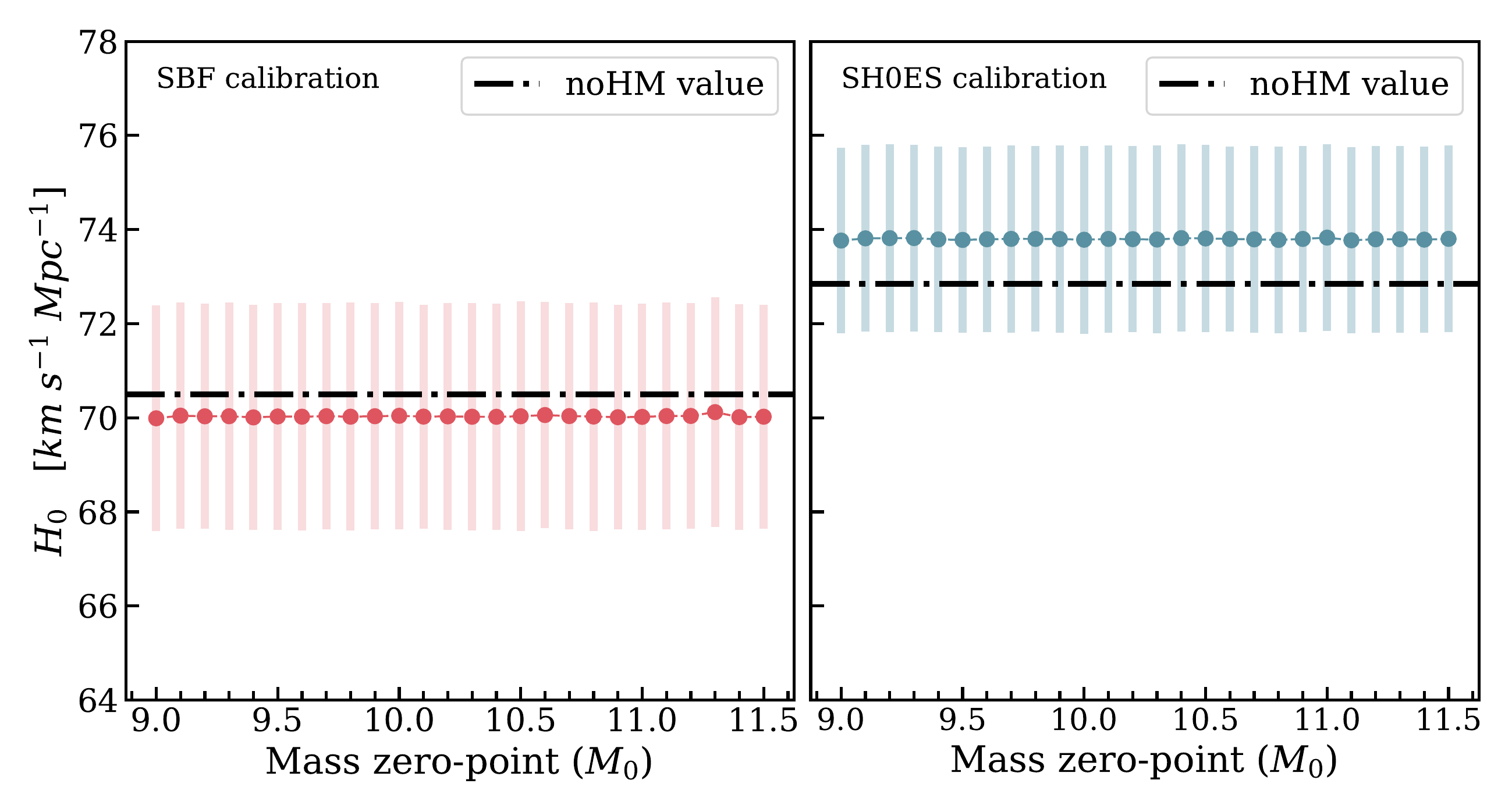}
\caption{\h0 values estimated including the linear mass correction in the calibration at different values of $M_0$. For the SBF calibration (left panel), the linear mass correction decreases the value of \h0 by $\sim 0.6\%$ with respect to the \h0 estimated without mass correction (noHM value, shown in dotted black line) for any chosen value of $M_0$. For the \shoes\ calibration (right panel), we see an increase of 1.3\% from the noHm value. This justifies an arbitrary choice for $M_0$.}

\label{fig:Ng1} 
\end{figure*}

\begin{figure*}[h!]
\centering
\includegraphics[width=12.0cm, height=7.0cm]{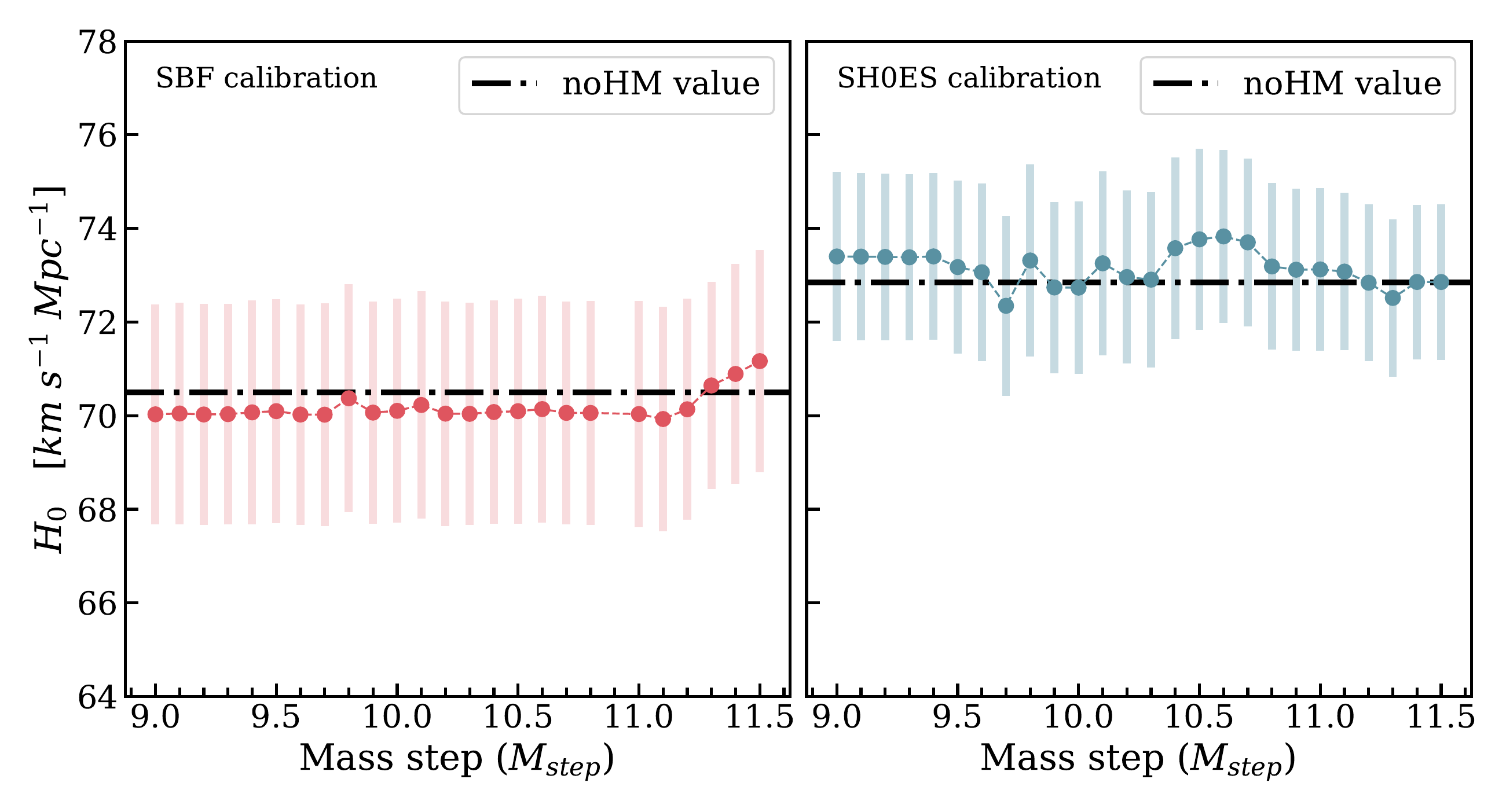}
\caption{\h0 values estimated including the step mass correction in the calibration at different values of $M_{step}$. The SBF calibration with a step mass correction decreases the \h0 value by $\sim 0.5\%$ almost consistently at each $M_{step}$ except at the extreme end. However, for the \shoes\ calibration, step based correction gives fluctuating values. The lowest value is found at $M_{step} = 9.7$ and is 0.7\% lower than the noHM value and the highest is found for $M_{step} = 10.6$, which is 1.4\% higher than the noHM value of \h0.}
\label{fig:Ng2}
\end{figure*}

We add each of the two HM corrections to the calibration relation as in equation \ref{eq:mass} and perform the Hierarchical Bayesian analysis described in Section~\ref{sec:bayesian} using the redshift-cut cosmological sample along with the SBF and \shoes calibration sets one by one. In order to see any possible effect of our arbitrary choices of the mass zero-point ($M_0$) and step mass values ($M_{step}$), we test it by varying these two quantities between 9 and 11.5 in steps of 0.1 and estimating \h0 at each step. Plots showing this test for the cases of linear and step mass correction are given in Figures \ref{fig:Ng1} and \ref{fig:Ng2}, respectively. We find that for the SBF calibration, both step and linear mass corrections give a $\sim 0.7\%$ decrease in \h0 compared to the \h0 estimated without mass correction (noHM) for any chosen value of $M_0$ and $M_{step}$. However, for the \shoes calibration, while the linear correction gives a $\sim 1.3\%$ increase for any value of $M_0$, the step correction gives fluctuating \h0 values when choosing different $M_{step}$ values. At $M_{step} = 9.7$, we find a decrease of 0.7\% in the \h0 value with respect to the noHM calibration, which is consistent with what was found in \citetalias{riess2016} assuming $M_{step} = 10$. Table \ref{tab:H0_mass} lists the corresponding \h0 values from different cases discussed here. 

\begin{table*}[h!]
\centering
\caption{Hubble constant values estimated with applying the host mass correction to redshift-cut cosmological sample calibrated with the SBF and \shoes samples. The first row shows the results of applying a linear mass correction, and the other rows show a mass step correction as described in the text. The last line shows the values of the \h0 estimated without mass correction for comparison.}
\label{tab:H0_mass}
\setlength{\tabcolsep}{20pt} % Default value: 6pt
\renewcommand{\arraystretch}{1.3} % Default value: 1
\begin{tabular}{cccccccccc}
\hline\hline
 \multirow{2}{5em}{Mass Correction } & \multicolumn{2}{c}{SBF Calibration} & \multicolumn{2}{c}{Cepheid Calibration} \\
  \cline{2-5}
 &  $H_0$ &  $\sigma_{H_0}$ & $H_0$ &  $\sigma_{H_0}$ \\
 \hline
 Linear ($M_0 = 11$) &  70.03 & 2.38 & 73.78 & 2.00 \\
  Step ($M_{step} = 10$) & 70.10 & 2.39 & 72.73 & 1.84  \\
  Step ($M_{step} = 9.7$) & 70.02 & 2.38 & 72.35 & 1.92 \\
  Step ($M_{step} = 10.6$) & 70.14 & 2.42 & 73.83 & 1.84 \\
  noHM correction & 70.50 & 2.37 & 72.84 & 1.66 \\
  \hline

\end{tabular}
\end{table*}

The mass corrections for both the  SBF and \shoes calibrations has a small effect on the \h0 estimates (ranging between 0.6\% to 1.4\%). The mass corrections does not resolve the difference among the \h0 estimates from SBF and \shoes calibrations. The \h0 estimate from the SBF calibration remains smaller than that from the SH0ES calibration. 
  
%%%%%%%%%%%%%%%%%%%%%%%%%%%%%%%%%%%%%%%%%%%%%%%%%%%%%%%%%%%%%%%%%%%%%%%%%%%%%%%%%%%%%%%%%%%%%%%%

\section{SN Ia distance comparison}
\label{sec6-comparison}

\begin{figure}[h]
\centering
\hspace*{-0.5cm}
\includegraphics[width=9.0cm,height=7.0cm]{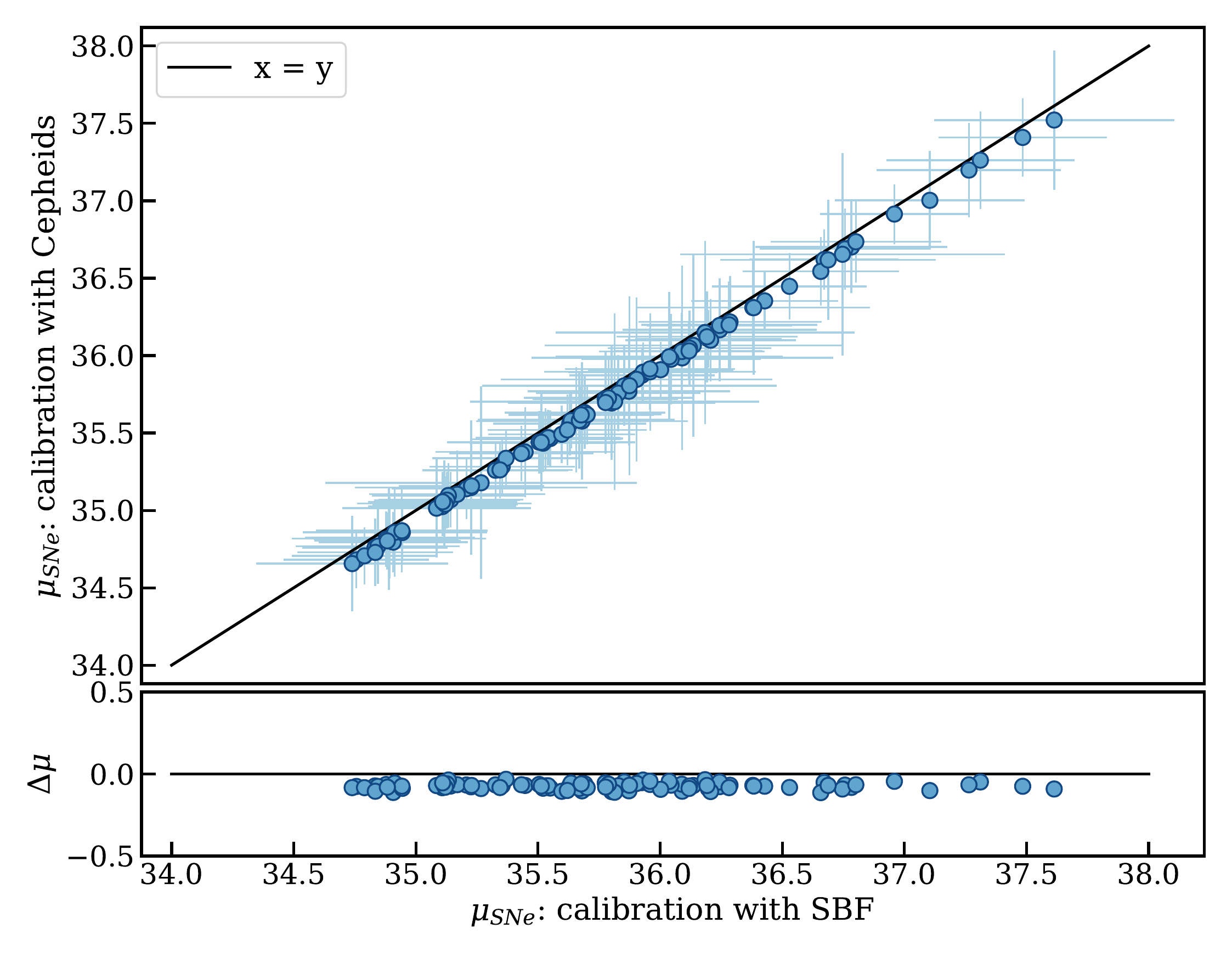}
\caption{Distance moduli of the \sne belonging to the redshift-cut cosmological sample estimated using the Cepheid (y-axis) and the SBF (x-axis) as calibrator. The calibration is performed as described in Sect~\ref{sec3-method}. It does not include the host galaxy mass correction. For a visual comparison, the line x=y is plotted. The bottom panel shows the residual difference among the distances calibrated by Cepheid and SBF, $\Delta\mu = \mu^{Ceph} - \mu^{SBF}$.}
\label{fig:noHm_comp}
\end{figure}

In order to understand the difference in the \h0 value derived using the Cepheid and the SBF calibrations, we now focus on the comparison of the SN Ia distance moduli obtained using the two calibrations. Using the luminosity correlation parameters inferred for SBF and \shoes calibrations (without the host galaxy mass correction), we evaluate the distances for the 96 SNe in the redshift-cut cosmological sample as:
\begin{equation}
\label{eq15}
    \mu = m_B - P^0 - P^1(s_{BV} - 1) - R(m_B - m_V),
\end{equation}
where the correlation parameter values are given in Table \ref{tab:tripp}. The uncertainty $\sigma_{\mu}^{SBF/Ceph}$ in the distance modulus of each object is computed via error propagation including the LC fitting errors and the errors in Tripp parameters computed by the Bayesian analysis. We also add 
%in quadrature 
the intrinsic variance of the calibrator sample $\sigma_{int, calib}$. Figure \ref{fig:noHm_comp} shows the comparison between the distance moduli of the \sne obtained using the SBF (x-axis) and Cepheid (y-axis) calibration about a slope-of-unity line. The residuals ($\mu^{Ceph} - \mu^{SBF}$) are plotted in the bottom panel. The distance moduli estimated with the SBF calibration result to be systematically larger than those estimated with Cepheid calibration (as shown in Fig.~\ref{fig:noHm_comp}). The different \h0 estimates correspond to a mean difference in distance moduli of $0.07$ mag. Adopting the latest LMC distance scale for the SBF and \shoes calibrators as described in Section~\ref{sec4-results}, the mean difference in the distance moduli is found to be 0.066 mag. \par
\begin{figure}[h]
\centering
\hspace*{-0.5cm}
\includegraphics[width=9.0cm,height=7.0cm]{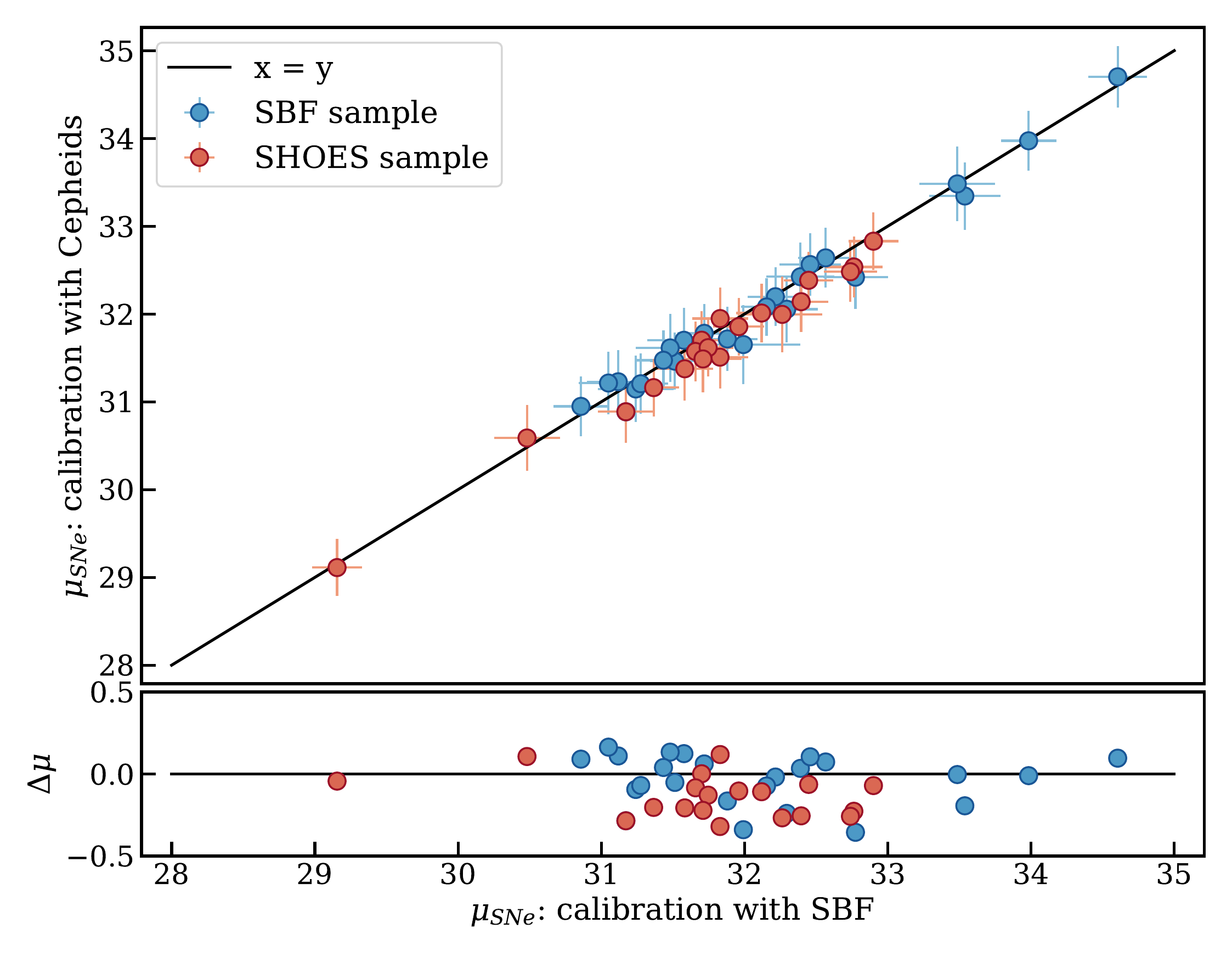}
\caption{Distance moduli of the \sne belonging to the two local samples (SBF and \shoes) estimated using the Cepheid (y-axis) or the SBF (x-axis) as calibrators. The two calibrations used here do not include the cosmological sample in the analysis. It does not include the host galaxy mass correction. For a visual comparison, the line x=y is plotted. The bottom panel shows the residual difference among the distances calibrated by Cepheid and SBF, $\Delta\mu = \mu^{Ceph} - \mu^{SBF}$.}
\label{fig:local}
\end{figure}

In order to examine the origin of this systematic difference, we inspect the SN distances of the two calibration samples (SBF and SH0ES). A direct comparison of distance moduli with SBF and Cepheid techniques requires that \sne happened in galaxies where  both  SBF  and  Cepheid  distance  measurements  are  available. Only one object SN2012cg among our two calibrator samples satisfy this requirement for which $\mu_{SBF} - \mu_{Cepheid}$ is $-0.06$ mag. Since this difference is not statistically significant, we compare the SN distance moduli of the SBF and \shoes samples measured by performing the same analysis as described in Section~\ref{sec3-method} but without including the cosmological sample (i.e., only $\mathcal{L}_{calib}$ in the likelihood). Figure~\ref{fig:local} shows the comparison of the distance moduli calibrated with the SBF and Cepheids for the two local calibration samples. For SNe of both the SBF and the \shoes sample, the SBF calibration gives a longer distance scale than the Cepheid calibration, indicating that the difference observed in the distance moduli of the cosmological sample comes directly from intrinsic differences in the local calibration samples. \par

\cite{ajhar} made a similar comparison using 14 galaxies that host SNe Ia. They compared the SBF distances of these galaxies with the SN distances estimated using the Cepheid calibration by \citetalias{freedman2001}, and found them in agreement. In our SBF sample there are nine \sne in common with their paper \citep[for five of them, the SBF measurement used in this paper comes directly from ][]{ajhar}. For these nine objects, we find good agreement ($\Delta \mu$ $\sim$ -0.01 mag) between the SN distances calibrated with \shoes Cepheids (performing our analysis using only the local sample) and the SN  distances calibrated with Cepheids by \cite{ajhar} taht are taken from Table 3, column 5 of their paper. This comparison limited to 9/24 galaxies seems to exclude a systematic offset associated with the calibration using \shoes Cepheids \citepalias{riess2016} and HST KP Cepheids \citepalias{freedman2001}.

\begin{figure}[]
\centering
\hspace{-2em}
\includegraphics[width=9.3cm, height=5cm]{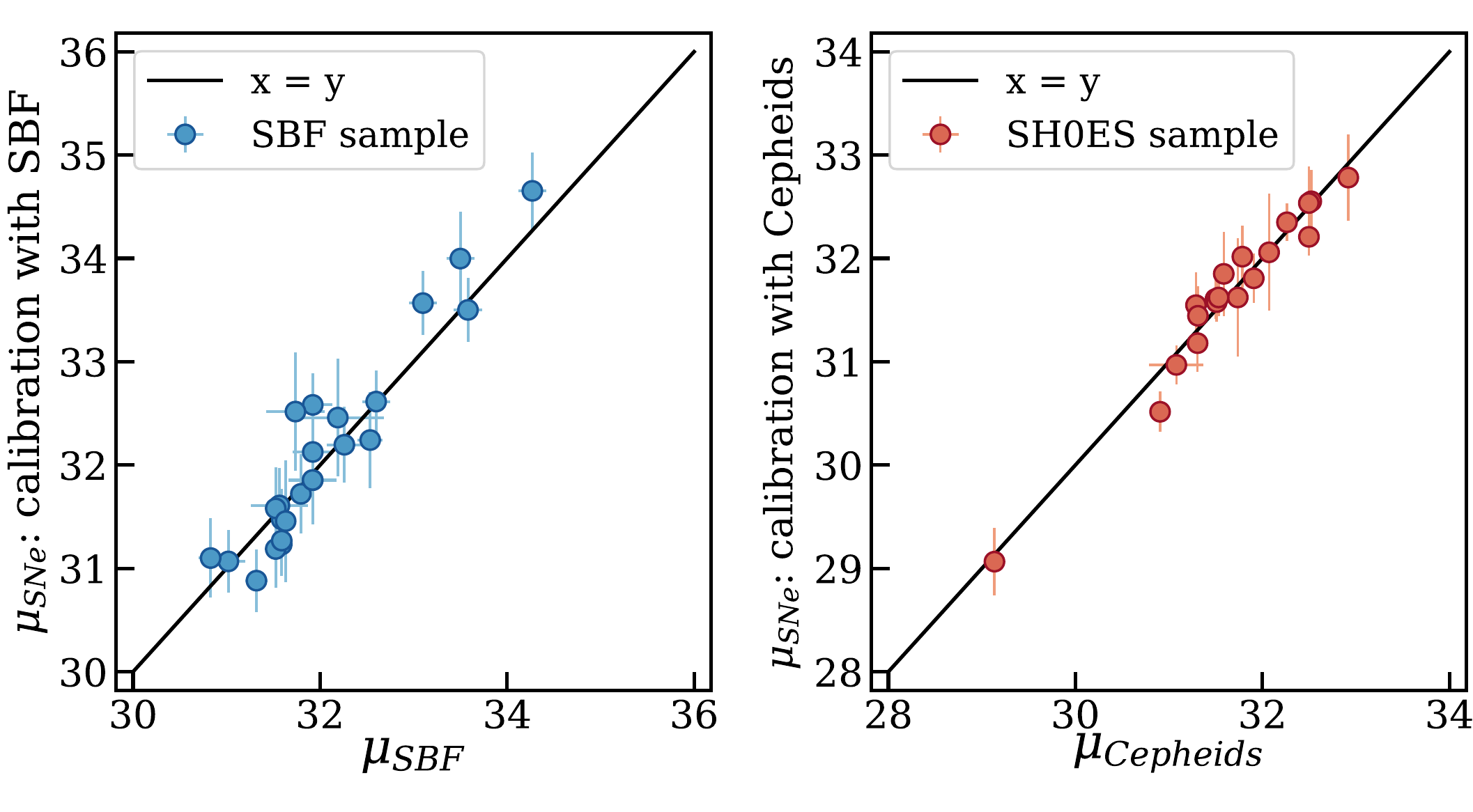}
\caption{Distance modulus comparisons of the two local calibration samples. The left plot shows the comparison between directly measured SBF distances ($\mu_{SBF}$, given in  Table~\ref{tab:tabel1}) for the SBF sample, with the SNe distances estimated using the SBF calibration of the cosmological redshift-cut sample of 96 SNe as described in Section~\ref{sec3-method}. The right plot shows similar comparison for the \shoes sample, comparing the measured Cepheid distances (from \citetalias{riess2016}) with the distances estimated using the Cepheid calibration of the cosmological redshift-cut sample. In both cases we find a good one-to-one agreement.}
\label{fig:app5} 
\end{figure}

In Figure~\ref{fig:app5} we show a comparison between the directly measured SBF distances ($\mu_{SBF}$, given in  Table~\ref{tab:tabel1})  versus the SN distances estimated using the SBF calibration (left plot) for the SBF sample. In the same figure we also show the plot for the \shoes sample, where we compare the directly measured Cepheids distances (from \citetalias{riess2016}) with the distances estimated using the Cepheid calibration. In both cases we find a good one-to-one agreement. In comparison to the Cepheids sample, the SBF sample shows a larger scatter (as indicated by $\sigma_{int,calib}$ for SBF).

\begin{figure}[h]
\centering
\includegraphics[width=9.0cm,height=7.0cm]{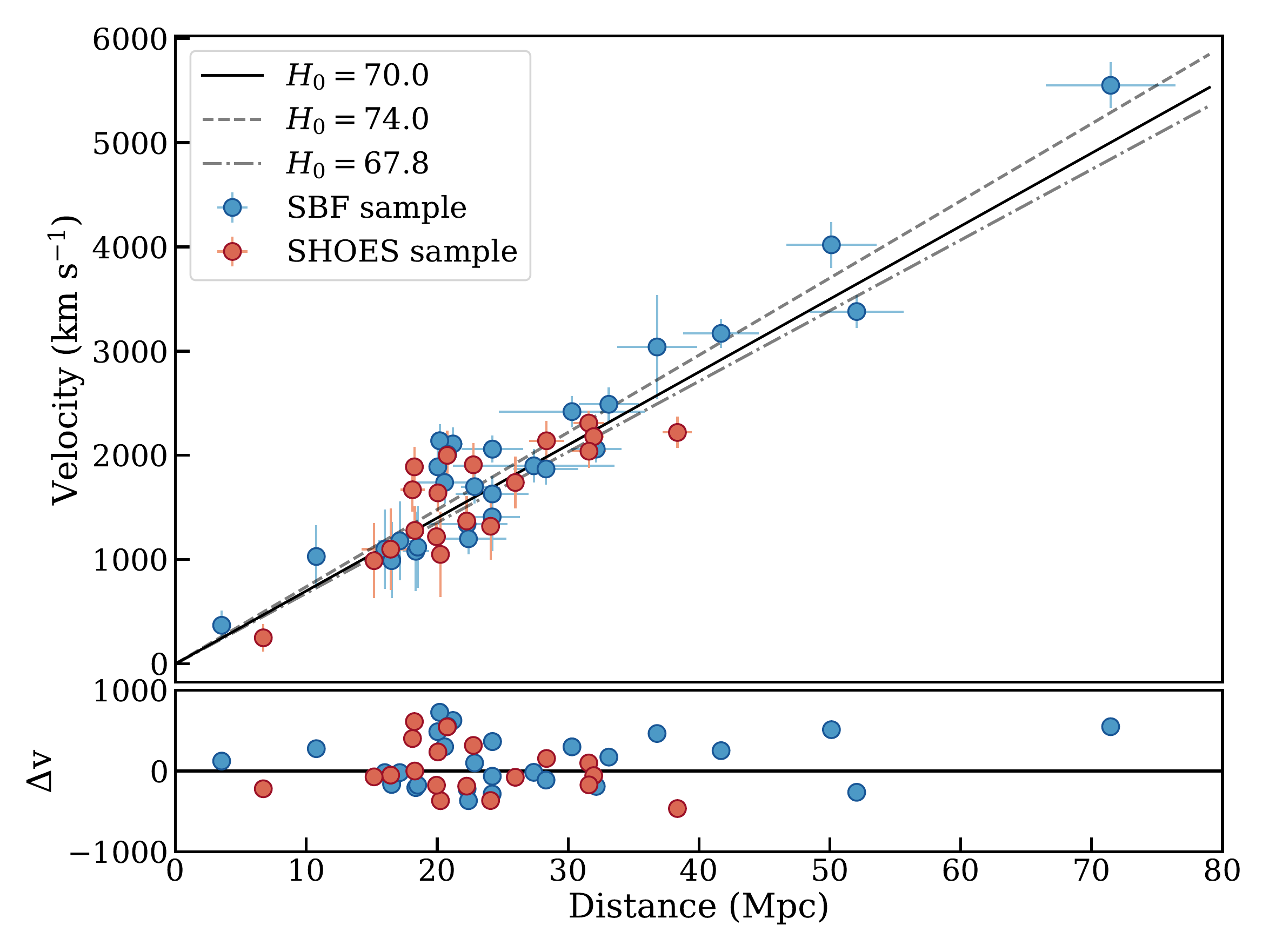}
\caption{Nearby Hubble diagram for galaxies with SBF and Cepheid distance measurements. The recessional velocities for all galaxies are corrected for peculiar motions with the CF model following \citet{Carrick2015}. As a reference, also shown is the Hubble law for the values of the Hubble constant of $H_0 = 70.0$ \hunit\ (solid line), $H_0 = 74.0$ \hunit\ (dashed line) and $H_0 = 67.8$ \hunit\ (dot-dashed line). The bottom panel shows the residuals for both the SBF and \shoes samples calculated assuming a $H_0$ of 70 \hunit.}
\label{fig:nearby}
\end{figure}

Figure~\ref{fig:nearby} shows the nearby Hubble diagram for the two calibrator samples, that is, the galaxy recessional velocity versus the distance estimated with SBF (blue data) and with the Cepheids (red data). At such distance scales, the peculiar velocities are significant with respect to the Hubble recessional velocity and need to be corrected. Here, the recessional velocity of each galaxy is corrected for peculiar velocities with the Cosmic Flow (CF) model following the analysis performed in \citet{Carrick2015}. This model takes into account the influence of the large-scale structures in the local Universe. The plot shows that the two samples are equally distributed and that the SBF sample reaches higher distances than the Cepheid one, enabling a larger distance range to calibrate cosmological distances.

%%%%%%%%%%%%%%%%%%%%%%%%%%%%%%%%%%%%%%%%%%%%%%%%%%%%%%%%%%%%%%%%%%%%%%%%%%%%%%%%%%%%%%%%%%%%%%%

\section{Discussion}
\label{sec7-discussion}

The SBF distances have been directly used to estimate $H_0$ as proposed for the first time by \cite{ferrarese2000}. This work presents the first attempt to use the SBF measurements to calibrate the peak luminosity of SNe Ia, and thus deriving SN distances and the Hubble constant value using this alternative calibration. In our analysis, we found a mean difference of 0.07 mag 
between the distance moduli of the cosmological samples estimated using the SBF calibration and the ones estimated using the Cepheid calibration (the corresponding difference is 3.3\% between the \h0 estimates from them). This difference cannot be accounted for by an identifiable offset on the Cepheid calibration used for the SBF measurements with respect to the \shoes sample (see discussion on LMC distance and Cepheid $P$--$L$ relations in Section~\ref{sec2-data} and comparison with \cite{ajhar} in Section~\ref{sec6-comparison}).

Even though we are not able to clearly identify the cause of the difference between the SBF and Cepheid calibrated distances, our results seem to indicate that there are intrinsic differences in \sne hosted in different types of environments, which are not accounted for by applying a simple host-mass correction. The different SN light curve behavior in the two samples could be attributed to differences in their SN Ia progenitors (e.g., \citealp{Mannucci2006,Maoz2014,Livio2018}) since the SBF sample is mainly composed of early E/S0 type galaxies, while the \shoes sample consists of late-type spiral galaxies. \cite{rigault2015} showed that \sne in locally star-forming environments are dimmer than \sne hosted in locally passive environments. In this scenario the larger distance moduli given by SBF could be due to \sne exploding in older environments. Considering the general evidence that in early-type galaxies we generally observe older Population II stars while in late-type galaxies we also observe young Population I stars, differences in the evolution of the lightcurve in the first $\sim 50$ days of the SN emission could be expected due to the different physical properties and composition of the SN ejecta, which can affect the amount of Fe-peak elements produced in the SN explosion.
Another ingredient is how dust extinction influences the SN Ia light curves. Comprehensive lightcurve modeling suggests that the main source of intrinsic scatter for the observed SN Ia emission is from the extinction parameter $R$, which reflects variation of the dust around the \sne \citep{Brout2020}, although a recent detailed work on NIR SN Ia lightcurves seems to exclude the dust as main driver of the host galaxy and SN Ia luminosity correlation \citep{Ponder2020}. In line with these results, we find slightly lower ($\Delta R \sim 0.15$) value for the SBF sample average extinction parameter $R$ with respect to the \shoes sample, see Table \ref{tab:H0}. However, this is not enough to explain the observed difference in the cosmological samples calibrated with the two methods, given that the average color ($m_B - m_V$) values of the three samples in this work are much smaller than 1 mag. The mean color for the SBF sample is 0.09 mag, for the \shoes sample it is 0.07 mag, and for the cosmological sample it is 0.1 mag.

\subsection*{Hubble tension}
\begin{figure}[h]
\hspace*{-0.5cm}
\centering
\includegraphics[width=9.0cm,height=7.0cm]{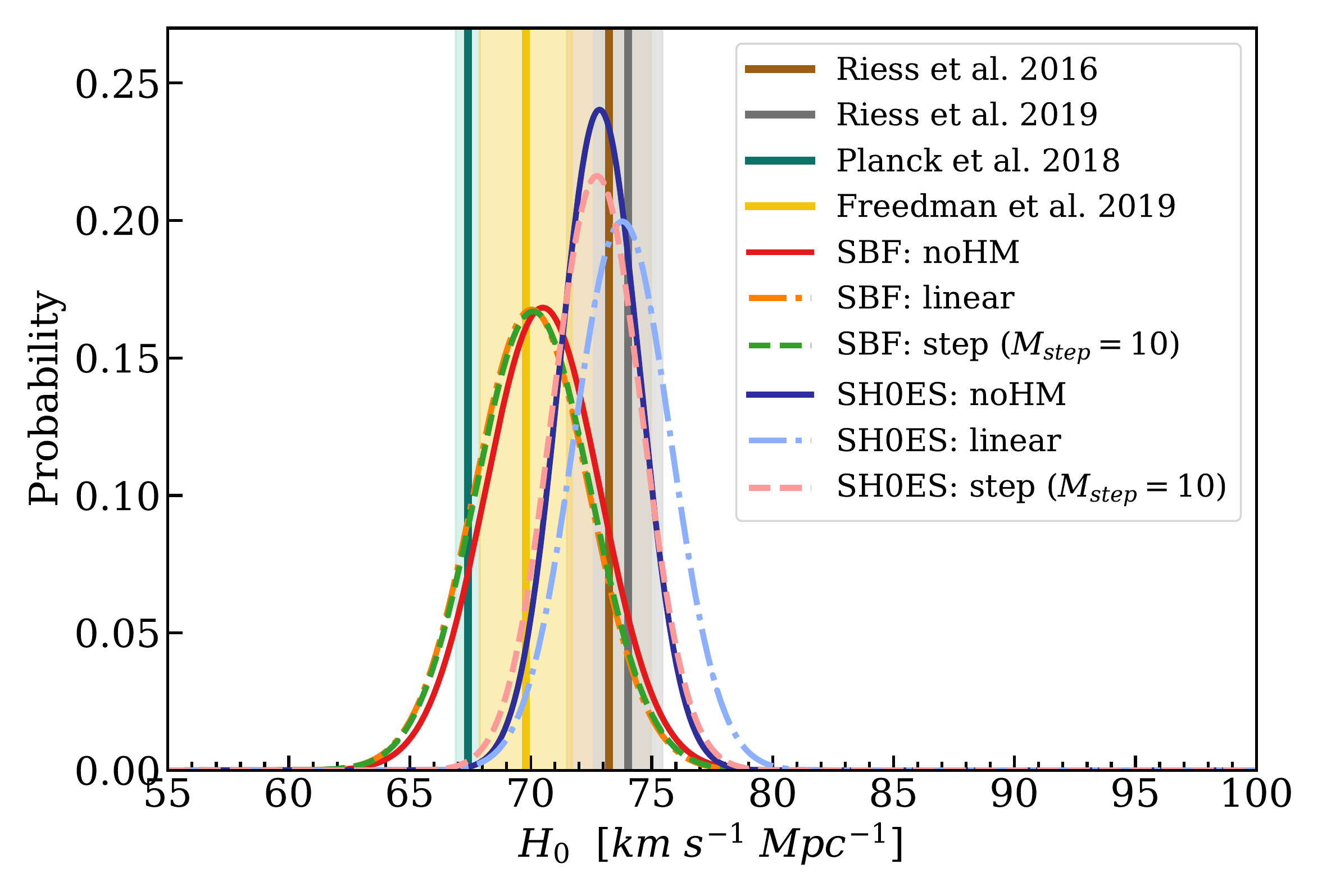}
\caption{Posterior distributions of \h0 estimated in this work with the SBF and \shoes\  calibrator samples. We give the distributions for noHM analysis (solid line), including a linear mass correction (dot-dashed line) and including a step-mass correction with $M_{step} = 10$ (dashed line).}
\label{fig:H0_dist}
\end{figure}

Our final \h0 value estimated from SBF calibration using 24 \sne applied to the redshift-cut cosmological sample is $H_0 = 70.50 \pm 2.37$ (stat) $\pm 3.38$ (sys) \hunit. This \h0 value, obtained with SBF calibration stands $\sim1.3 \sigma$ away from both the Planck \h0 estimate of the early Universe \citep{planck2018} and the \shoes program \h0 estimate \citep{riess2019}, when we only take into account the statistical errors. Our \h0 value is in good agreement with the recent estimate by \cite{trgb} based on the TRGB calibration of SNe Ia. As pointed out by \cite{Freedman2020}, TRGB stars populate the gas- and dust- free halos of the host galaxy in contrast to the Cepheids, which are found in the higher-surface-brightness disk regions. TRGB stars sample environments more similar to the SBF galaxies. Figure \ref{fig:H0_dist} summarizes the \h0 measurements in the present work for the various cases of noHM and HM corrections with SBF and \shoes calibrators, and shows their comparison with measurements by \cite{planck2018}, \cite{trgb}, \cite{riess2016}, and \cite{riess2019}. 

\subsection*{Perspectives for the SN Ia SBF calibration}

The present work uses both ground based and HST SBF optical data. In the future, we expect major improvements in this regard by using dedicated observations by the \textit{James Webb Space Telescope (JWST)}. The SBF method works better in the NIR because the main source of the brightness fluctuations comes from red giant branch stars which are brighter at redder wavelengths \citep{blakeslee2010} and less affected by dust extinction. The red giants are excellent targets for \textit{JWST}. However, the SBF calibration is presently not well-constrained in the NIR bands compared to the optical ones \citep{jensen2015}. SBF offers a complementary tool to calibrate SN Ia luminosity with respect to the Cepheids by sampling a set of different type of host galaxy environment. Although SBF is currently a secondary distance indicator, as it is dependent on Cepheid calibration, the theoretical calibrations that will eventually make it an independent technique in the distance ladder \citep{cantiello2005} are improving.

Furthermore, SBF represents an experimental methodology that is able to anchor the distance ladder up to larger distances with respect to the Cepheid calibrations (see e.g., Fig.\ \ref{fig:nearby}). As SNe Ia are rare events, reaching larger distances will provide more galaxies that host a SN Ia, giving us larger number of calibrators, which is very important to decrease the statistical errors and reach a percent level precision goal. In this work, the measured scatter in the B-band absolute magnitudes of the fiducial calibrating sample is 0.27 mag. With 24 SBF calibrators, the uncertainty in mean absolute magnitude is 0.05 mag, which corresponds to about 2.5\% uncertainty in distance. While the SBF sample is expected to largely increase with the future instruments and newly discovered \sne, the increase in the number of Cepheid calibrators will be limited by the smaller distance necessary for Cepheid measurements, and thus the smaller number of galaxies possibly hosting a SN explosion. The importance of having a larger number of SN Ia calibrators is also highlighted in \cite{trgb}, and by \cite{Huang2020} where Mira variables have been used to calibrate \sne and to measure the Hubble constant. 

%%%%%%%%%%%%%%%%%%%%%%%%%%%%%%%%%%%%%%%%%%%%%%%%%%%%%%%%%%%%%%%%%%%%%%%%%%%%%%%%%%%%%%%%%%%%%%%%

\section{Conclusion and perspectives}
\label{sec8-conclusion}

The primary goal of this work is to show the potential of the SBF method to provide an alternative distance scale for the local Universe aimed at calibrating the absolute magnitudes of \sne and measuring the Hubble constant. We built a set of 24 \sne calibrators that have distance measurements to their hosts, mostly early-type galaxies, obtained with the SBF technique. We applied the SBF calibration to a sample of 96 \sne with redshifts between $z=0.02$ and $z=0.075$ (obtained from the Combined Pantheon Sample) and derived a value of $H_0 = 70.50 \pm 2.37\ \rm (stat) \pm 3.38\ (sys)$ \hunit\ (i.e., $3.4\% stat, 4.8\% sys$). This value lies in between the value obtained with SNe Ia calibrated with Cepheids and that inferred from the analysis of the CMB; it is consistent with both of them within the errors.

\begin{figure*}[h]
%\hspace*{-5cm}
\centering
\includegraphics[width=12.7cm,height=11.2cm]{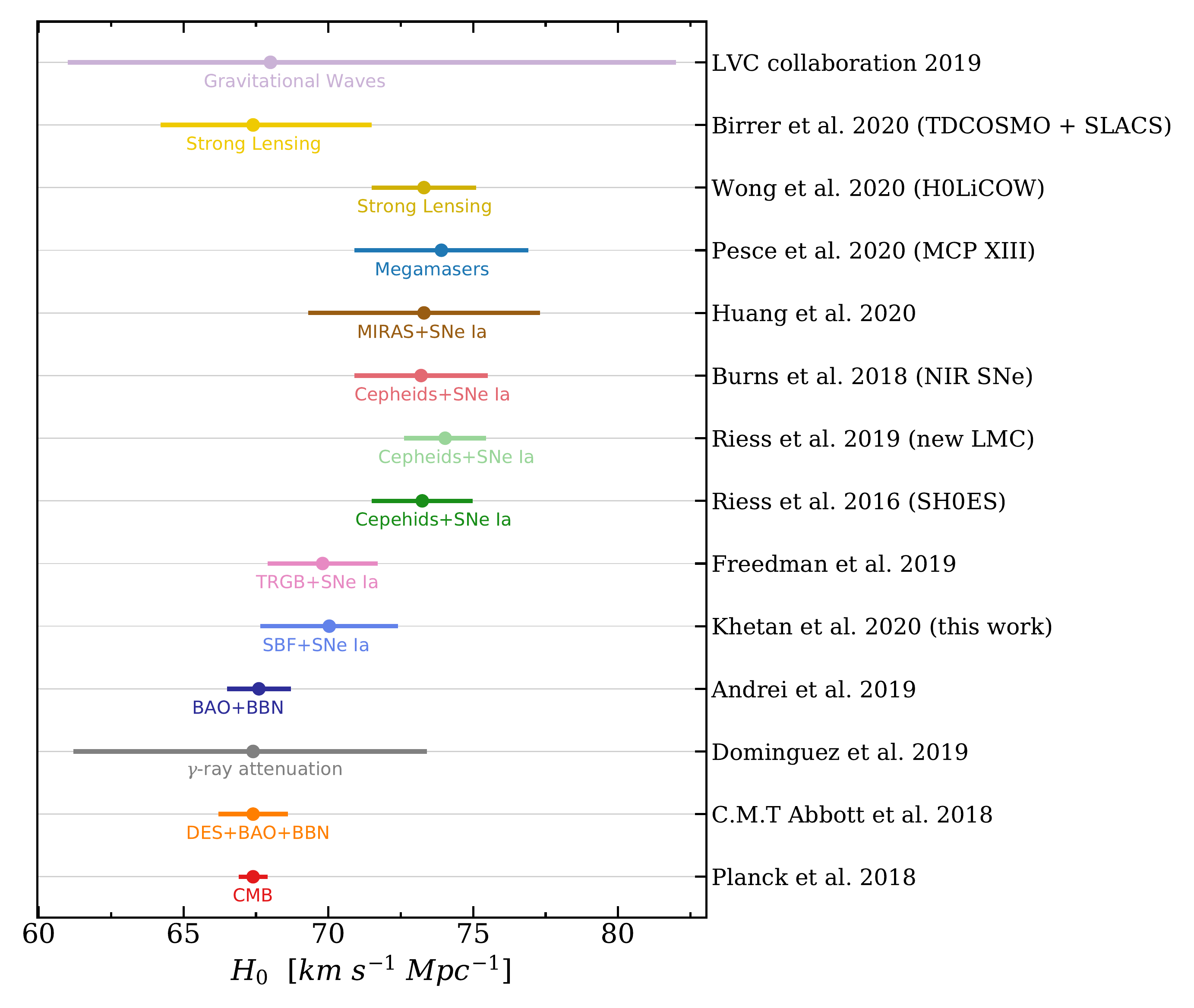}
\caption{Compilation of Hubble constant values obtained using different observations and techniques from the recent literature including the value from this work. The literature references are written on the y-axis. Two independent estimates from early Universe \citep{planck2015,cmtabbott2018} are shown at the bottom. The next is an estimate using extragalactic background light $\gamma$ ray attenuation \citep{GammaH02019} and another from BAO at all redshifts + BBN estimate \citep{BAO_h02019}. Then we show measurements from \sne calibrated with TRGB \citep{trgb}, \sne calibrated with Cepheids \citep[\shoes\ sample,][]{riess2016, riess2019}, using near-infrared (NIR) filters \citep{burns2018} and \sne with Mira variables \citep{Huang2020}. Then we show the \h0 values estimated using 6 masers in the Hubble flow \citep{Pesce2020}. The next \h0 value shown is inferred via gravitational lensing time delays using six lensed quasars \citep{wong2020} and a more recent value obtained using 40 strong lenses \citep{Birrer2020}. Finally we shows the \h0 derived with gravitational-wave signals from binary compact object mergers \citep{LVCH02019}.}
\label{fig:H0_ref}
\end{figure*}

We found a systematic difference of 0.07 mag among the distances estimated using the SBF calibration and the ones using the Cepheid calibration (see Figure~\ref{fig:noHm_comp}). This accounts for the $sim 3.3$\% smaller \h0 value obtained using SBF calibration with respect the one using the \shoes sample as calibrator. This also explains the $\sim 5$\% larger \h0 value of \cite{riess2019}, which uses \shoes sample as calibrators, compared to the SBF.

Although we are not able to completely exclude a hidden primary calibration offset, the observed difference could be attributed to the different host properties of the SBF and \shoes calibrator samples. Cepheids are usually observed in late-type galaxies while SBF can be measured only for homogeneous, passive environments, such as early-type and lenticular $S0$ galaxies. In terms of SNe Ia, different host galaxy types can translate into: a) a difference in the intrinsic dust reddening or immediate extinction, possibly due to the presence of local dense circumstellar medium; b) a different stellar population for the underlying SN progenitor, for example, due to the existence of multiple channels for the formation of the binary systems leading to a SN Ia explosion \citep{Mannucci2006,Foley2020}; c) a difference in the metallicity or chemical composition of the underlying progenitor, which can lead to a different light curve evolution \citep{Maoz2014,Livio2018}. At the moment we can neither confirm nor exclude any of these possible scenarios. We believe that additional observations and analysis, in particular at NIR wavelengths, are needed to shed light on this problem. Compared to optical LCs, NIR \sne LCs have a narrow luminosity distribution and are less sensitive to host galaxy dust extinction~\citep{Avelino2019}. Moreover, the possibility of investigating the immediate environments, using, for example, integral field spectrographs, of nearby \sne can provide important clues to the fundamental physical properties of the circumstellar gas surrounding SN progenitors.

Our analysis shows that applying a correction for the host-galaxy stellar mass in the luminosity calibration relation does not reduce or correct for the possible SN Ia luminosity dependence on galaxy types (see Section ~\ref{sec5-masscorrection}).
This suggests the need for alternative parameter(s) that could account for the variation in the luminosity of \sne hosted in different environments. This is particularly timely taking into account the upcoming observations of innovative observatories (e.g., Vera C. Rubin Observatory, \textit{JWST}) that are expected to increase the number of detected \sne, in particular at larger redshift.

Today the value of local Universe \h0 is known with an uncertainty of less than 10\%. However, Figure \ref{fig:H0_ref}, which shows the current status of \h0 estimates with different methodologies, reveals the existence of a dichotomy in the \h0 measurements; a first group of measures characterized by a central value below 70 \hunit\, and a second one centered above 73 \hunit. The current \textquotedblleft{tension}\textquotedblright\, on the \h0 measurements is not only limited to CMB and Cepheid measurements but instead involves a dozen of different methods, mostly independent of each other. Our results together with the other data reported in Figure \ref{fig:H0_ref} suggest that there is a certain margin to  interpret the discrepant results in terms of systematics while relaxing the quest for new physics.

%%%%%%%%%%%%%%%%%%%%%%%%%%%%%%%%%%%%%%%%%%%%%%%%%%
   
\begin{acknowledgements}
We are grateful to Christopher Burns for his very useful comments and help with Light curve fitting using \texttt{SNooPy}. We thank the SN group at the Observatory of Padova, INAF for hosting NK as short-term visitor and for the valuable discussions on the nature of SNe Ia and their observation. We also extend our thanks to the researchers at DARK, Copenhagen for hosting NK as short-term visitor and discussing various details of the analysis and comments on the manuscript. NK is grateful to Jeff Cooke and CAS (Swinburne University) Melbourne for hosting her as a visiting PhD student. NK would further like to acknowledge helpful discussions with Chris Blake and Suhail Dhawan. MB, EB, EC and MC acknowledge financial support from MIUR (PRIN 2017 grant 20179ZF5KS). This work was supported by a VILLUM FONDEN Investigator grant to JH (project number 16599) and by a VILLUM FONDEN Young Investigator grant to CG (project number 25501). This project is funded by the Danish Council for Independent Research under the project `Fundamentals of Dark Matter Structures', DFF - 6108-00470. We would also like to extend our thanks to Adam Riess and Dan Scolnic for useful discussions. We are grateful to the referee for their comments.
\end{acknowledgements}

   \bibliographystyle{aa} % style aa.bst
   \bibliography{ref} % your references Yourfile.bib

\begin{appendix}
\newgeometry{top=20mm, bottom=20mm, left = 15mm, right= 15mm}
\section{Light curve fitting with SNooPy}
\label{app:LCfitting}

\begin{table*}[h!]
\centering
\caption{Literature references for the optical photometry data of the \sne in the SBF calibrator sample.}
\label{tab:LCref}
% tab_LCreference.tex created at UT 2019-11-25 17:11:13
\begingroup
\setlength{\tabcolsep}{25pt} % Default value: 6pt
\renewcommand{\arraystretch}{1.2} % Default value: 1
\begin{tabular}{|c|c|c|}
\hline\hline
Supernova & Host Galaxy & Photometry Reference \\
\hline
%SN2012fr   & NGC1365   & \cite{contreras2018} \\
SN2000cx   & NGC524    & \cite{li2001} \\
%SN1991T    & NGC4527   & \cite{lira1998} \\
SN1994D    & NGC4526   & \cite{richmond1995} \\
SN2007on   & NGC1404   & \cite{stritzinger2011} \\
SN2012cg   & NGC4424   & \cite{vinko2018} \\
SN1980N    & NGC1316   & \cite{hamuy1991} \\
%SN1981B    & NGC4536   & \cite{tsvetkov1982} \\
SN2003hv   & NGC1201   & \cite{silverman2012} \\
SN2008Q    & NGC524    & \cite{brown2014} \\
SN1970J    & NGC7619   & \cite{cadonau1990} \\
SN1983G    & NGC4753   & \cite{cadonau1990} \\
SN2014bv   & NGC4386   & \cite{brown2014} \\
SN2015bp   & NGC5839   & \cite{brown2014} \\
SN2016coj  & NGC4125   & \cite{richmond2017} \\
SN1981D    & NGC1316   & \cite{hamuy1991} \\
SN1992A    & NGC1380   & \cite{altavilla2004} \\
SN2018aoz  & NGC3923   & Ni et al. \textit{in prep} \\
SN2011iv   & NGC1404   & \cite{gall2018} \\
SN2006dd   & NGC1316   & \cite{stritzinger2010} \\
SN1992bo   & E352-057  & \cite{SN1992bo} \\
SN1997E    & NGC2258   & \cite{jha2006} \\
SN1995D    & NGC2962   & \cite{riess1999} \\
SN1996X    & NGC5061   & \cite{riess1999} \\
SN1998bp   & NGC6495   & \cite{jha2006} \\
SN2017fgc  & NGC0474   & Burke et al. \textit{in prep}\\
SN2020ue   & NGC4636   & Khetan et al. \textit{in prep}\\
%SN1998bu   & NGC3368   & \cite{jha1999} \\
%SN2017ejb  & NGC4696   & GSP + DLT40 \\
%SN2006X    & NGC4321   & \cite{contreras2010} \\
%SN1991bg   & NGC4374   & \cite{leibundgut1993} \\
%SN1986G    & NGC5128   & \cite{phillips1987} \\
%SN2011eh   & NGC3613   & \cite{brown2014} \\
%SN2005ke   & NGC1371   & \cite{contreras2010} \\
%SN2003gs   & NGC936    & \cite{silverman2012} \\
\hline
\end{tabular}
\endgroup
\end{table*}
\restoregeometry 

\newgeometry{top=20mm, bottom=20mm, left = 15mm, right= 15mm}     % use whatever margins you want for left, right, top and bottom.

\begin{table*}[h!]
\centering
\caption{B-band LC fits of the 24 \sne in the SBF calibrator sample. The fits are done using \texttt{SNooPy}.}
\label{tab:lcfits}
% table4.tex created at UT 2019-11-14 18:17:16
\newcommand{\addpic}{\includegraphics[width=4.5cm,height=3.0cm]}
\begin{tabular}{c c c c c c c c}\label{app:SBF}

\addpic{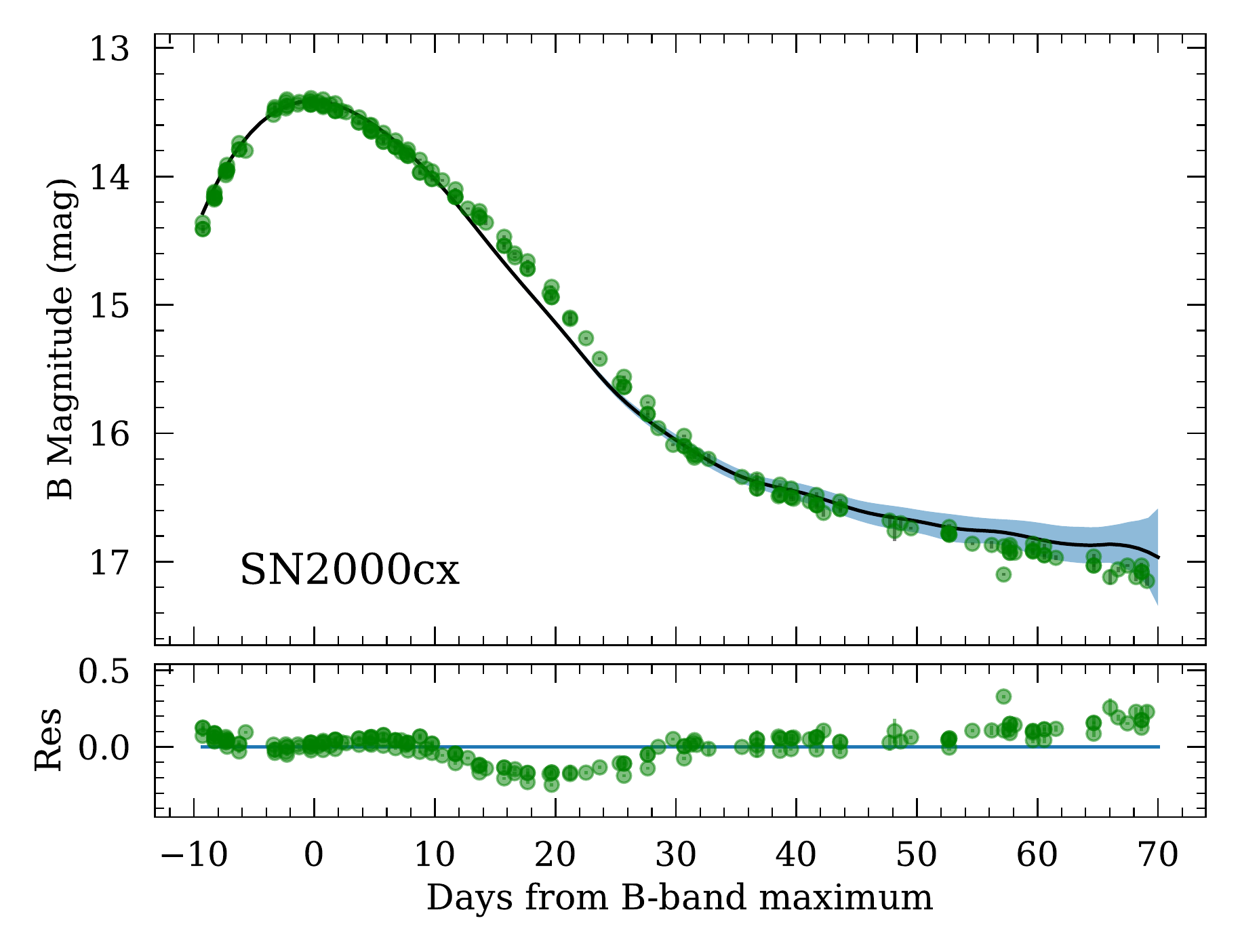} &
%\addpic{lcfits/SN1991T_Bmag.pdf} &
\addpic{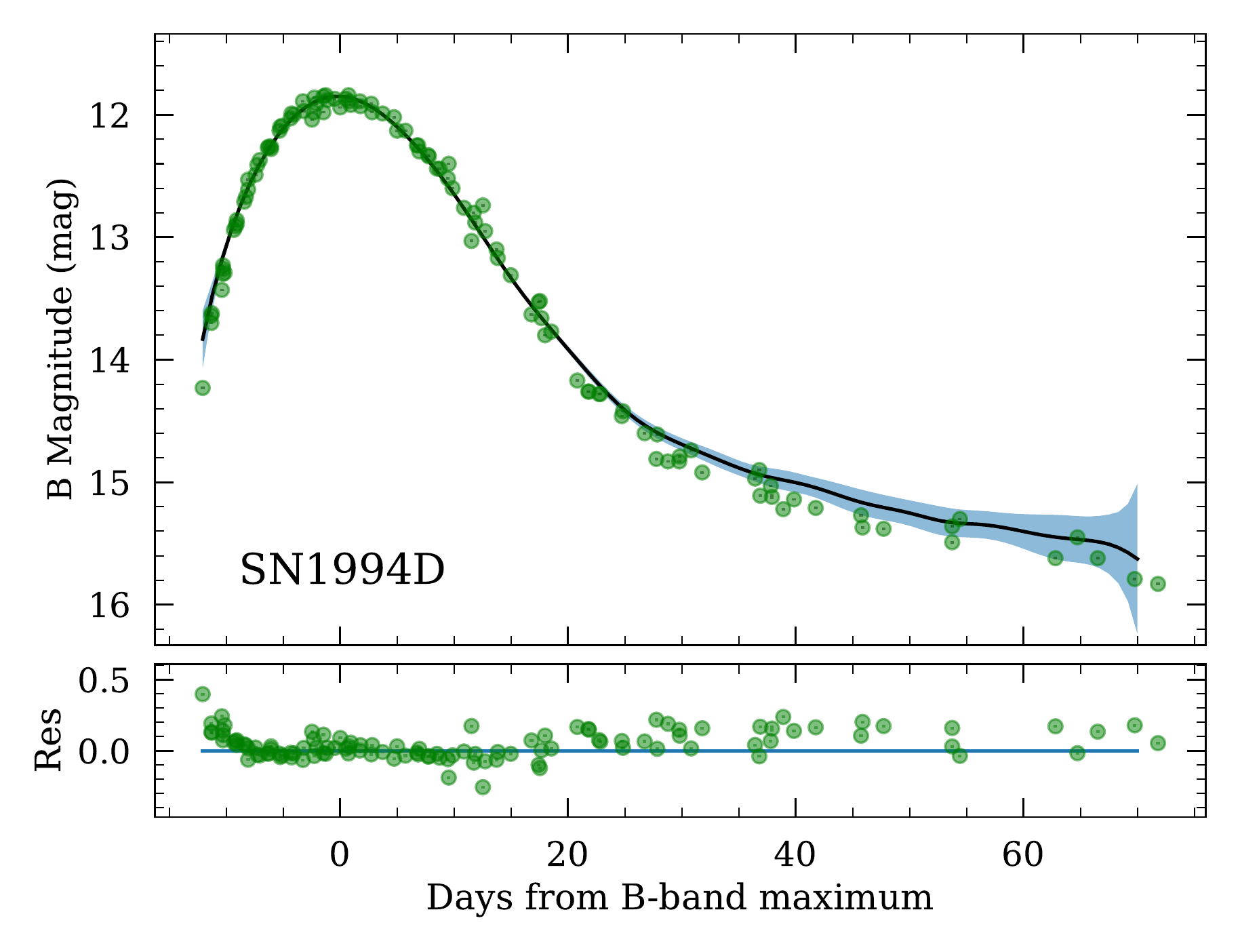} &
\addpic{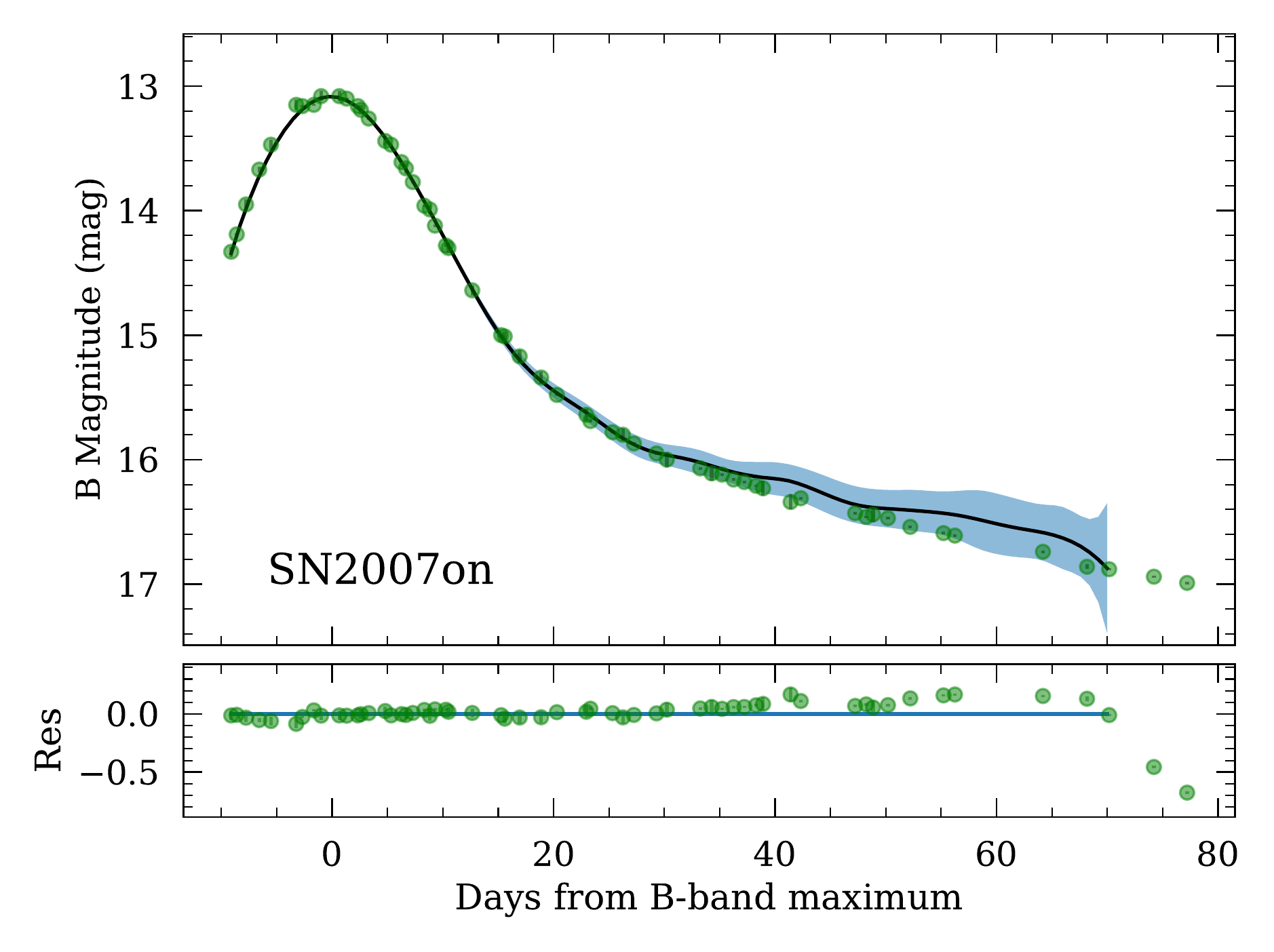} &
\\
\addpic{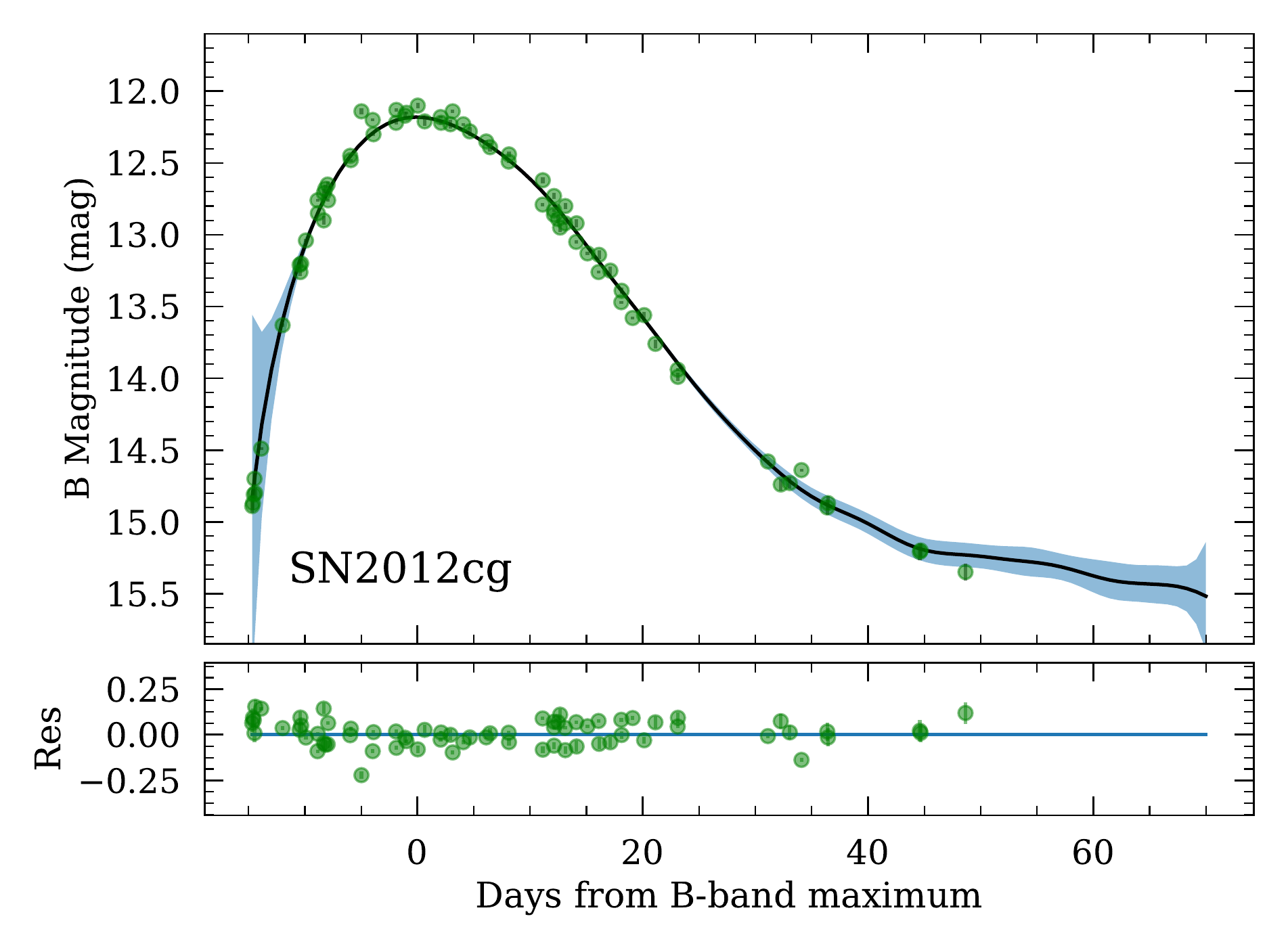} &
\addpic{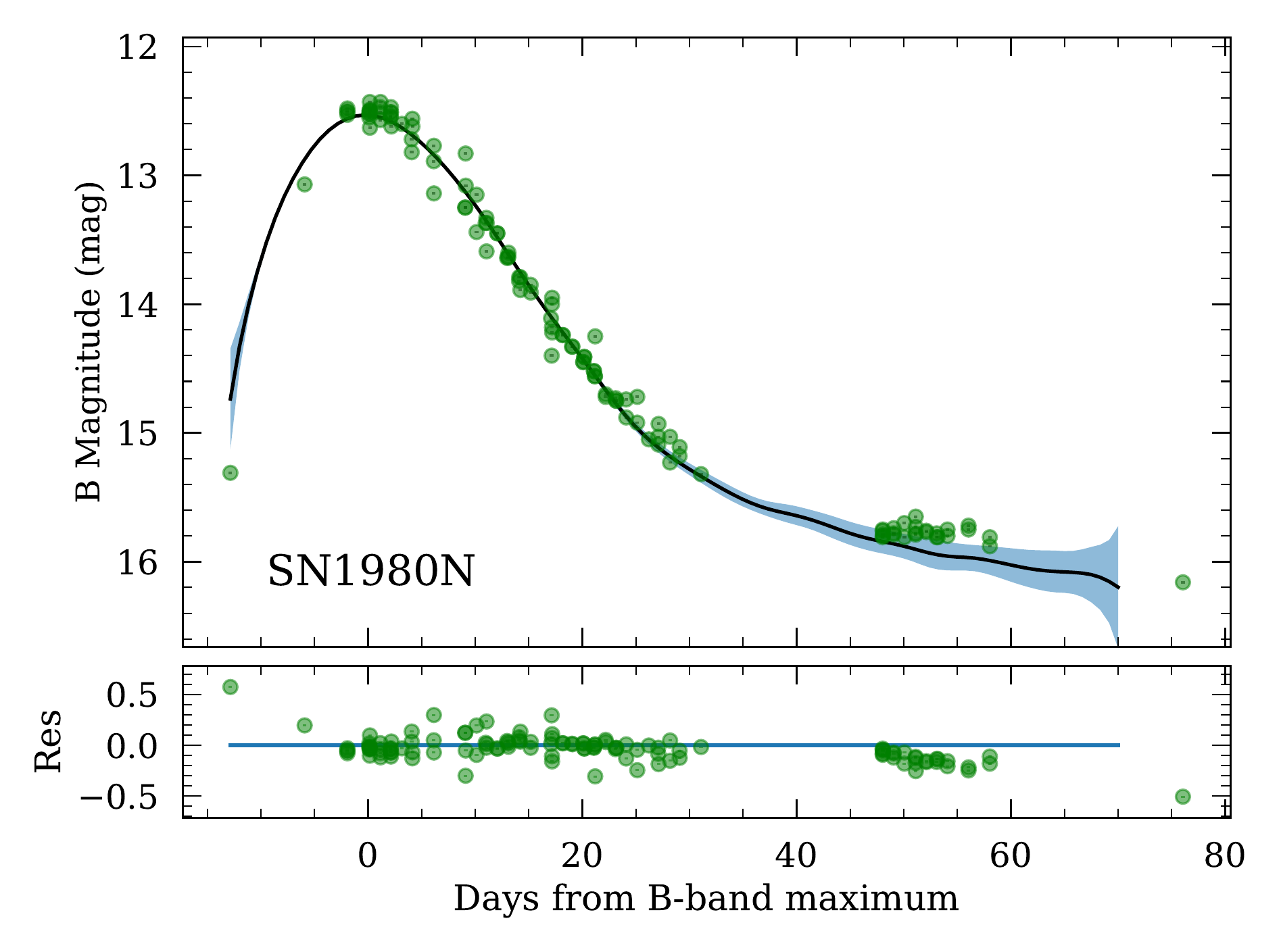} &
%\\
%\addpic{lcfits/SN2012fr_Bmag.pdf} &
%\addpic{lcfits/SN1981B_Bmag.pdf} &
\addpic{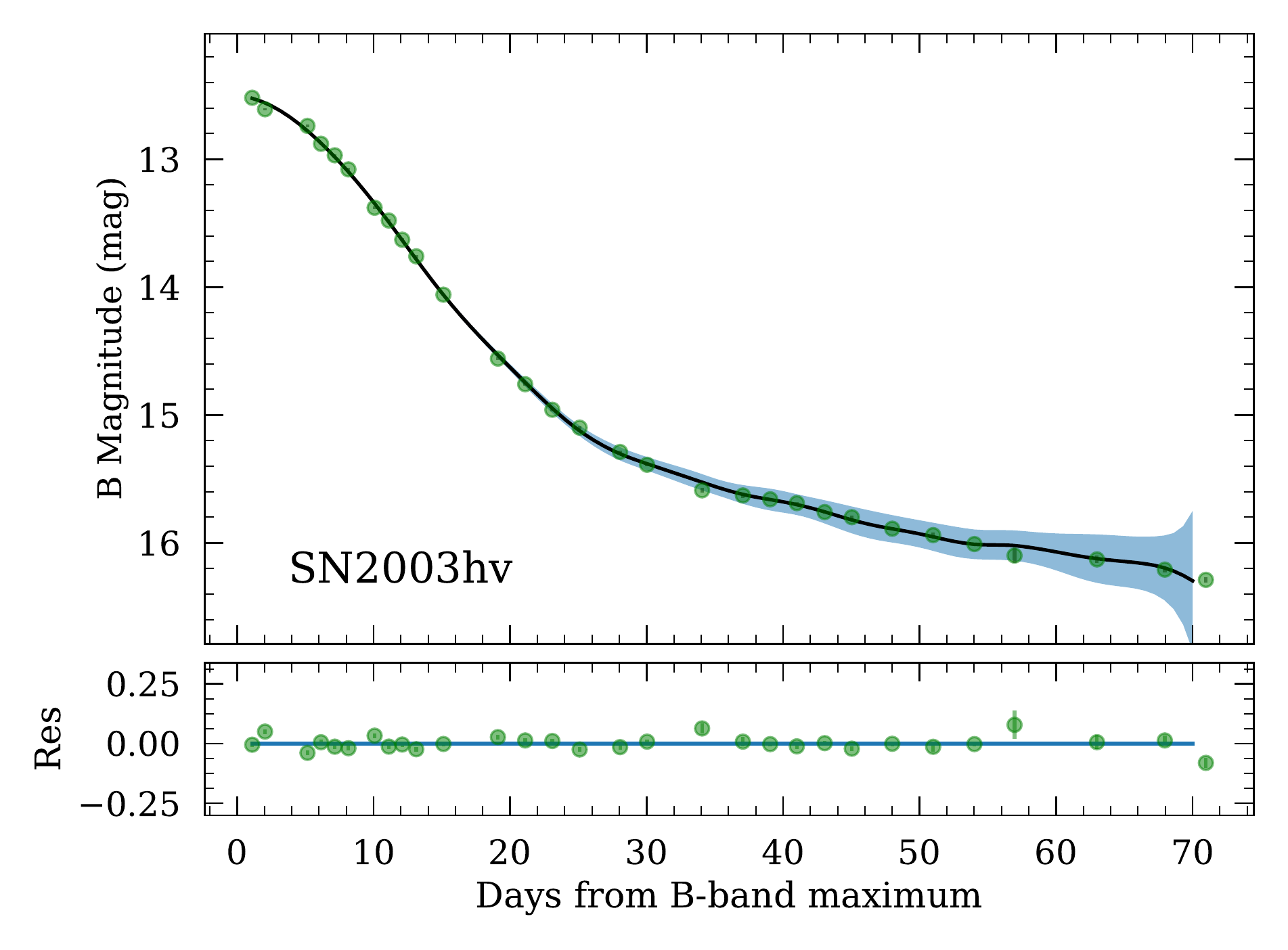} &
\\
\addpic{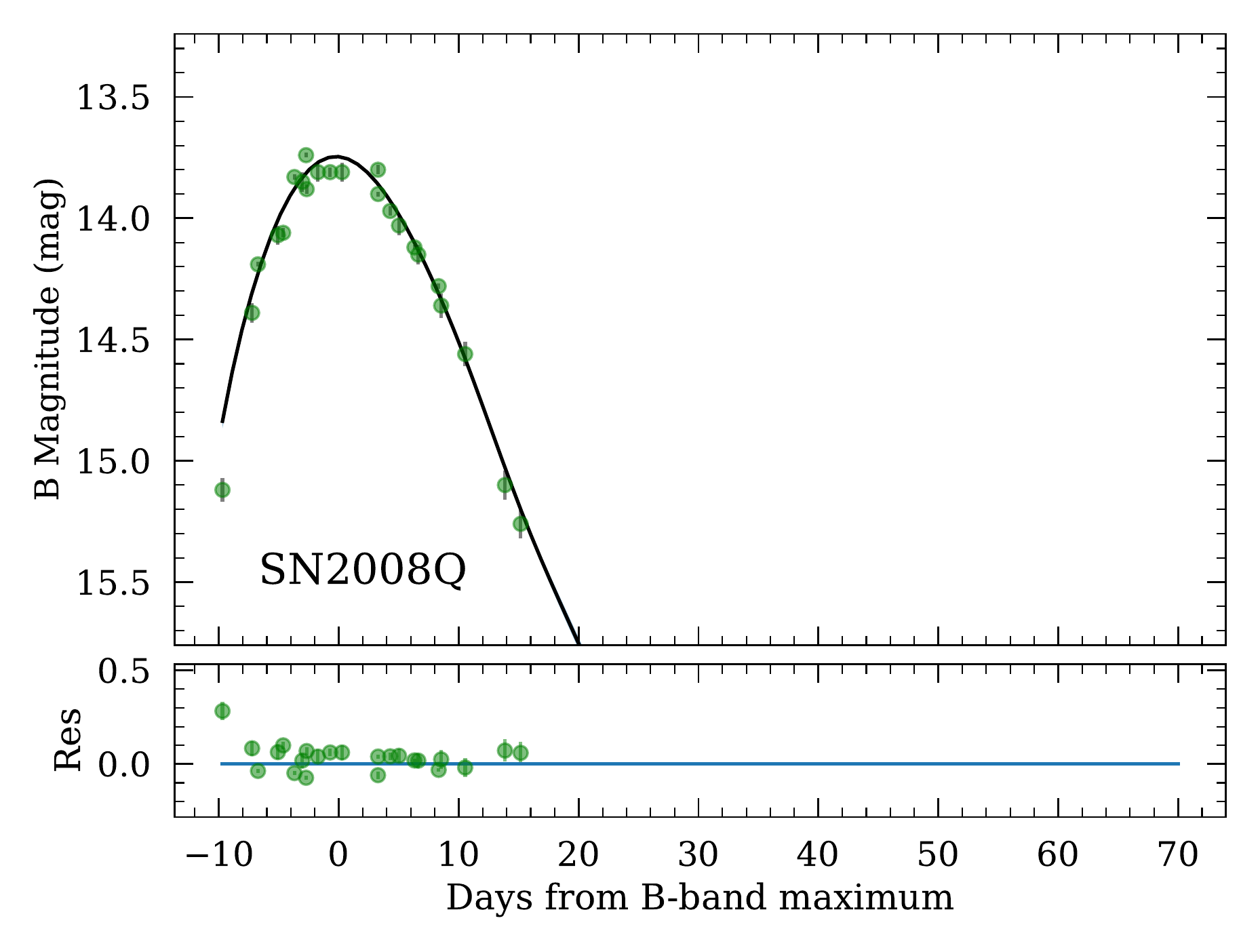} &
\addpic{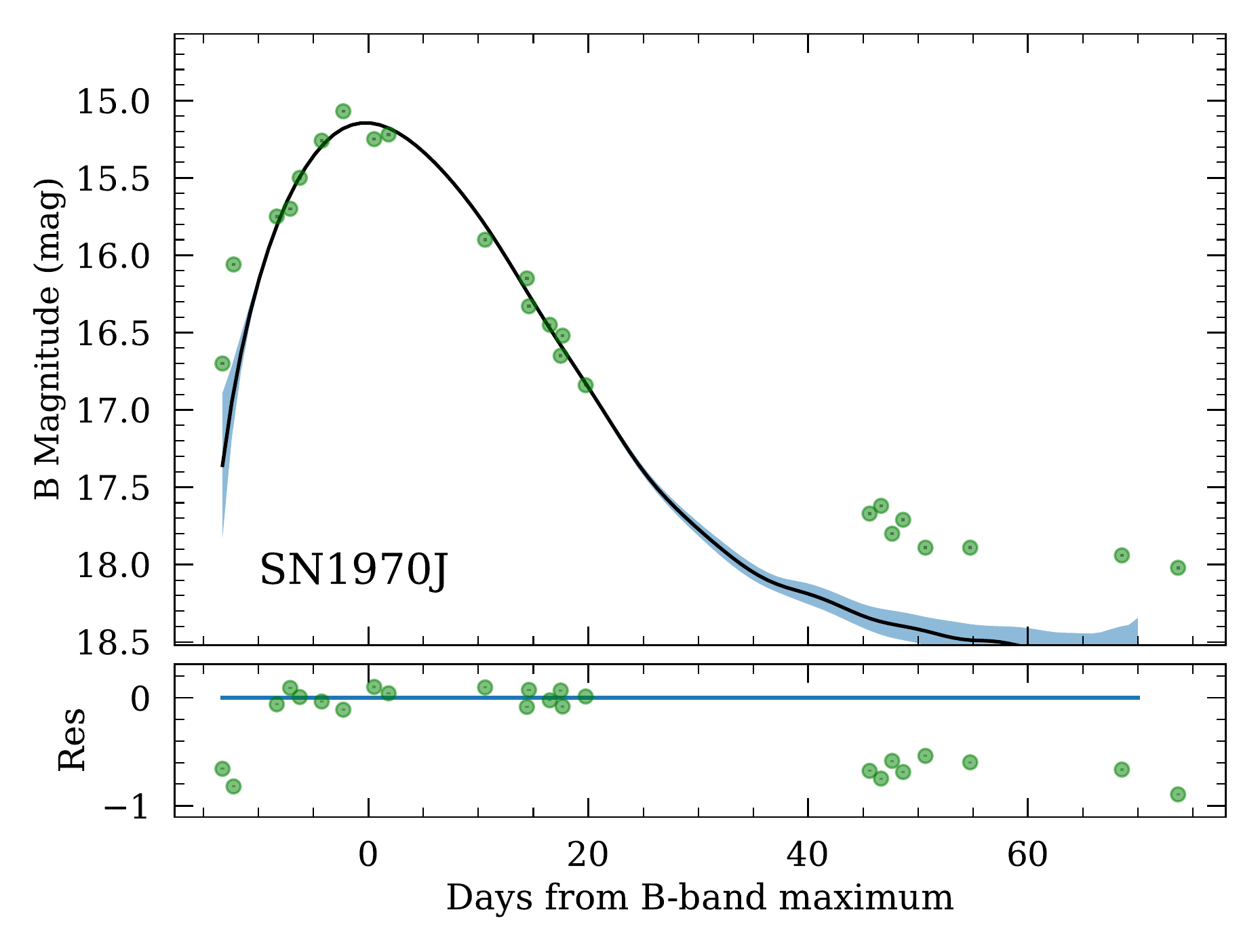} &
\addpic{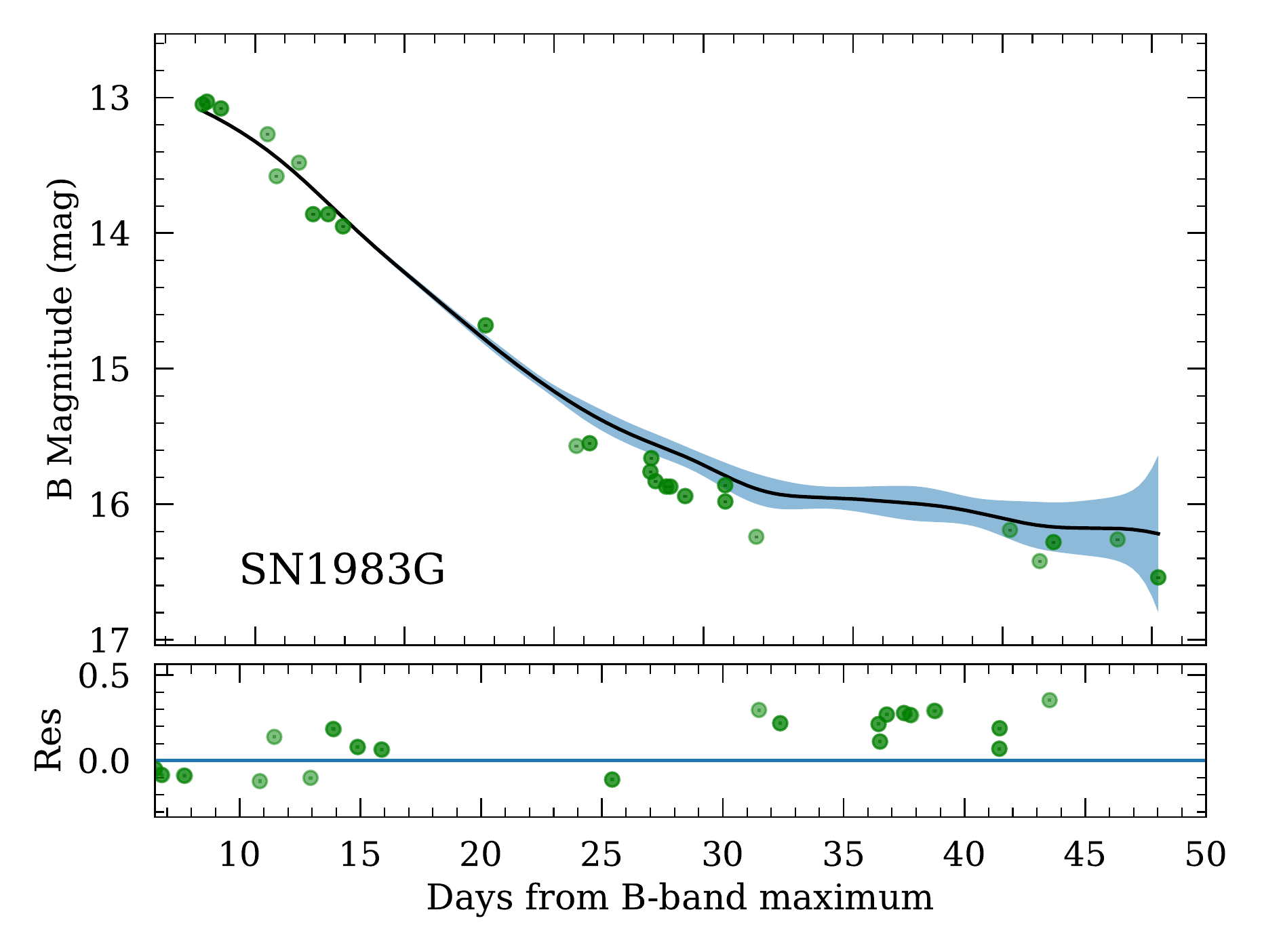} &
\\
\addpic{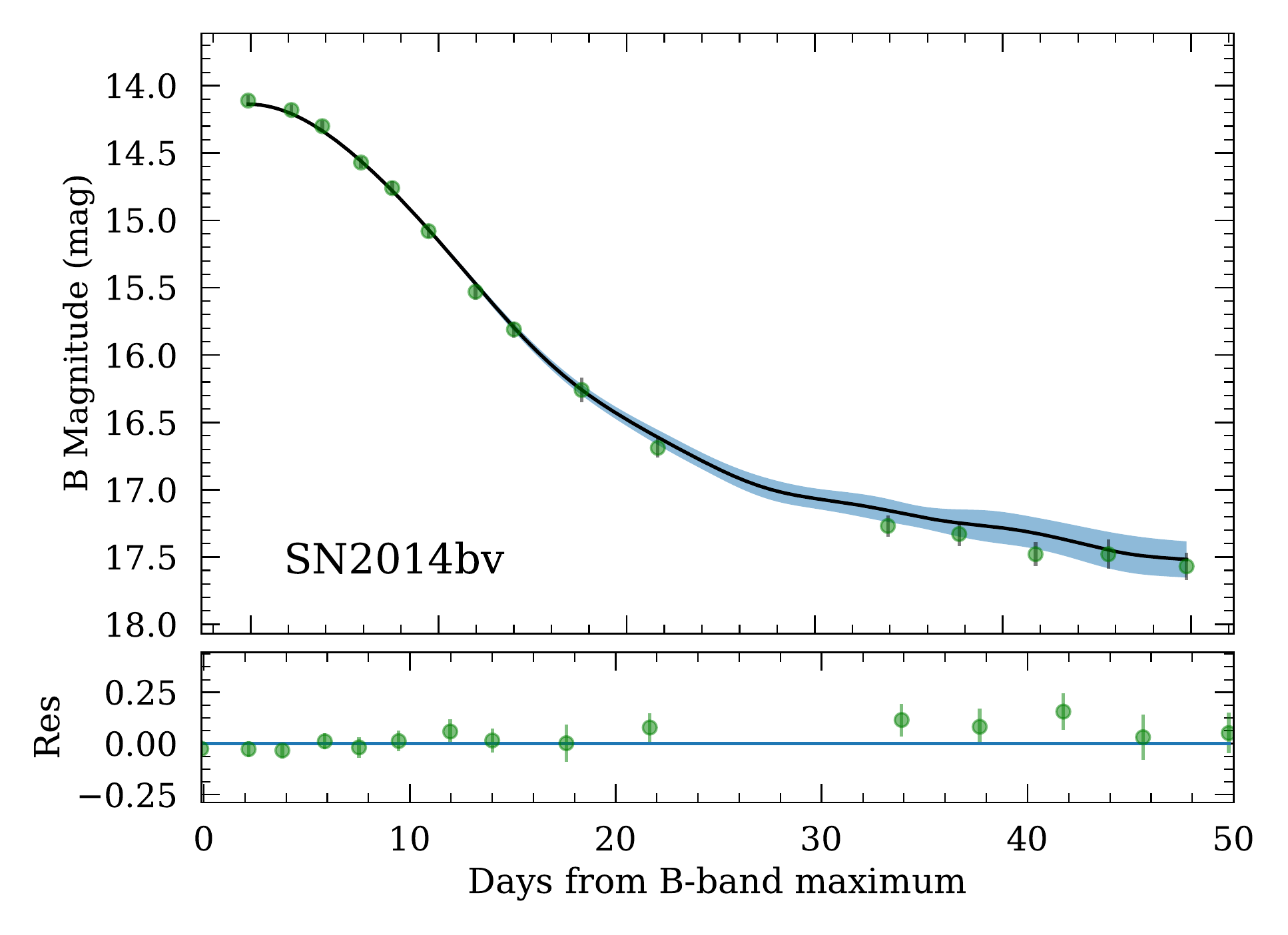} &
\addpic{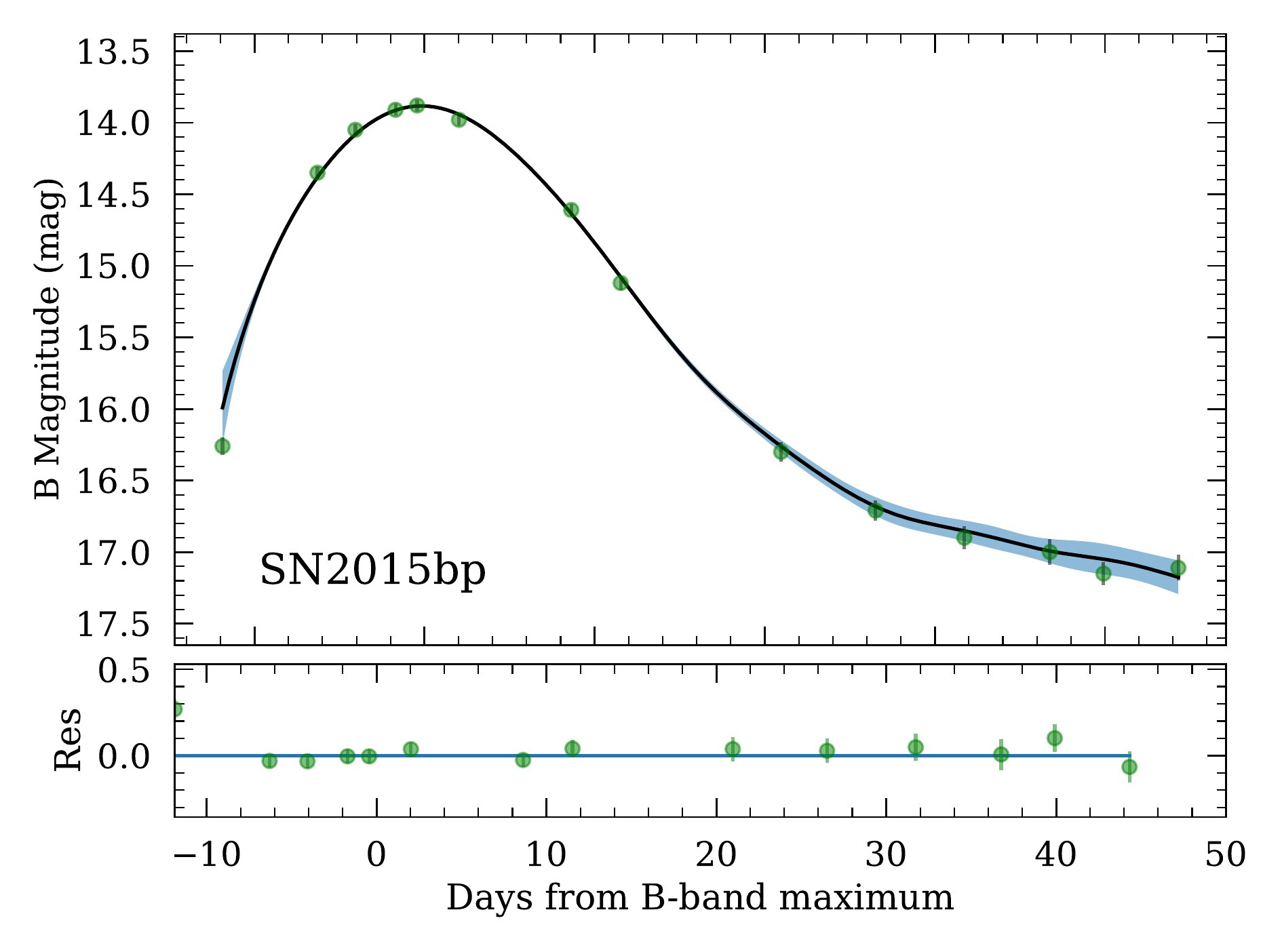} &
\addpic{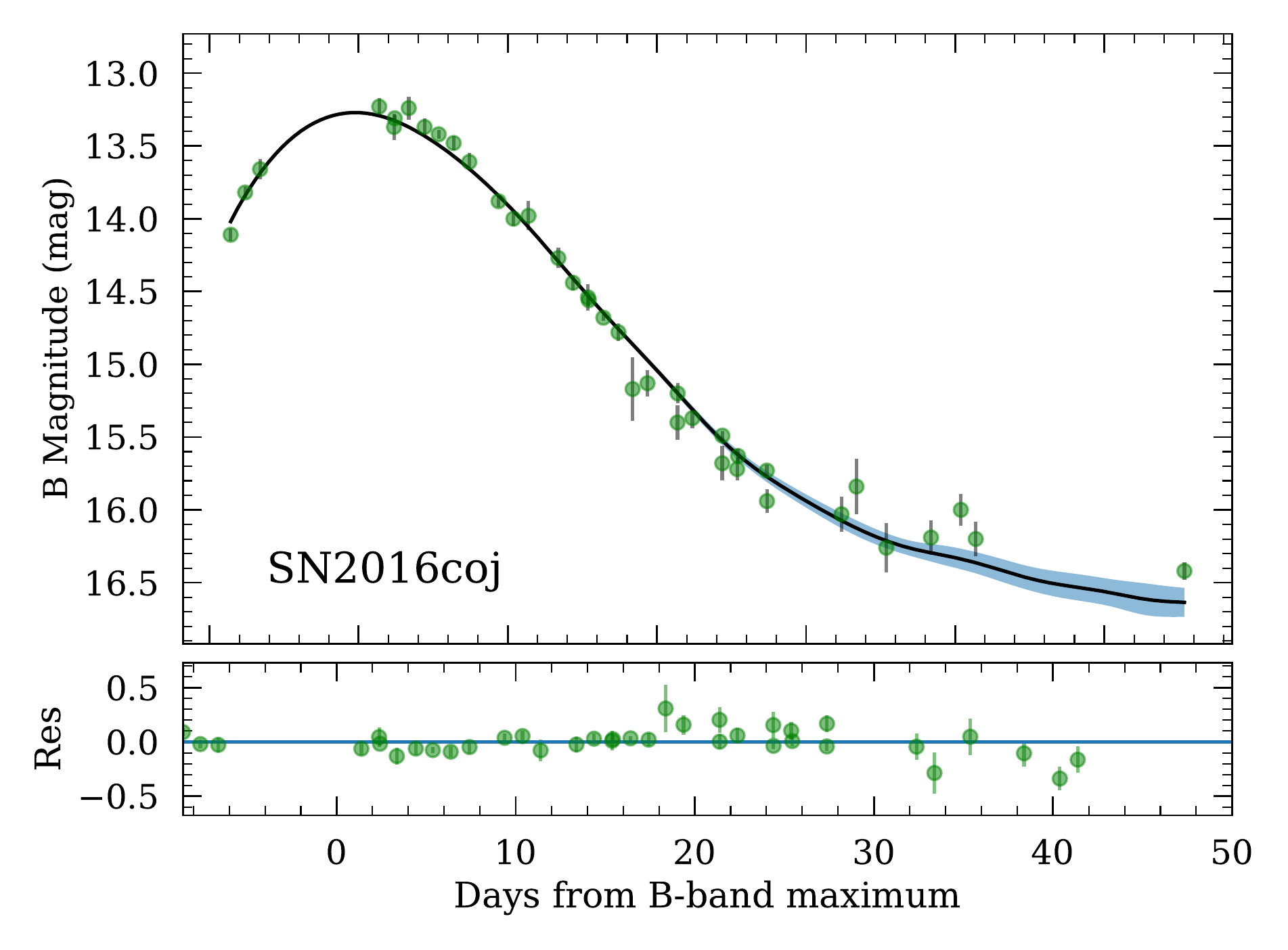} &
\\
\addpic{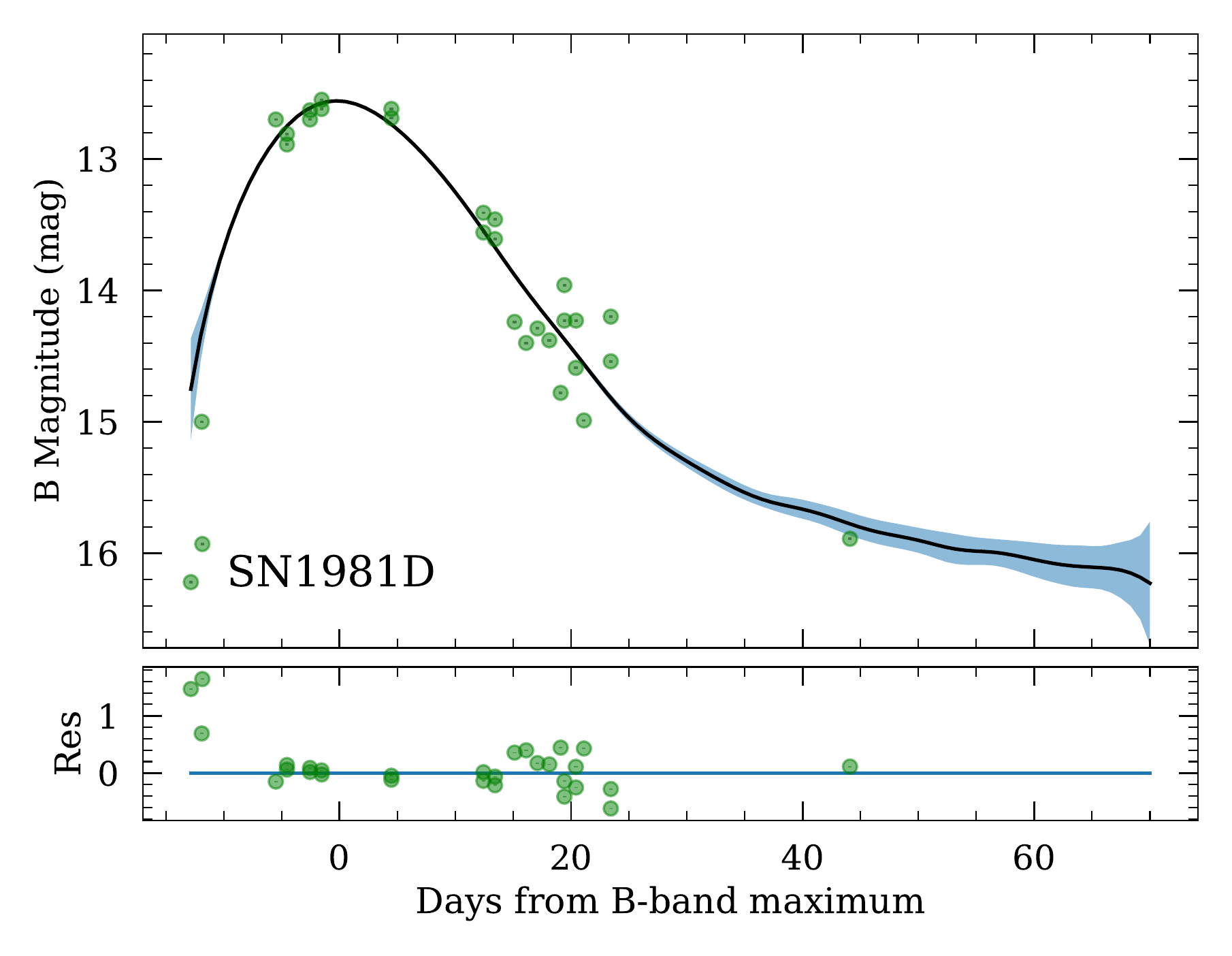} &
\addpic{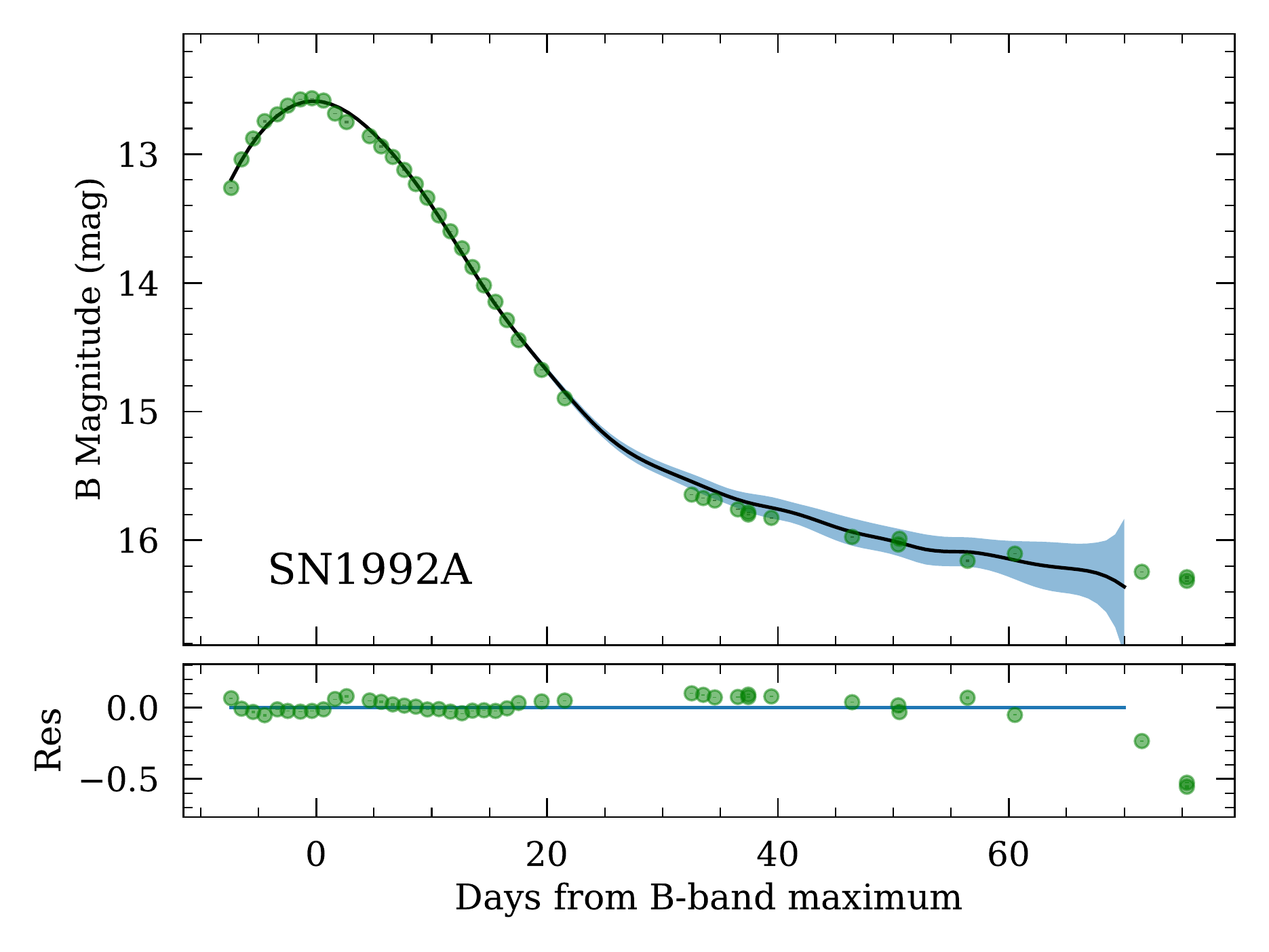} &
\addpic{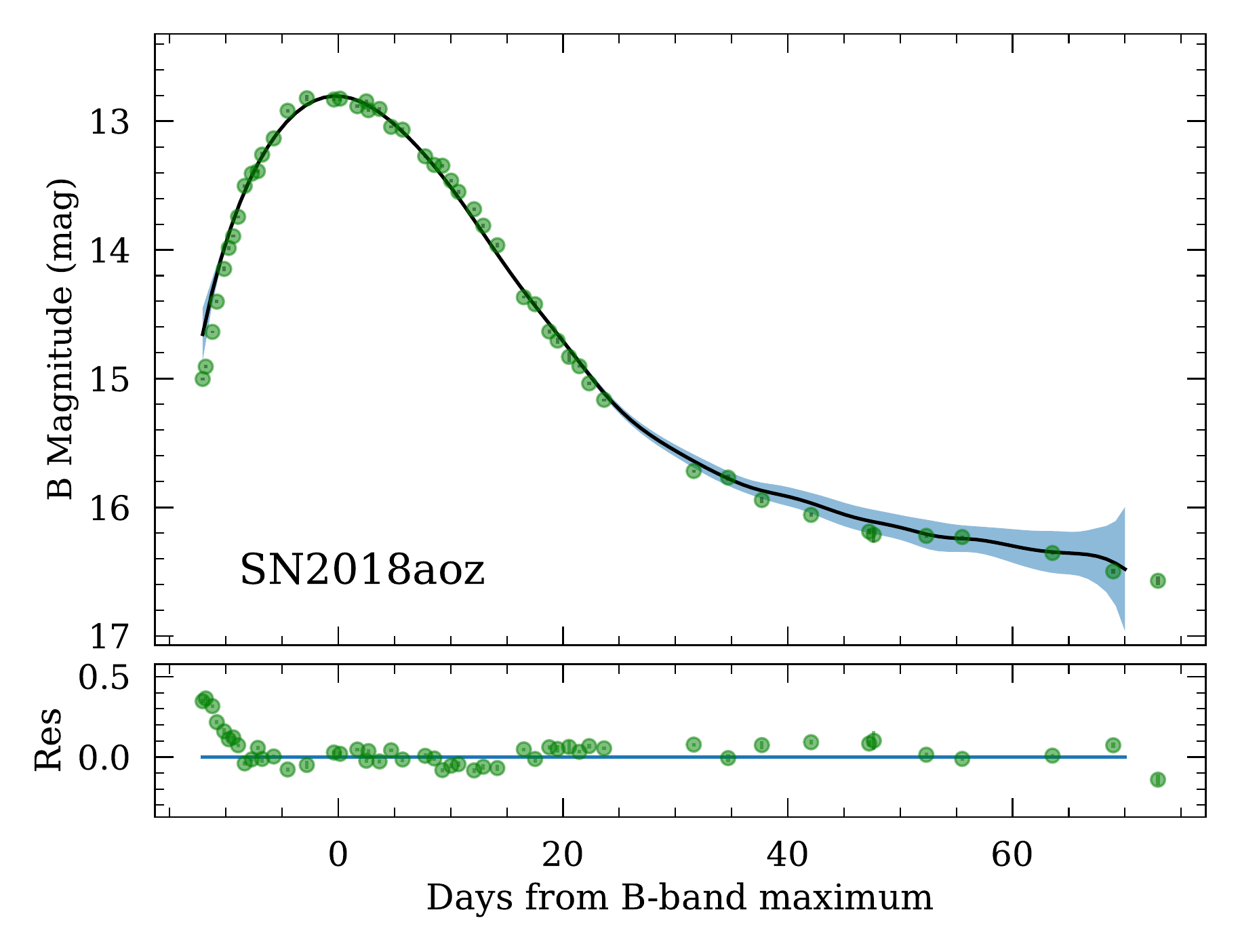} &
\\
\addpic{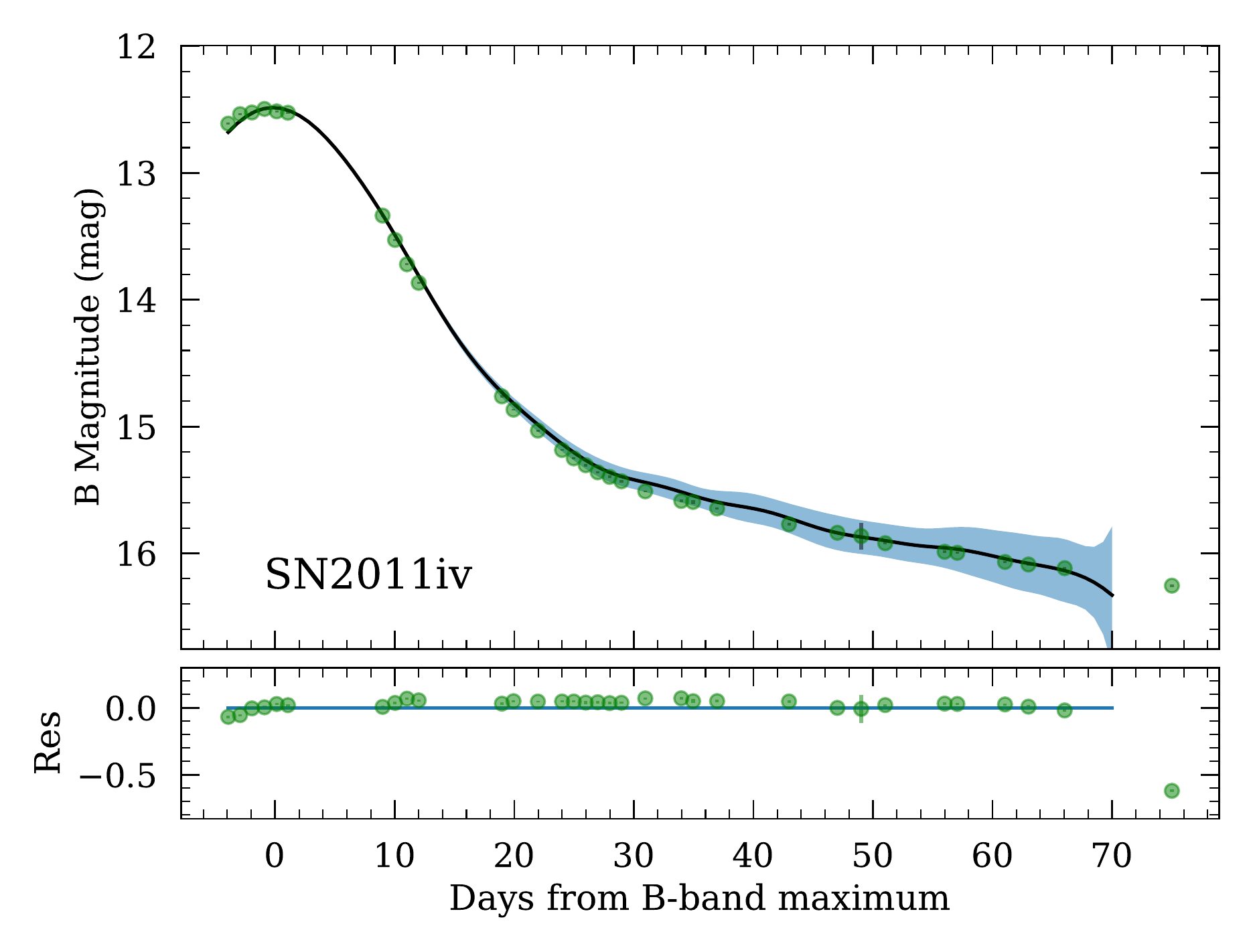} &
\addpic{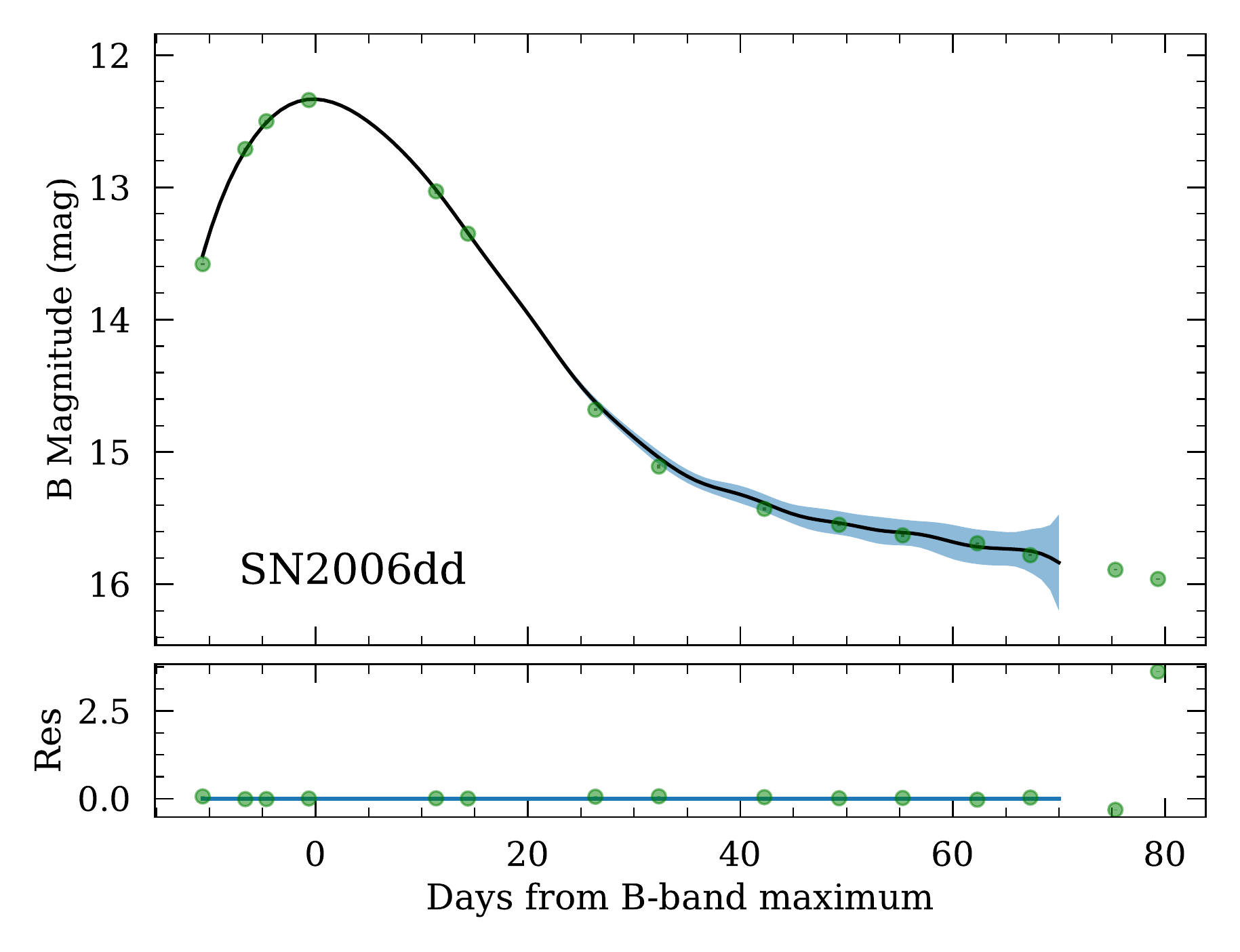} &
\addpic{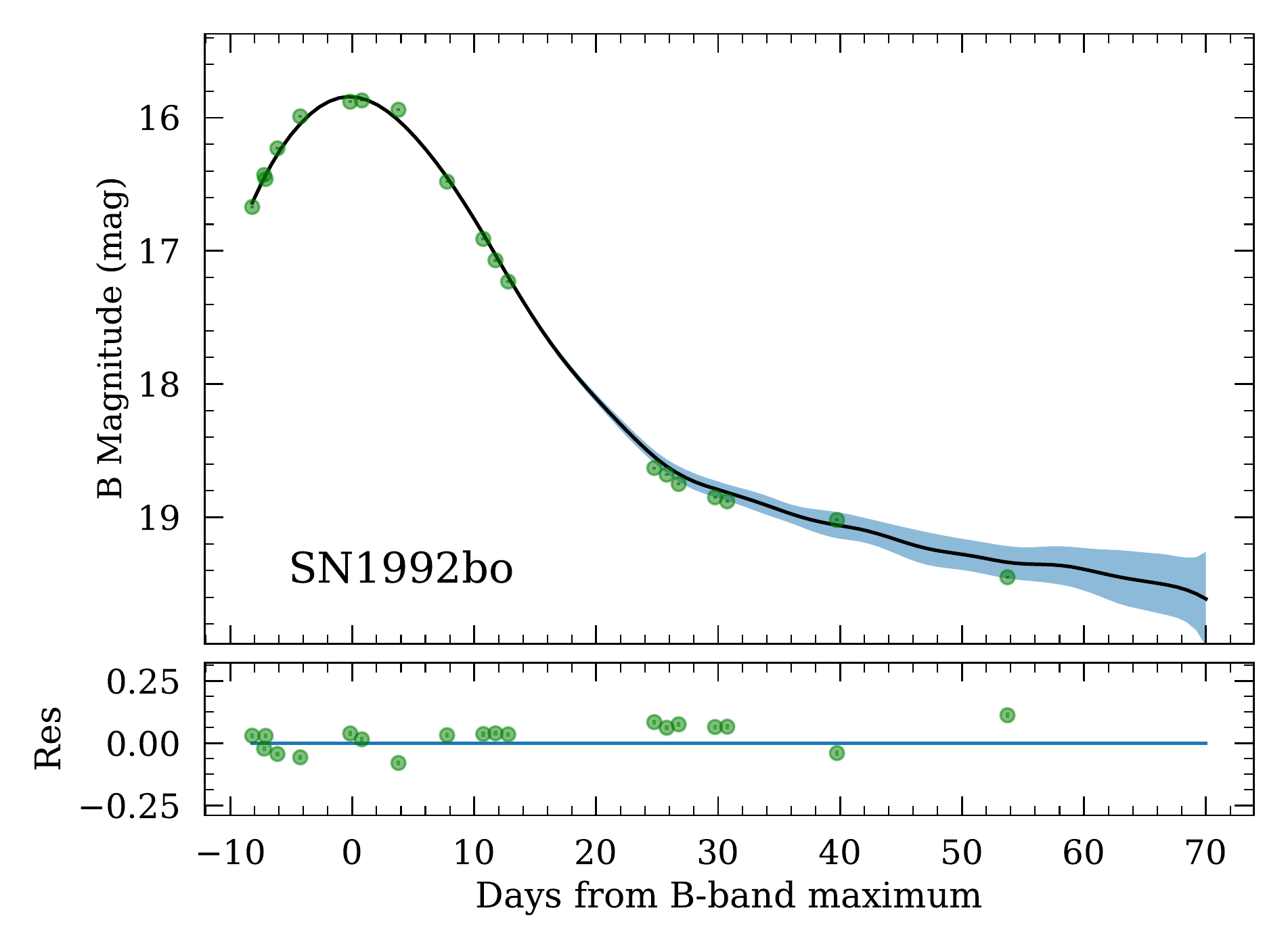} &
\\
\addpic{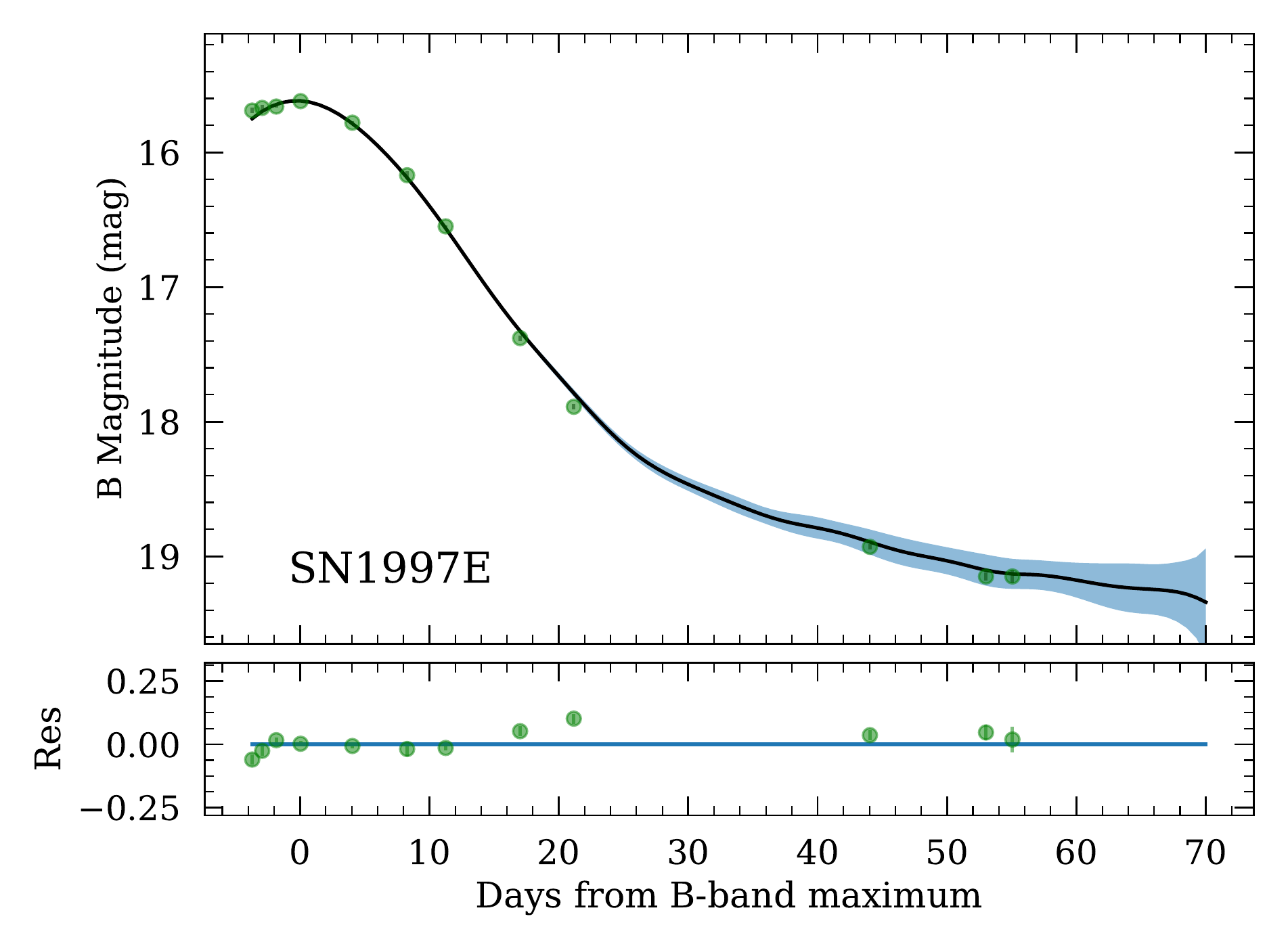} &
\addpic{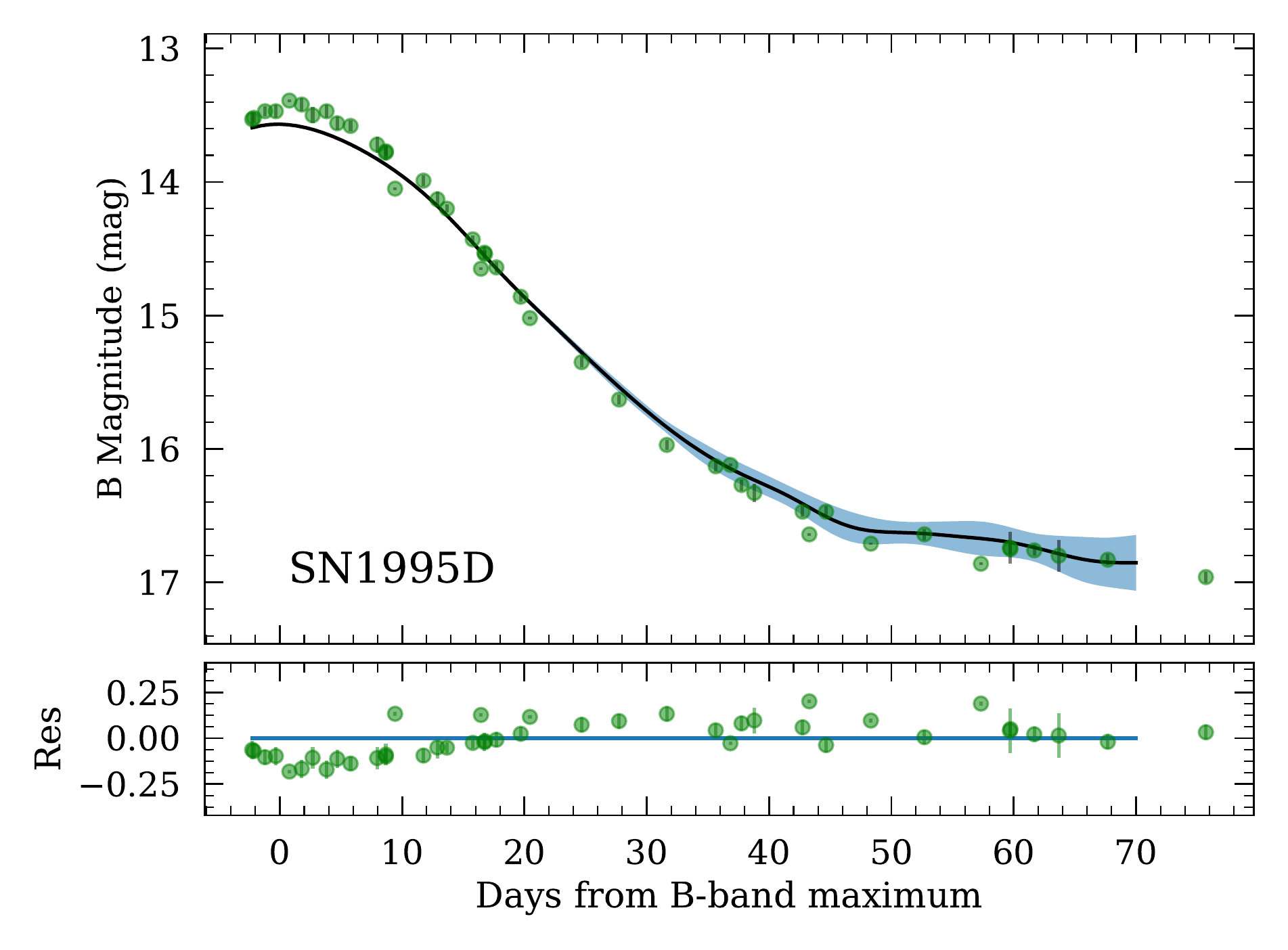} &
\addpic{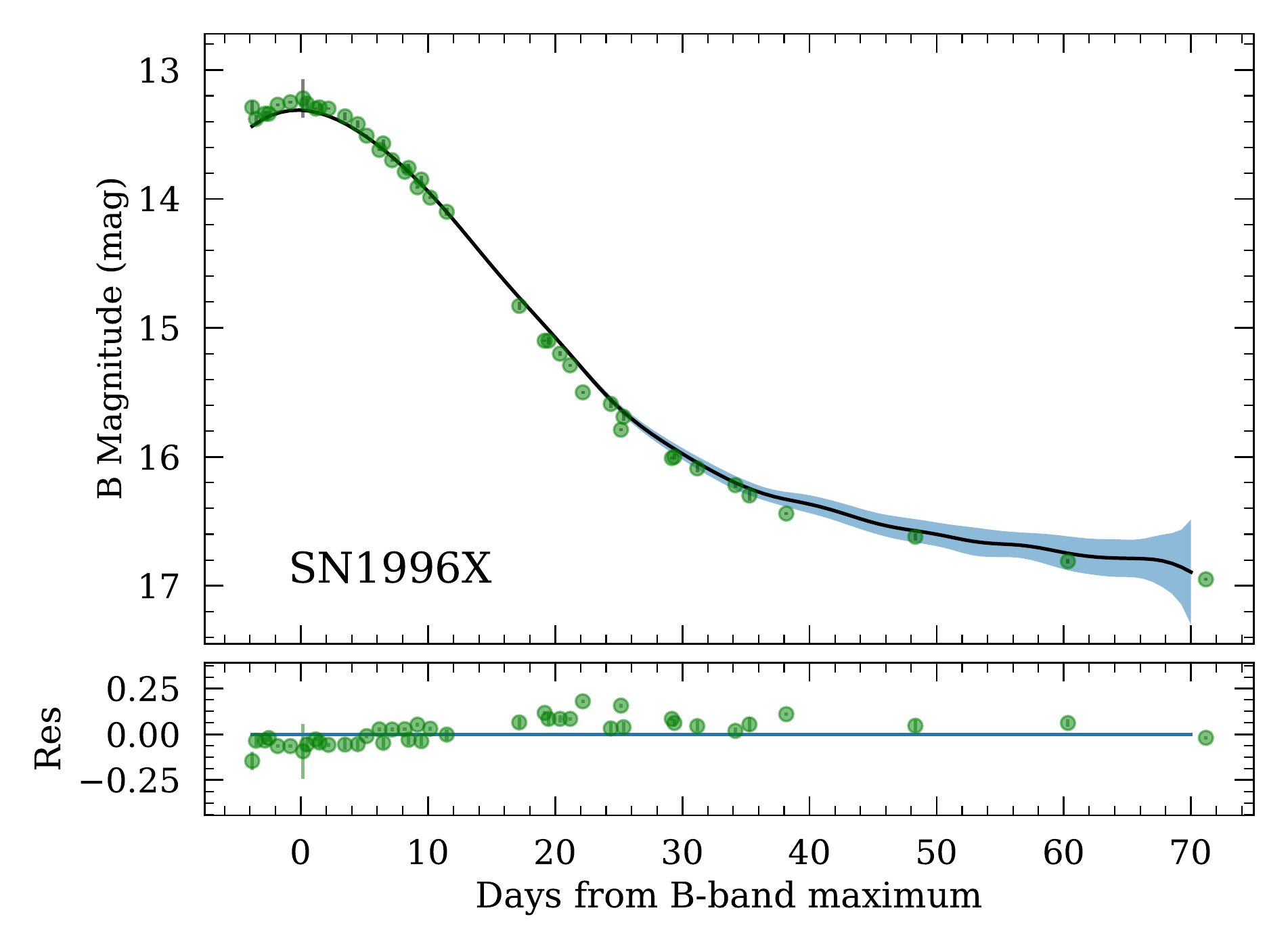} &
\\
\addpic{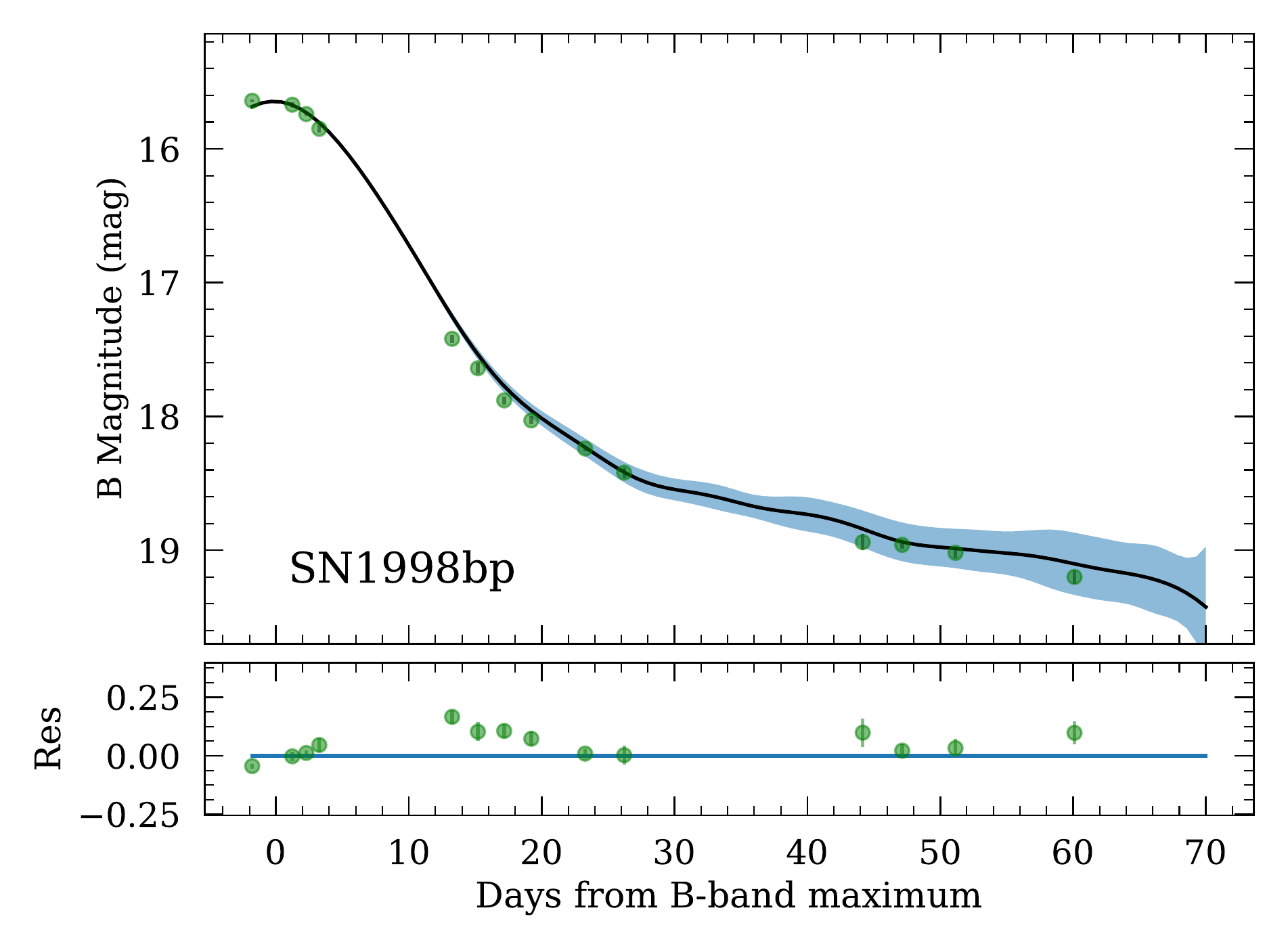} &
\addpic{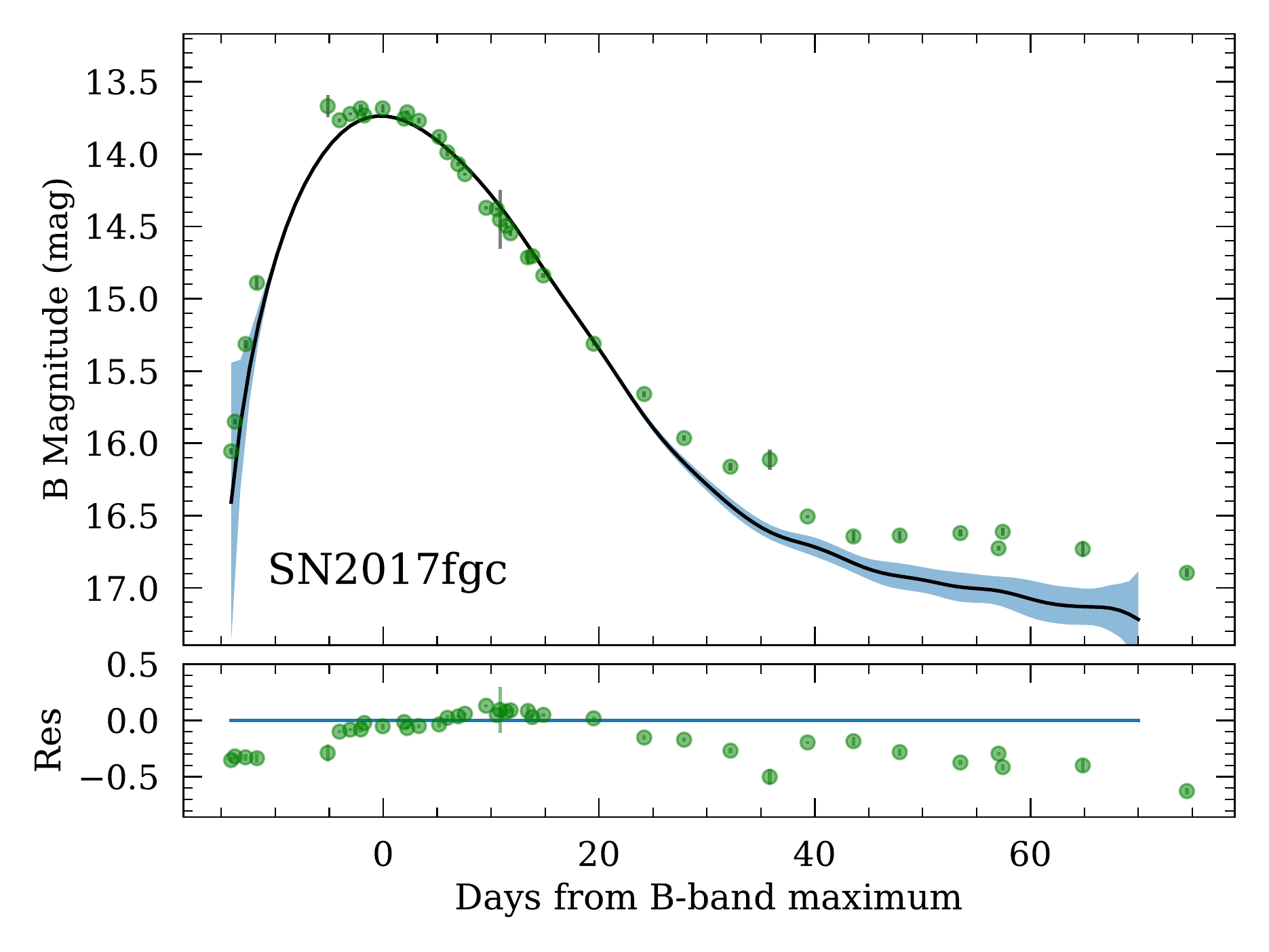} &
\addpic{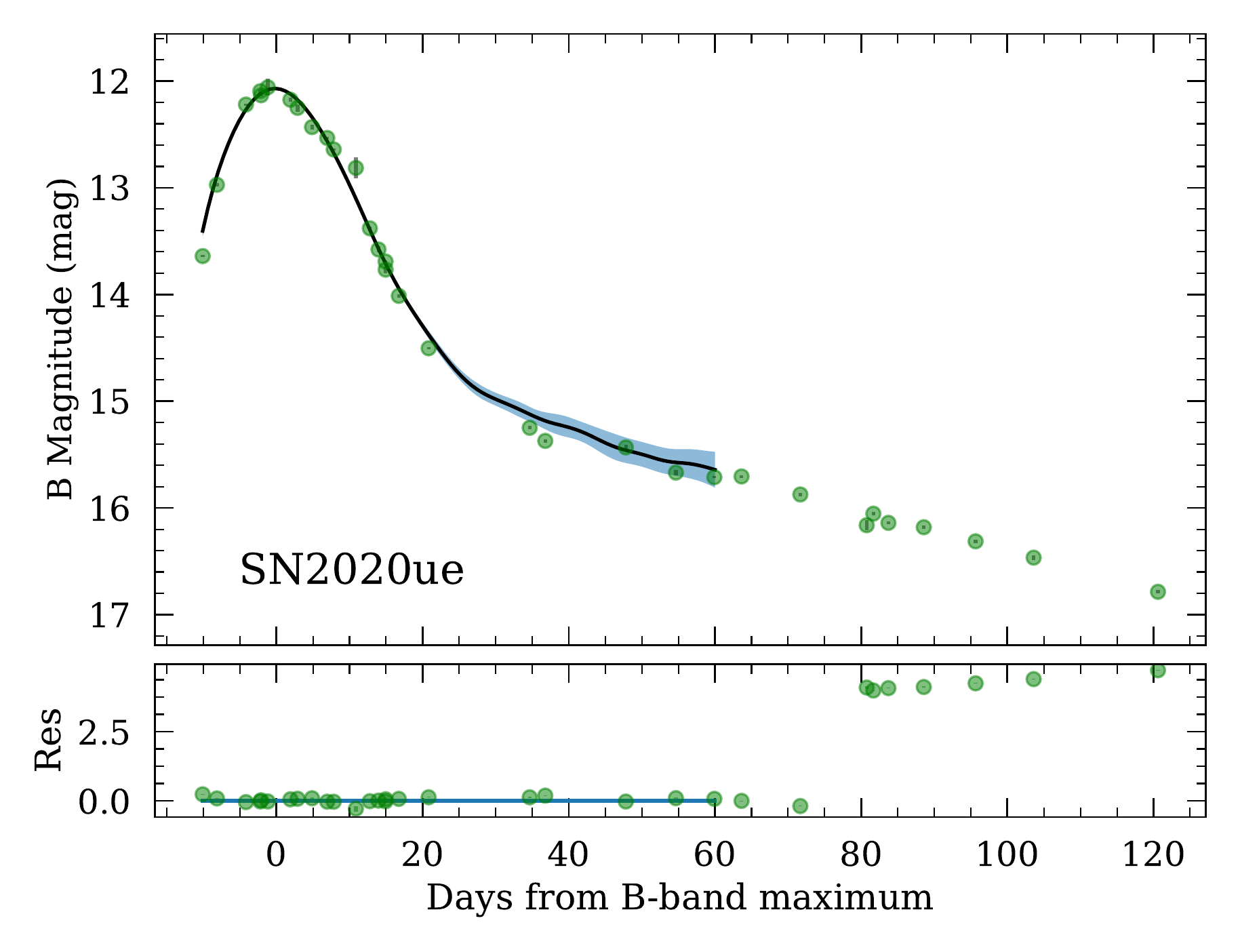} &
%\addpic{lcfits/SN1998bu_Bmag.pdf} &
%\addpic{lcfits/SN2017ejb_Bmag.pdf} &
\\
%\addpic{lcfits/SN2006X_Bmag.pdf} &
%\addpic{lcfits/SN1991bg_Bmag.pdf} &
%\addpic{lcfits/SN1986G_Bmag.pdf} &
%\addpic{lcfits/SN2011eh_Bmag.pdf} &
%\addpic{lcfits/SN2005ke_Bmag.pdf} &
%\addpic{lcfits/SN2003gs_Bmag.pdf} &
\end{tabular}
\end{table*}
\restoregeometry
%=========================================================================
% \newgeometry{top=20mm, bottom=15mm, left = 15mm, right= 15mm}    
% \section{corner plots}
% \label{app:corner_plots}

% \begin{figure*}[h!]
% \centering
% \hspace*{-0.5cm}
% \includegraphics[width=14.0cm,height=14.0cm]{plots/SBF_noHM_corner_12thAug.pdf}
% \caption{Corner plot showing posterior distributions for the parameters $P_0, P_1, R$ and \h0 along with the intrinsic scatters obtained using the SBF sample (24 \sne) with the redshift cut cosmological sample (96 \sne). The title on each histogram shows the median value of the respective posterior distribution. The luminosity correction does not include any dependence in host galaxy stellar mass.}
% \label{fig:corner_noHM_sbf}
% \end{figure*}

% \begin{figure*}[h!]
% \centering
% \hspace*{-0.5cm}
% \includegraphics[width=14.0cm,height=14.0cm]{plots/SHOES_noHM_corner.pdf}
% \caption{Corner plot showing posterior distributions for the parameters $P_0, P_1, R$ and \h0 along with the intrinsic scatters obtained using the \shoes sample with the redshift cut cosmological sample. The title on each histogram shows the median value of the respective posterior distribution. The luminosity correction does not include any dependence in host galaxy stellar mass.}
% \label{fig:corner_noHM_shoes}
% \end{figure*}

\restoregeometry 

%=========================================================================

\section{Host galaxy stellar-mass evaluation}
\label{app:hostmass}

The stellar mass of the host galaxies of the two calibrator samples, SBF and \shoes and the cosmological sample are evaluated using the  approach described in the following. The mass for each galaxy is determined using the 2MASS extended source catalog \citep{skrutskie2006}. We use the $K_{S}$-band magnitude for each galaxy and correct it for the extinction. Then, assuming a constant mass-to-light ratio, the stellar mass of the host galaxy is evaluated using an empirical relation derived by \cite{wen2013}:

\begin{equation}
\begin{aligned} \log_{10}\left(\frac{M{*}}{\mathrm{M}{\odot}}\right) = \ &(-0.498 \pm 0.002)+(1.105 \pm 0.001) \\ & \times \log _{10}\left(\frac{\nu L_{\nu}\left(K_{s}\right)}{\mathrm{L}_{\odot}}\right) \end{aligned} \ 
\end{equation}

where $L_{\nu}\left(K_{s}\right)$ is the $K_{S}$-band luminosity and 1.105 is the mass-to-light ratio. However, the calculation of $L_{\nu}\left(K_{s}\right)$ requires knowledge of the distance modulus of the galaxy and hence it introduces a covarience in host mass with the estimated distances (Hubble residual) and should be dealt carefully. Solving the above equation by translating $L_{\nu}\left(K_{s}\right)$ into $\mu$, one finds that $\log_{10}(M_{\ast}/M_{\odot}) \propto 0.4\mu $, and therefore we include $ 0.4\delta \mu^2$ error in our calibration calculations where $\delta \mu$ is the error on the distance modulus. To estimate the error on the stellar mass, we use the standard error propagation.
For the \shoes calibrator sample, two galaxies, NGC4038 and UGC9391 (corresponding to SN2007sr and SN2003du, respectively) are not in the 2MASS catalog. For their mass calculation we evaluate the magnitude directly from the $K_{S}$-band images. 
We first flag the foreground stars and replace with random neighboring background pixels using imedit in IRAF \citep{tody_iraf}, and then we subtract the sky background. The total flux within an ellipse of appropriate size is measured, which is then converted to stellar mass using the same method and mass-to-light ratio as described above.

\end{appendix}
%\caption{Distance Moduli comparisons of the two local calibrator samples. The top row shows comparisons between directly measured SBF distances ($\mu_{SBF}$, given in  Table~\ref{tab:tabel1}) for the SBF sample, first with the SNe distances estimated using the SBF calibration (left panel) and then with those estimated using the Cepheid calibration (right panel). The bottom row shows similar plots for the \shoes\  sample , comparing the measured cepheid distances (from \citepalias{riess2016}) with the distances estimated using the SBF calibration (left panel) and using the Cepheid calibration (right panel). In all cases we find a one-to-one agreement.}

\end{document}